\documentclass[fleqn,usenatbib]{mnras}

\usepackage{newtxtext,newtxmath}

\usepackage{subcaption}

\usepackage[T1]{fontenc}

\DeclareRobustCommand{\VAN}[3]{#2}
\let\VANthebibliography\thebibliography
\def\thebibliography{\DeclareRobustCommand{\VAN}[3]{##3}\VANthebibliography}

\newcommand{\degree}{$^{\circ}$}
\newcommand{\km}{\mathrm{km}}
\newcommand{\mJy}{\mathrm{mJy}}
\newcommand{\uJy}{\muup\mathrm{Jy}}
\newcommand{\MHz}{\mathrm{MHz}}
\newcommand{\GHz}{\mathrm{GHz}}

\usepackage{graphicx,xcolor,tabularx}	
\usepackage{amsmath}	
\usepackage{amssymb}
\usepackage{bm}
\setcitestyle{citesep={,}}
\usepackage{caption}
\usepackage{subcaption}

\title[Angular Clustering in LoTSS-DR2]{Cosmology from LOFAR Two-metre Sky Survey Data Release 2: \\Angular Clustering of Radio Sources}

\author[C.L.\ Hale et al.]{C.L.\ Hale,$^{1}$ \thanks{E-mail: Catherine.Hale@ed.ac.uk}
D.J.\ Schwarz,$^{2}$
P.N.\ Best,$^{1}$
S.J.\ Nakoneczny,$^{3,4}$
D.\ Alonso,$^{5}$
D.\ Bacon,$^{6}$
L.\ B\"ohme,$^{2}$
N.\ Bhardwaj,$^{2}$\newauthor
M.\ Bilicki,$^{7}$ 
S.\ Camera,$^{8,9,10,11}$
C.S.\ Heneka,$^{12}$
M.\ Pashapour-Ahmadabadi,$^{2}$
P.\ Tiwari,$^{13}$ \newauthor
J.\ Zheng,$^{2}$
K.J.\ Duncan,$^{1}$
M.J.\ Jarvis,$^{3,11}$
R.\ Kondapally,$^{1}$
M.\ Magliocchetti,$^{14}$
H.J.A.\ Rottgering,$^{15}$ \newauthor
and T.W.\ Shimwell$^{16,15}$
\\
$^{1}$Institute for Astronomy, School of Physics and Astronomy, University of Edinburgh, Royal Observatory Edinburgh, Blackford Hill, Edinburgh, EH9 3HJ, UK\\
$^{2}$Fakult\"at f\"ur Physik, Universit\"at Bielefeld, Postfach 100131, 33501 Bielefeld, Germany \\
$^{3}$ Division of Physics, Mathematics, and Astronomy, California Institute of Technology, 1200 E California Blvd, Pasadena, CA, 91125, USA \\
$^{4}$ Department of Astrophysics, National Centre for Nuclear Research, Pasteura 7, 02-093 Warsaw, Poland \\
$^{5}$ Department of Physics, University of Oxford, Denys Wilkinson Building, Keble Road, Oxford OX1 3RH, UK\\
$^{6}$ Institute of Cosmology and Gravitation, University of Portsmouth, Dennis Sciama Building, Burnaby Road, Portsmouth PO1 3FX, United Kingdom \\
$^{7}$ Center for Theoretical Physics, Polish Academy of Sciences, al. Lotnik\'{o}w 32/46, 02-668 Warsaw, Poland\\
$^{8}$ Dipartimento di Fisica, Universit\`a degli Studi di Torino, Via P.\ Giuria 1, 10125 Torino, Italy\\
$^{9}$ INFN -- Istituto Nazionale di Fisica {Nucleare}, Sezione di Torino, Via P.\ Giuria 1, 10125 Torino, Italy\\
$^{10}$ INAF -- Istituto Nazionale di Astrofisica, Osservatorio Astrofisico di Torino, Strada Osservatorio 20, 10025 Pino Torinese, Italy\\
$^{11}$ Department of Physics \& Astronomy, University of the Western Cape, Cape Town 7535, South Africa\\
$^{12}$ Institut f\"ur Theoretische Physik, Fakult\"at f\"ur Physik und Astronomie, Universit\"at Heidelberg, Philosophenweg 16, 69120 Heidelberg, Germany\\
$^{13}$ National Astronomical Observatories, Chinese Academy of Science, Beijing, 100101, P.R.China\\
$^{14}$ INAF-IAPS, via Fosso del Cavaliere 100, 00133, Rome, Italy \\
$^{15}$ Leiden Observatory, Leiden University, PO Box 9513, NL-2300 RA Leiden, The Netherlands \\
$^{16}$ ASTRON, Netherlands Institute for Radio Astronomy, Oude Hoogeveensedijk 4, 7991 PD, Dwingeloo, The Netherlands \\}

\date{Accepted XXX. Received YYY; in original form ZZZ}

\pubyear{2022}

\begin{document}
\label{firstpage}
\pagerange{\pageref{firstpage}--\pageref{lastpage}}
\maketitle

\begin{abstract}
Covering {$\sim$5600$\,\deg^2$} to {rms sensitivities} of $\sim$70$-$100 $\muup$Jy beam$^{-1}$, the LOFAR Two-metre Sky Survey Data Release 2 (LoTSS-DR2) provides the {largest} low-frequency {($\sim$150 MHz)} radio catalogue to {date, making} it an excellent tool for large-area radio cosmology studies. In this work, we use LoTSS-DR2 sources to investigate the angular two-point correlation function of galaxies within the survey. We discuss systematics in the data and an improved methodology for {generating random catalogues}, compared to that used for LoTSS-DR1, before presenting the angular clustering for $\sim$900\,000 sources $\geq$$1.5$ mJy and a peak signal-to-noise $\geq$$7.5$ across $\sim$$80\%$ of the {observed area.} {Using the clustering we infer the bias assuming two evolutionary models.} {When fitting {angular scales of} {$0.5 \leq\theta<5\,\deg$}, {using a linear {bias} model}, {we find LoTSS-DR2 sources are biased tracers of the underlying matter,} with a bias of {$b_{C}= 2.14^{+0.22}_{-0.20}$} (assuming constant bias) and {$b_{E}(z=0)= 1.79^{+0.15}_{-0.14}$} (for an evolving model, inversely proportional to the {growth factor}), {corresponding to} {$b_E= 2.81^{+0.24}_{-0.22}$} at the median redshift of our sample, assuming the LoTSS Deep Fields redshift distribution is representative of our data. This reduces to {$b_{C}= 2.02^{+0.17}_{-0.16}$} and {$b_{E}(z=0)= 1.67^{+0.12}_{-0.12}$} when allowing preferential redshift distributions from the Deep Fields to model our data. Whilst the clustering amplitude is slightly lower than LoTSS-DR1 ($\geq$2 mJy), our study benefits from larger samples and improved redshift estimates.}
\end{abstract}




\section{Introduction}
\label{sec:intro}

The LOw Frequency ARray \citep[LOFAR;][]{vanHaarlem2013} is a key {radio} {telescope array, transforming views of the low-frequency radio skies}. Based in Europe, its full array combines a dense core of stations in the {Netherlands with additional} stations {that have} much larger baselines both across the Netherlands and Europe. This allows baselines of up to $\sim100\,\km$ across the Netherlands and $\sim2000\,\km$ across Europe, {producing 6\arcsec \ resolution using the Dutch stations {only} and sub-arcsecond resolution imaging using the full array {\citep{Morabito2022, Sweijen2022}}, at $150\,\MHz$}. These stations combine two types of antennas to operate in two low frequency ranges: the Low-Band Antennas (LBA; $10-80\,\MHz$) and High-Band Antennas (HBA; $120-240\,\MHz$). Such low frequency observations lead to a large field of view for each LOFAR observation, making it an excellent instrument for survey science. As part of this, LOFAR is {currently focusing on} {several} large-area survey projects{, including}: {the} LOFAR LBA Sky Survey \citep[LoLSS;][]{Lolss} and the LOFAR Two-metre Sky Survey \citep[LoTSS;][]{Shimwell2017, Shimwell2019, Shimwell2022} with the HBA, which is what we use for this work. LoTSS aims to observe the entire northern hemisphere at $144\,\MHz$ to a typical {rms} sensitivity of {$\sigma_{144\,\MHz}\sim70-100\,\uJy\,\mathrm{beam}^{-1}$} and trace a combination of Active Galactic Nuclei (AGN) and Star-Forming Galaxies (SFGs) across large periods of cosmic time. At such frequencies, the dominant {radiative} mechanism {is synchrotron emission from} relativistic electrons spiraling in the magnetic fields. This leads to a typically power-law-like distribution for flux densities as a function of frequency ($S_{\nu} \propto \nu^{-\alpha}$) with a {range of spectral indices, typically assumed to be $\alpha\sim0.7-0.8$} for {an average} radio population \citep{Kellermann1969, Mauch2003, Smolcic2017, deGasperin2018}{, though much larger or smaller values can be observed for individual {sources with} flat or peaked spectra \citep[e.g.][]{Massaro2014,Callingham2017,ODea2021}}.

{LoTSS has developed over a series of data releases, improving {in properties such as} {angular} resolution, sensitivity, {image fidelity} and areal coverage}. Initially{, observations covering} $350\,\deg^2$ {were} released with direction-independent calibration only at a resolution of $25\,\arcsec$, {detecting} $\sim$44 000 sources with a typical noise of $\sim0.5\,\mJy\,\mathrm{beam}^{-1}$. This was then improved upon {in both} resolution and sensitivity with the first fully direction-dependent calibrated data release for LoTSS: LoTSS-DR1 \citep{Shimwell2019}. This data release {covered} $424\,\deg^2$ over the {The Hobby-Eberly Telescope Dark Energy Experiment} (HETDEX) Spring Field {\citep{Hill2008}} {with} a corresponding catalogue of $\sim$325 000 sources, with a {1$\sigma$} sensitivity of $\sim70-100\,\uJy\,\mathrm{beam}^{-1}$ {at 6\arcsec \ angular resolution}. This sky coverage has now been {enlarged in} the latest data release, LoTSS-DR2 \citep{Shimwell2022}, which covers $\sim5600\,\deg^2$ with an accompanying catalogue of $\sim4.4$ million sources. This is the largest catalogue {of radio sources} within an individual radio survey to date. Such a combination of area and large source numbers means that LoTSS-DR2 provides an excellent dataset for radio cosmology studies, {allowing} for a more detailed understanding of the distribution of radio sources {in the Universe.} 

{The study of the {distribution of sources observed in galaxy surveys throughout} the Universe is important for a number of reasons. Most importantly, it allows us to understand more about how galaxies trace the large-scale structure of the {Universe and the underlying dark matter distribution. Starting from initial primordial over-densities, dense regions of matter have {come together and} evolved over {time.}  This has {resulted in} the large-scale distribution of matter we observe today }\citep[][]{Colless2001, Doroshkevich2004, Springel2006}. This coming together {of dark matter forms} {haloes in these initially {over-dense} regions, and leaves an absence of dark matter, known as voids, in regions of initial under-densities}. Filaments then connect dense regions together. {Luminous} matter, that we observe in astrophysical objects such as stars and galaxies, is also attracted together under the effects of gravity but {is} further influenced by factors such as the {effect of feedback} associated with both star formation and from active galactic nuclei {\citep[see e.g.][]{Ceverino2009, Hopkins2012, Fabian2012, Morganti2017}}. {Since galaxies form in dense regions, they trace peaks in the underlying matter distribution, leading galaxies to be known as biased tracers of the matter distribution in the Universe} \citep[see e.g.][]{Peebles1980,Kaiser1984, Mo1996, Desjacques2018}.}

{On large scales, the galaxy overdensity, {$\delta_g({\bf x},z)$}, can be considered to trace the matter overdensity, {$\delta_m({\bf x},z)$}, related by a quantity known as ``galaxy bias'', $b(z)$:}
\begin{equation}
  \delta_g({\bf x},z)=b(z)\,\delta_m({\bf x},z).  
\end{equation}
{To quantify galaxy bias, a common method is {to first determine} the excess probability to observe galaxies within different spatial separations, compared to if they were randomly distributed. This is known as the spatial two-point correlation function, $\xi(r, {z})$. The redshift dependent linear bias, $b(z)$,} can {then} be measured and is related to the ratio of spatial clustering of galaxies, {$\xi(r, {z})$,} to the {clustering of matter, {$\xi_{M}({r, z})$}}, as given by:
\begin{equation}
b^2{(z)} = \frac{\xi_{{g}}(r,z)}{\xi_{{M}}(r,z)}.
\label{eq:bias}
\end{equation}
{The spatial clustering of galaxies, $\xi_g(r)$, defines} the excess clustering of galaxies observed at a given spatial separation, compared to if they were randomly distributed. {Such} measurements of the spatial clustering rely on accurate redshifts {and corrections due to peculiar velocities}. {Where highly accurate redshifts are not available} for sources in a survey, it is still possible to estimate the spatial clustering {by} combining the observed projected angular clustering of sources with their redshift distributions using methods such as Limber inversion {\citep{Limber1953, Limber1954}}. Radio surveys provide excellent catalogues to measure the large-scale structure of the Universe {as they predominately trace extragalactic sources over a broad redshift range and over large areas, but typically rely on angular clustering measurements instead of spatial measurements}. 

{The angular two-point correlation function \citep[$\omega(\theta)$, see e.g.][]{Totsuji1969, Peebles1980, Cress1996, Blake2002, Overzier2003, Wang2013} {does not rely on redshifts for its calculation and} quantifies the excess probability ($dP$) of {pairs of} sources observed within a survey catalogue at a given {projected angular separation, $\theta$,} compared to if the sources were randomly distributed on the sky, with no intrinsic large-scale structure. {This is defined by:}}

\begin{equation}
    dP = N \left[ 1+\omega(\theta) \right] d\Omega, 
\end{equation}

\noindent {where $d\Omega$ is {the solid angle of the observations} and $N$ is the mean {number of} sources per unit area. }

{Radio continuum surveys rely on multi-wavelength information for redshifts \citep[see e.g.][]{Smolcic2017b, Prescott2018, Algera2020}, which are typically {dominated by} less accurate photometric redshifts for a large fraction of the sources}. {For LOFAR, in the first LoTSS data release \citep{Shimwell2019}, sources were cross-matched to sources in surveys such as Pan-STARSS \citep{PanSTARRS} and WISE \citep{Wright2010,Williams2019}, with $\sim$50\% of {LoTSS-DR1} sources having redshift information \citep[see][]{Duncan2019}.} {Similarly for the LoTSS Deep Fields, the wealth of multi-wavelength data has been used to obtain redshifts for 97\% of sources across the multi-wavelength defined regions in the three fields {LoTSS} Deep Fields \citep[see][]{Sabater2021, Tasse2021, Kondapally2021, Duncan2021} which was used to help classify such sources \citep[see][]{Best2023}. {The accuracy {of redshifts} for such {radio} sources will be improved upon with future spectroscopic surveys \citep[such as WEAVE-LOFAR;][]{Smith2016}. }}

{Combining measurements of the angular clustering and redshift distribution, {the spatial clustering for a population of sources can be inferred. {The spatial clustering can then be used to estimate {the} galaxy bias} of radio sources (as in Equation \ref{eq:bias}), {this will be discussed further in Section \ref{sec:bias}.} Such clustering and bias measurements have been presented in a number of works \citep[see e.g.][]{Magliocchetti1999, Magliocchetti2004, Negrello2006,Lindsay2014,Nusser2015,Magliocchetti2017, Hale2018, Siewert2020, Tiwari:2021, Mazumder2022}}. {A number of such studies suggest an evolving bias model for radio sources, suggesting radio sources are more biased tracers of the underlying matter distribution at higher redshift. Moreover, studies which further consider the bias for radio SFGs and AGN separately have shown that these sources have different bias distributions and trace different mass haloes \citep[see e.g.][]{Magliocchetti2017, Hale2018, Chakraborty2020,Mazumder2022}. Such studies have shown that AGN appear to inhabit more massive haloes than for SFGs at similar redshifts, reflecting the fact that they preferentially inhabit massive ellipticals. Further studies which classify AGN suggest that the haloes hosting radio AGN may be related to the accretion mode of AGN \citep[using high redshift analogues to high/low excitation radio galaxies, see][]{Hale2018}}. {Such differences in the bias of different source populations can be advantageous for cosmological analysis, using the multi-tracer techniques \citep[see e.g.][]{Ferramacho2014, Raccanelli2012, Gomes2020}. These techniques require understanding of the bias evolution for different source populations and make use of such difference to help place constraints on, for example, non-Gaussianity.}}

{Further cross-correlating radio data with other cosmological tracers \citep[see e.g.][]{Allison2015, Alonso2021} {can also help} remove some of the systematics which remain in the data and have added further constraints on the galaxy bias evolution of radio sources, and \cite{Alonso2021} further used this to place constraints on the redshift distributions for radio sources, where no redshift information was available.}  {Measurements of bias have been used in numerous studies to {relate} such measurements to the typical mass of the dark matter haloes which are hosting such sources \citep[see e.g. those described in][]{Mo1996, Tinker2010}, but there are caveats to such measurements, {especially if full halo occupation models are not taken into account} \citep[see e.g.][]{Aird2021}.} 

In this paper, we investigate the angular clustering of radio sources within $\sim4500\,\deg^2$ of the LoTSS-DR2 survey and {use this to infer} the average bias of LoTSS-DR2 sources. The paper is arranged as follows: in Section \ref{sec:data} we describe the LoTSS-DR2 data used in this {analysis,} as well as the methods to measure the angular clustering of radio galaxies in Section \ref{sec:methods}. This includes a detailed description of the methods used {in order} to obtain accurate random sources that mimic the distribution of observational biases across the field of view, which {develops} the techniques used for LoTSS-DR1 \citep{Siewert2020}. Then, in Section \ref{sec:tpcf_results} we present our measurements of the angular clustering of sources and our validation of these measurements {before presenting our methods to determine galaxy bias in Section \ref{sec:bias}. {This allows us to place constraint on how such sources trace the underlying matter and dark matter haloes across cosmic time.} We then discuss our results in Section \ref{sec:discussion}.} We then go on to  draw final conclusions in Section \ref{sec:conclusions}. For this paper we assume standard cosmological parameters from \cite{Planck2020} in a flat model Universe, specifically: {$H_0=67.4$ km s$^{-1}$ Mpc$^{-1}$, $\Omega_b$=0.0493, $\Omega_c$=0.264, $\Omega_m = \Omega_b+\Omega_c$, $\Omega_{\Lambda}=1-\Omega_m$, $n_s$=0.965, $\sigma_8$=0.811, unless otherwise stated.}

\section{Data}
\label{sec:data}
For this work we make use of the data and associated data products from two LOFAR survey projects: (i) the large area LoTSS-DR2 {survey} {\citep{Shimwell2022}} and (ii) {the associated redshift information from sources in the smaller {LoTSS Deep} fields \citep{Duncan2021}.} 

\subsection{LoTSS-DR2}
\label{sec:data_lotss}
\begin{figure*}
    \centering
    \includegraphics[width=12cm]{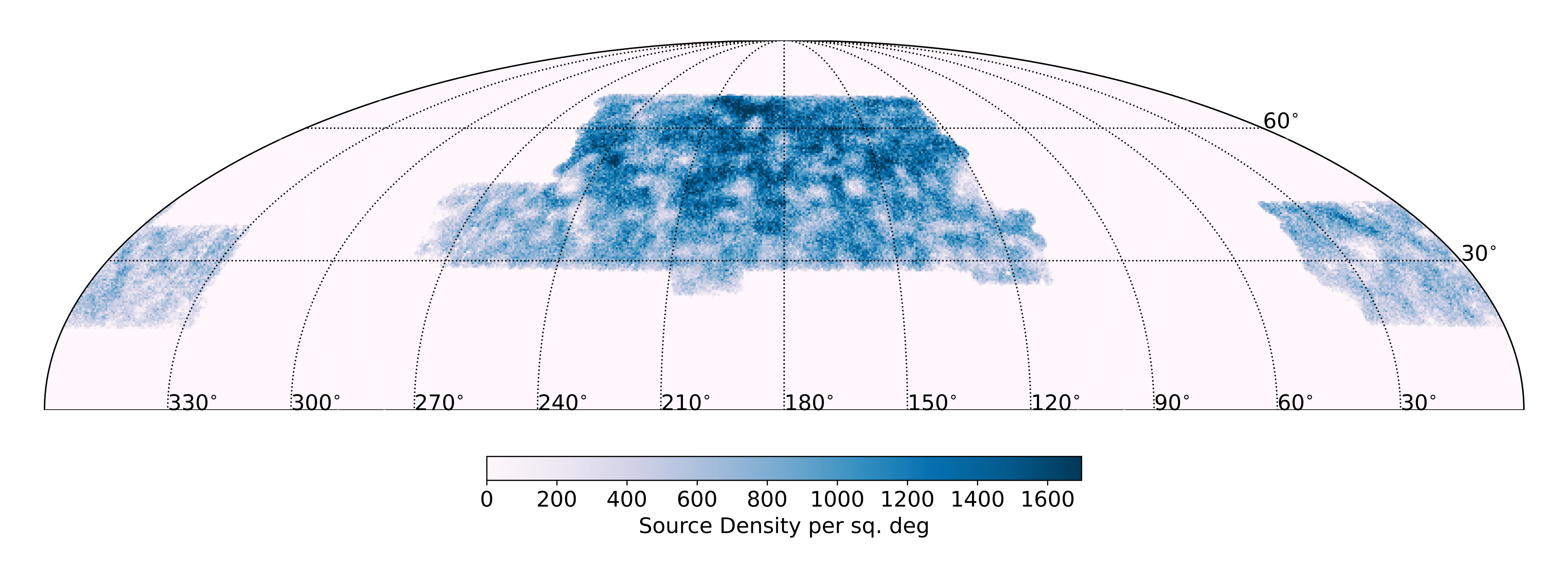}
    \includegraphics[width=12cm]{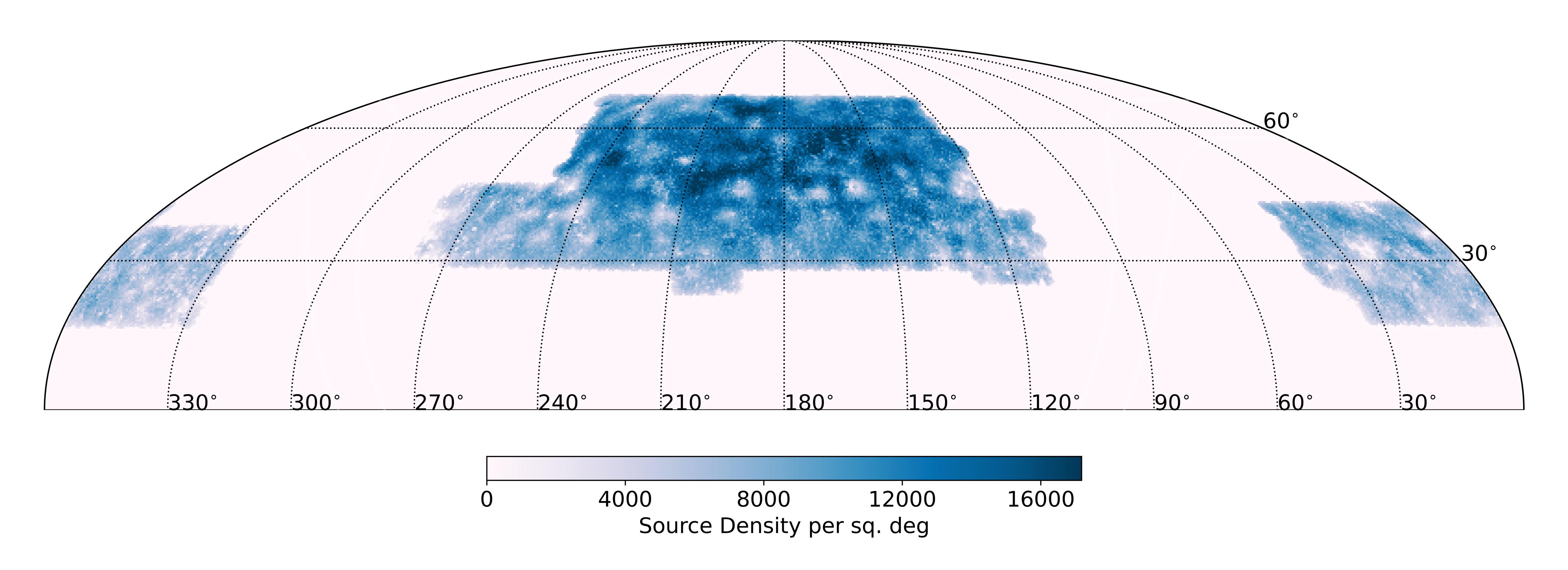}
    \caption{{Sky {density} distribution of all sources in the LoTSS-DR2 survey (upper) from \protect \cite{Shimwell2022} and for the random catalogues generated for this work (lower; prior to any flux density, SNR or spatial cuts). This shows the two large regions covered by the survey, centred on right ascensions of 1h (15 $\deg$) and 13h (195 $\deg$). The figure is plotted in the {Mollweide} projection using HealPix \protect \citep{healpix, healpy} {with an N$_{\textrm{side}}$=256}. The colour scale indicates the source density per sq. $\deg$ across the field of view.} }
    \label{fig:data_distribution}
\end{figure*}

The majority of data {used} in this work {consists of images and catalogues from the {mosaics generated from combining 841 individual pointings} of LoTSS-DR2 \citep{Shimwell2022} covering $\sim$5600$\,\deg^2$ over two regions}. The first of these is centered at {13h in RA, covering $4178\,\deg^2$, and the second region is centred at an RA of 1h, covering $1457\,\deg^2$}. {The data were} reduced {in a two stage process} which consists of both a direction-{independent} and a direction-dependent calibration pipeline. The {former} flags, calibrates and averages the data in order to reduce the large data volumes, {whilst the latter does further calibration and imaging to account for direction-dependent effects}. This includes the effect of the varying ionosphere across the field of view, which is more prominent at the {observing frequencies that telescopes such as LOFAR operate at}, compared to higher-frequency radio observations. As presented in {works such as} \cite{vanWeeren2016, Williams2016, Shimwell2019, Tasse2021}, such direction-dependent calibration {of LOFAR data} is crucial for improving image fidelity {and for} producing higher resolution imaging of the field {at 6\arcsec \ angular resolution}, compared to 25\arcsec \ without this accounted for \citep[see e.g.][]{Shimwell2017}, {when using {only the} Dutch LOFAR stations}. {Source catalogues were} {generated} using the source finder \textsc{PyBDSF} \citep{Mohan2015} which detected a total of $\sim$4.4 million sources {across the full LoTSS-DR2 coverage}. The distribution of these sources over the northern hemisphere can be seen in Figure \ref{fig:data_distribution}. {This distribution varies significantly across the field of view due to a combination of factors. {These include} intrinsic large-scale structure, {and non-uniform} detection across the field of view resulting from instrumental, calibration and source finding effects}. {Understanding the factors which cause such non-uniformity in the data is important} in order to accurately measure the {true} angular clustering of sources and will be discussed further in Section \ref{sec:randoms}. {Unless otherwise stated, any mention of images and pointings from LoTSS-DR2 refer to the mosaic images which are available from \url{https://lofar-surveys.org}, and are the mosaiced region closest to the pointing centre. }

\subsection{LoTSS Deep Fields}
\label{sec:data_deepfields}

{In} order to relate any observed angular clustering to the spatial clustering and bias, it is crucial to have knowledge of the redshift distribution of the sources within the field. As there are not direct measurements {of redshifts for the full population of LoTSS-DR2 sources\footnote{{Redshifts for a number of sources will be available in the value added catalogue of {\cite{Hardcastle2023}} which is cross-matching sources $\geq$4 mJy, to ensure accurate host positions for source $\geq$8 mJy. However, there will be significant incompleteness compared to the full population of sources used in this work.}}} we make use of the {LoTSS} Deep Fields data \citep{Sabater2021, Tasse2021} {which targets {a handful of} fields in the northern hemisphere with an abundance of multi-wavelength data, {these are observed to deeper sensitivities} than in LoTSS-DR2. {Observations} within these fields are important to help} infer the redshift distribution of the sources observed within {LoTSS-DR2}. The first {LoTSS} Deep Fields data release consisted of three fields: Bo\"otes, Lockman Hole {and the European Large-Area ISO Survey Northern Field 1 (ELAIS-N1) field}. These were observed for a total of 80, 164 and 112 hours {respectively, covering $\sim20\,\deg^2$ in each field}. 

For each field, a smaller region was defined for which there exists deep multi-wavelength information. In such regions, the source catalogues from \textsc{PyBDSF} were cross-matched to {host galaxies} \citep{Kondapally2021} using a wealth of ancillary data. This cross-matched area constituted a total area of $8.6\,\deg^2$ in the Bo\"otes field, $6.7\,\deg^2$ in ELAIS-N1 and $10.3\,\deg^2$ in the Lockman Hole field, totalling $25.6\,\deg^2$ across the three fields. For the cross-matched sources, a redshift was also associated to the source using a {combination of} template fitting {to the multi-wavelength data} as well as machine learning  methods in order to obtain probability density functions (PDFs) {for the redshift distributions, denoted $p(z)$}. A `best redshift' was then assigned to each source based on the PDF, or a spectroscopic redshift if such was available for the sources. More detail on this can be found in \cite{Duncan2021}. {We use these redshift distributions} to estimate {the} redshift distribution, $p(z)$, for sources in the {wider} LoTSS-DR2 survey. {This will be} discussed further in Section \ref{sec:nz}.

\section{Angular clustering and Randoms Generation}
\label{sec:methods}
\subsection{Angular Clustering}
\label{sec:angclus}
As discussed in Section \ref{sec:intro}, one way to investigate the clustering of sources within a galaxy catalogue is through measuring the angular two point correlation function (TPCF) {, denoted by $\omega(\theta)$}. The TPCF quantifies the excess clustering observed at a given angular separation in the catalogue data, compared to what would be observed over the field of view if there was no large-scale structure within the {data.} Naively, such excess probability to detect galaxies in the data at a given angular separation compared to the distribution from random sources is given by:
\begin{equation}
    \omega(\theta) = \frac{\overline{{DD}}(\theta) }{\overline{{RR}}(\theta)} -1 .
    \label{eq:tpcf1}
\end{equation}
\noindent In this estimator, $\overline{DD}(\theta)$ is the counts of pairs {of galaxies within the data catalogue at a given angular separation $\theta$ (normalised such that $\Sigma_{\theta} \overline{DD}(\theta)=1$)} and $\overline{RR}(\theta)$ is the corresponding normalised pair counts within a random catalogue. This random catalogue is generated to mimic observational {effects} across the field of view. If the data {were} indeed randomly distributed and exhibited no large-scale structure behaviour, $\omega(\theta)$ would {fluctuate} around a value of 0. Any deviation from this suggests intrinsic large-scale structure. {A number of predictions for galaxies as well as observations have suggested that this angular clustering behaves as a power law for galaxies and specifically radio sources \citep[see e.g.][but see Section \ref{sec:tpcf_results}]{Peebles1980, Blake2002, Lindsay2014, Magliocchetti2017}}. Whilst Equation \ref{eq:tpcf1} could be used to estimate $\omega(\theta)$, work by \cite{Landy1993} has shown that a more accurate estimator of $\omega(\theta)$ is given by
\begin{equation}
    \omega(\theta) = \frac{\overline{{DD}}(\theta) - 2 \overline{{DR}}(\theta) + \overline{{RR}}(\theta)}{\overline{{RR}}(\theta)}.
    \label{eq:LS}
\end{equation}

\noindent In this estimator, $\overline{DR}(\theta)$ is the corresponding normalised pair counts between the data and random catalogues within a given angular separation. {This} estimator has been shown to {have minimal variance and} be less biased {than other estimators such as Equation \ref{eq:tpcf1}} \citep[see][]{Landy1993}. {As such, we use Equation \ref{eq:LS} to calculate $\omega(\theta)$ in this work. }

{To calculate $\omega(\theta)$}, a random catalogue must {first} be generated {to compare to the data}. If {source} detection across the field of view were uniform, such {a random catalogue} could be generated through sampling random positions across the observed field of view. However, {{the} detection of sources is not uniform (see Figure \ref{fig:data_distribution}) and will be affected by a number of {observational} effects across the sky.} {Thus}, the generation of randoms which accurately mimic the detection of sources across the sky is crucial {to avoid observational effects being mistaken} for intrinsic large-scale structure. {We therefore} employ a number of methods (discussed in Section \ref{sec:randoms}) to mimic such observations across the field of view.

To measure $\omega(\theta)$, we make use of the package \texttt{TreeCorr} \citep{TreeCorr} to calculate the pairs of galaxies within angular separation {bins that are uniformly spaced bins in $\ln(\theta)$ and cover the range of angular scales possible with the data}. Due to the large area coverage of LoTSS-DR2, we ensure that the metric for calculating separations within \texttt{TreeCorr} is set to \texttt{`Arc'}. {This helps to more accurately calculate separations across large fields of view, {using great circle distances}. We also set the parameter \texttt{bin\_slop} to $0$ which enforces that exact calculations are made to calculate the number of pairs of sources within each angular separation bin, as opposed to the default method which has some flexibility between the separation bins in order to help speed up the calculation of pairs}. Such parameters were determined to be important in the work of \cite{Siewert2020}, {where} a non-zero \texttt{bin\_slop} {was} found to introduce larger errors in the measurement of {$\omega(\theta)$. The} associated uncertainties in $\omega(\theta)$ will be discussed in greater detail in \ref{sec:jack} {and its connection to linear bias also discussed in Sections \ref{sec:b_pyccl}-\ref{sec:limber}}.
 
\subsection{Randoms}
\label{sec:randoms}
As discussed in Section \ref{sec:angclus}, in order to measure the angular clustering {from LoTSS-DR2 we need to have a catalogue of random sources which {mimics} the detection of data across the field of view.} Figure \ref{fig:data_distribution} {highlights} {the non-uniform detection of radio sources {across the field of view}, due to a combination of factors including {sensitivity variations across the field of view due to bright sources, reduced sensitivity with declination and smearing of points sources across the field of view. In building our random catalogue we will take a series of steps to account for these effects. An outline of these steps, as well as the section in which these shall be applied is as follows:}}

{\begin{enumerate}
         \item \textit{Survey Area}  - we generate randoms across the survey {field} of view, ensuring we remove any masked regions within pointings which are masked out due to failures within the data reduction process. We consider this in Section \ref{sec:randoms_input}.
  	 \item {\textit{Smearing} - There may be position-dependent smearing effects} across the field of view of a pointing, as well across the 5600 sq. deg. Smearing will affect the detection of sources (which is based on {signal-to-noise ratio `SNR', defined {here as peak} flux density/rms{ (root mean square noise), for which the \texttt{Isl\_rms} column is used for rms of the data}}\footnote{{For the randoms, we use the pixel rms value at the source centre. Using a central rms value for the data makes a negligible difference to the number of sources when the final flux density and SNR cuts are applied are described in Section \ref{sec:additional_constraints}}}), {and could arise from effects such as residual calibration uncertainties and uncorrected smearing effects inherent to the data averaging. We model smearing across the field of view and its dependence on field elevation and correct for this, which is discussed in Section \ref{sec:smearing}.}
  	  \item \textit{Incompleteness and measurement errors -} The sensitivity {(rms) will vary across the {survey area}, such as with elevation or declination} \citep[see Fig. 2 of][]{Shimwell2019} {or location within the mosaic and proximity to} bright {sources, where the noise is known to be elevated}. {Variations may also exist} towards the edge of the field, where {there are fewer neighbouring pointings that can be mosaiced together} (as mosaicing would reduce the noise). This will affect source detection and hence the completeness. Furthermore, the source finder may have a completeness dependence with SNR and its measurement errors can affect the properties such as flux density associated with sources. We account for completeness as a function of source input SNR and the effect that noise and the source finder may have on the measured flux properties of sources in Section \ref{sec:comp_meas}.
         \item \textit{Additional spatial masking - } Finally, there may be additional spatial regions which should be masked to avoid regions such as the unmosaiced edges of pointings; this is described in Section \ref{sec:mask}.
\end{enumerate}}

{We note, though, that there may be limitations to generating the randoms which may be more challenging to account for, especially over the large area of LoTSS-DR2. This includes residual primary beam uncertainties which are unknown and that mosaicking pointings together may cause additional smearing which can very spatially due to pointing dependent astrometric offsets. To minimise the effects of these, additional flux limit and SNR limits can be applied to both the data and random samples. Specifically, for our final analysis we limit the sample to $\geq$1.5 mJy and $\geq$ 7.5$\sigma$. We discuss these and additional cuts in Sections \ref{sec:additional_constraints} - \ref{sec:finaldata}.}

\subsubsection{Input Simulated Catalogue}
\label{sec:randoms_input}
The first step in generating accurate random catalogues for the LoTSS-DR2 survey is to generate a sample of input positions which are uniformly {distributed} across the field of view of LoTSS-DR2, accounting for {masked regions} within the fields. For this work, we generated random positions in the range: RA from 0\degree\ to 360\degree\ and Dec from 20\degree\ to 80\degree. This wide area encompasses the full LoTSS-DR2 {footprint}, but a significant fraction of such a region is not covered by LoTSS-DR2. Therefore, we use the associated rms maps of each individual pointing {to identify the} sources within the LoTSS-DR2 area. {We assign each random position an rms value, based on the pixel value at the source location, using the rms map for the closest pointing. This also allows sources within masked regions, or regions not surveyed in LoTSS-DR2 to be identified}. Random sources falling within the {surveyed} region are retained and consist of $\sim200$ million input simulated positions across the field of view of LoTSS-DR2. 

{To account} for sensitivity variations and the effect {that this has} on the detection of sources, we take a number of iterative steps. Firstly, we assign simulated properties of radio sources to each of the $\sim200$ million random positions. Such properties include the  flux density of the simulated source, as well as source shape information. To do this, we make use of the SKA Design Studies Simulated Skies \citep[{hereafter SKADS}][]{Wilman2008, Wilman2010}, which provide a simulated catalogue of sources covering $100\,\deg^2$ {with} multiple observable properties {for each simulated} source. These properties include an associated redshift, flux density measurements at several frequencies in the range $151\,\MHz - 18\,\GHz$, shape information and source type (e.g.\ AGN or SFG). {Recent observations suggest that SKADS underestimated the number of SFGs at the faintest flux densities} \citep[see e.g.][]{Bonaldi2016,Smolcic2017, vandervlugt2021, Matthews2021, Hale2023, Best2023}. Therefore, we employ a modified version of the SKADS catalogue {where the number of SFGs in the original catalogue are doubled, as also done in \cite{Hale2023}}. {The source counts from the modified SKADS catalogue better} reflects deep data from the {LoTSS} Deep Fields \citep{Mandal2021}, source counts presented for LoTSS-DR2 \citep{Shimwell2022} and data from other wavelengths scaled to $144\,\MHz$, assuming a spectral index\footnote{We use this value for the spectral index unless otherwise stated, under the convention $S_{\nu} \propto \nu^{-\alpha}$.} of $\alpha=0.7$, We initially use a minimum flux density of $0.1\,\mJy$ for the SKADS sources {to validate the randoms}, but increase this to $0.2\,\mJy$ once flux density cuts are applied (see Section \ref{sec:additional_constraints}). {We note that the relatively limited area of SKADS compared to LoTSS-DR2 means that the contribution of the much rarer, bright sources may be undersampled and so may differ from LOFAR observations. However such bright sources are {rare in the observations and simulations and so will not contribute largely to the clustering. Moreover, those sources will not be sensitivity limited}. {Due to the nature of the large area of LoTSS-DR2, SKADS sources will need to be repeated in our random sample, to ensure both spatial coverage and to allow the random sample to be significantly larger than the data. Whilst other simulated radio catalogues exist, such as T-RECS \citep[][]{Bonaldi2019, Bonaldi2023}, we will demonstrate later that the source counts used from this modified SKADS model can accurately represent the source counts of our data and other deeper observations, and have been shown to be successful in estimating completeness in other studies \citep{Hale2023}. Therefore, we {feel} we are able to adopt SKADS for use in this work. With future studies which split by source type and redshift, it will become increasingly important to use simulated catalogues which both have overall flux distributions which reflect the data as well as reflect the evolving luminosity functions for different populations.}}

{{As \textsc{PyBDSF}} {relies} on {peak SNR} in order to determine whether a source is detected {above} the local noise, we need {a} peak flux density for the simulated sources}. For a given integrated flux density, a point source is more likely to be detected than an extended source, due to the decreasing {peak} SNR for more extended sources. {To assign a peak flux density to our simulated sources}, we use the component catalogue {which corresponds to the modified SKADS catalogue.} {The catalogue used for this work has} a flux density limit of $5\,\uJy$ at $1.4\,\GHz$ ($\sim25\,\uJy$ at $144\,\MHz$), and includes the shapes and orientations of components that make up the individual sources in the SKADS catalogue. {Following} \cite{Hale2021, Hale2023}, we model each SKADS source through combining the emission related to the modelled components of a source. For each component, we model this as an ellipse {randomly positioned within a pixel of the same pixel scale as the LOFAR observations}. {We convolve this ellipse} with a Gaussian kernel representing the {restoring beam which is an approximation to the point spread function (PSF)} of the LOFAR observations ({6\arcsec}) {and sum these {components} together}\footnote{{We note that the knowledge of the true underlying source size distribution is challenging to understand from current observations, due to complexities such as source deconvolution and smearing in the image. Whilst SKADS provides one source size model, knowledge of these for the data will be improved with deep, high-resolution imaging of galaxies, such as with observations from the LOFAR International stations \citep[see e.g.][]{Morabito2022, Sweijen2022}.}}. {This {procedure} provides} an input catalogue of sources which have information on the integrated flux density, redshift, source type and peak flux density, {which we can assign to our random catalogues}. Unlike in \cite{Hale2021, Hale2023}, though, we do not inject sources into the images and re-extract sources using the source finder, \textsc{PyBDSF}. This is due to the large area of the field being considered, for which a significant computational effort would be required to create sufficient random sources to measure the clustering. Instead we make use of information from the simulations performed in \cite{Shimwell2022} to account for incompleteness across the sky. However, we must firstly account for smearing across the field of view.

\subsubsection{Smearing}
\label{sec:smearing}
{Smearing effects can reduce the peak flux densities of sources, and hence their detection.} This smearing can originate from a range of {factors including}: bandwidth and time {smearing \citep{BridleSchwab}; residual calibration errors; {the size of the facets used in the reduction}; and residual effects from the ionosphere interacting with the radio signals.} The first of these, bandwidth and time smearing, is described in detail in \citet{BridleSchwab} and is related to the averaging of data, {which causes an increasing smearing with distance from the pointing centre}. {In LoTSS-DR1, \cite{Shimwell2019} suggested that the use of \texttt{DDFacet} reduced the effects of such smearing at the largest angular separations compared to \cite{BridleSchwab} \citep[see Fig. 10 of][]{Shimwell2019}. {This is because DDFacet uses a different PSF in each facet which can be used to account for smearing in the data. The 6\arcsec restoring beam of LOFAR images is then used uniformly across the images. However, such a process leads to residual effects. For example, sources which are not fully deconvolved may still exhibit smearing and as only one PSF per facet is assumed, this can also lead to residual effects. We do not adopt the relation for smearing as presented in Fig 10 of \cite{Shimwell2019}, but instead investigate the smearing for the LoTSS-DR2 data and how it varies with observational properties.}}

Given the large {survey area of LoTSS-DR2} ($\sim$5600$\,\deg^2$), we consider whether there is a possibility {of smearing} being a function of position across the {survey, in particular with} the {elevation of the observations, as the {primary beam size} of {an individual pointing increases} at low elevation with LOFAR as it is} not a steerable telescope, {and as there are larger ionospheric effects, because more of the Earth's atmosphere is along the line of sight}. This leads to larger {and more elongated PSF} sizes and observational area at lower declination \citep[{see LOFAR observations at lower declinations in}][]{Hale2019}. Therefore, we consider the dependence of the observed smearing as a function of these parameters. 

To investigate {the relationship of the position-dependent smearing} we make use of sources from the Faint Images of the Radio Sky at Twenty-cm survey \citep[FIRST;][]{Becker1995,Helfand2015} {where} we have overlap between the two surveys {(mostly in the 13h field)}. FIRST is a 1.4 GHz survey with the VLA which observed the northern sky to $\sigma_{1.4\textrm{GHz}} \sim 0.15\,\mJy$ at 5\arcsec \ resolution. To study the smearing, it is important to identify sources which are believed to be unresolved. Such sources should have a ratio of integrated to peak flux densities ($\frac{S_I}{S_P}$) of 1, though scatter will exist due to the effects of noise at lower signal-to-noise (SNR). Due to the {higher angular} resolution in FIRST compared to LoTSS-DR2, we make the assumption that those sources which are unresolved in FIRST will also be unresolved in LoTSS-DR2. To identify unresolved sources in FIRST, we took those which are isolated (no neighbours within 12\arcsec) and are high signal-to-noise (SNR$\geq$10). For those sources we follow the methods of {previous works such as} \cite{Smolcic2017, Shimwell2019, Hale2021} and {use} a 95\% SNR {envelope of the form}:

\begin{equation}
\frac{S_I}{S_P} = A \pm B \times \textrm{SNR}^{-C},
\end{equation}

\begin{figure}
    \centering
    \includegraphics[width=8cm]{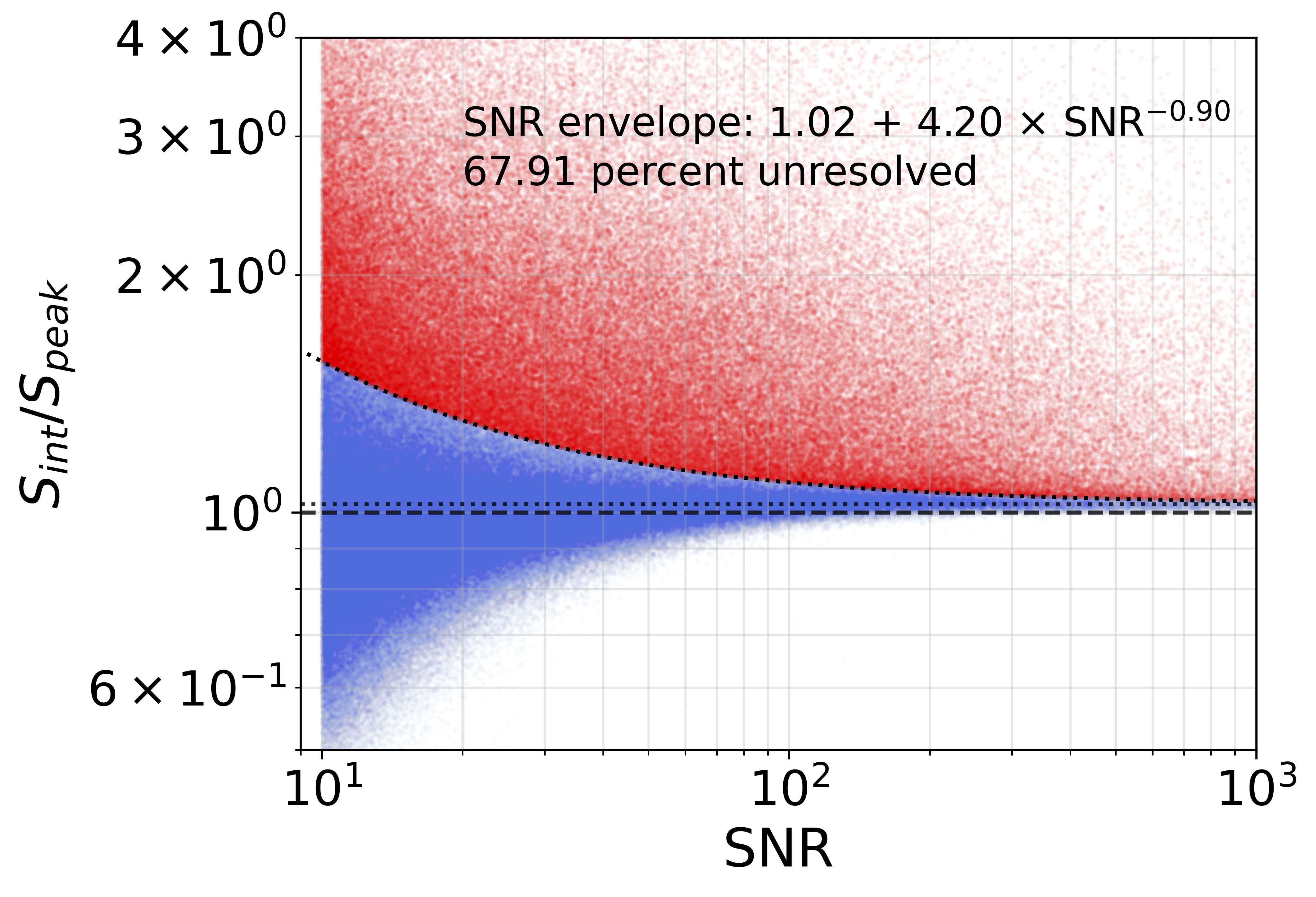}
    \caption{SNR envelope for integrated to peak flux density ratio as a function of SNR that is determined for isolated, high SNR sources in FIRST {(see Section \ref{sec:smearing})} Sources in blue are considered to be unresolved and in red are resolved. The model for the envelope is also provided. }
    \label{fig:SNRenvelope_FIRST}
\end{figure}

\noindent where the $\pm$ reflects the upper/lower envelopes. {$A$ is found using the value of $\frac{S_I}{S_P}$ at high SNR, and sources with {$\frac{S_I}{S_P}$ below $A$ are used to fit for $B$ and $C$ in order to define the envelope. The form of the envelope fit for these sources can be seen in Figure \ref{fig:SNRenvelope_FIRST}.} Those FIRST sources which are below the upper envelope are considered to be unresolved. These unresolved FIRST sources are then {cross-matched} {within a 3\arcsec matching radius to {LoTSS-DR2 sources which are} isolated (again, within 12\arcsec), high-SNR sources \citep[{SNR$\geq$20, to ensure sources are less} affected by Eddington bias, see]{Eddington1913}, and {those} sources which were considered single sources by \textsc{PyBDSF} (i.e.\ \texttt{S\_Code}=`S'). }}

\begin{figure*} 
    \includegraphics[width=18cm]{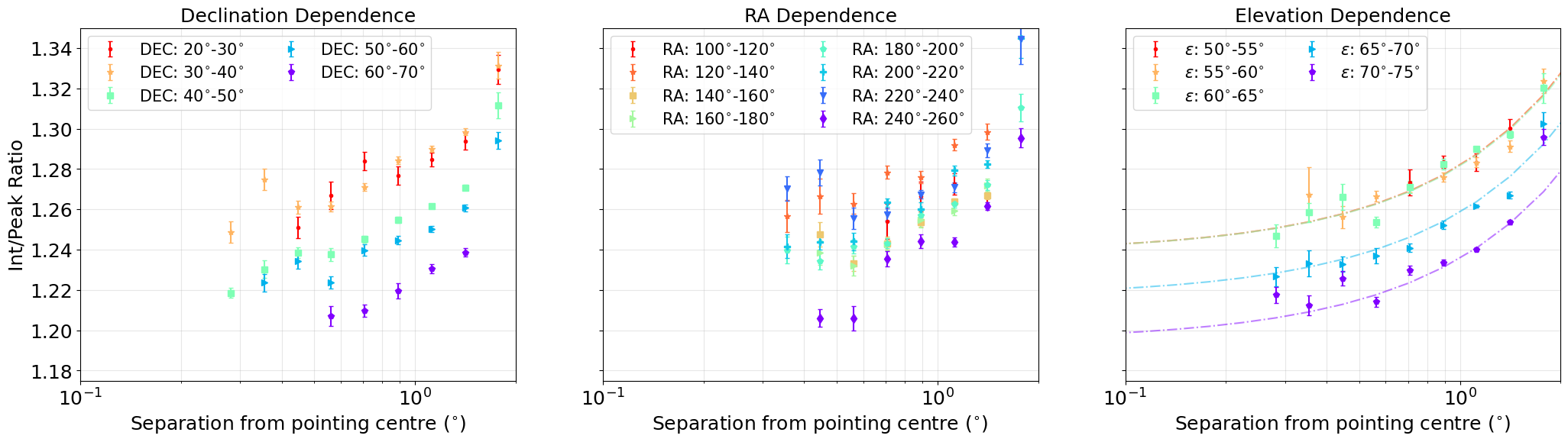}
    \caption{{The measured integrated to peak flux density ratio (an indicator of source smearing, y-axis) as a function of separation from the closest pointing centre (x-axis). The dependence of such smearing is shown as a function of declination (left), {RA (centre) and elevation (right). The dashed-dot line in the right hand panel indicates the elevation dependent smearing model which will be used in this work. For elevation bins $\leq$65\degree \ a constant model is used {(green, orange and red data)}.}}}
    \label{fig:smear_comparisons}
\end{figure*}

We then consider the position-dependent median ratio of the integrated-to-peak flux densities as a function of distance to the nearest pointing centre {and its dependence on RA, Dec and mean elevation of the field observation}. Only those separation bins that have at least 200 sources within them are presented in Figure \ref{fig:smear_comparisons} and errorbars are generated by bootstrap resampling the sources within the bin 100 times after resampling one third of the sources. 

\begin{figure}
    \centering
    \includegraphics[width=8cm]{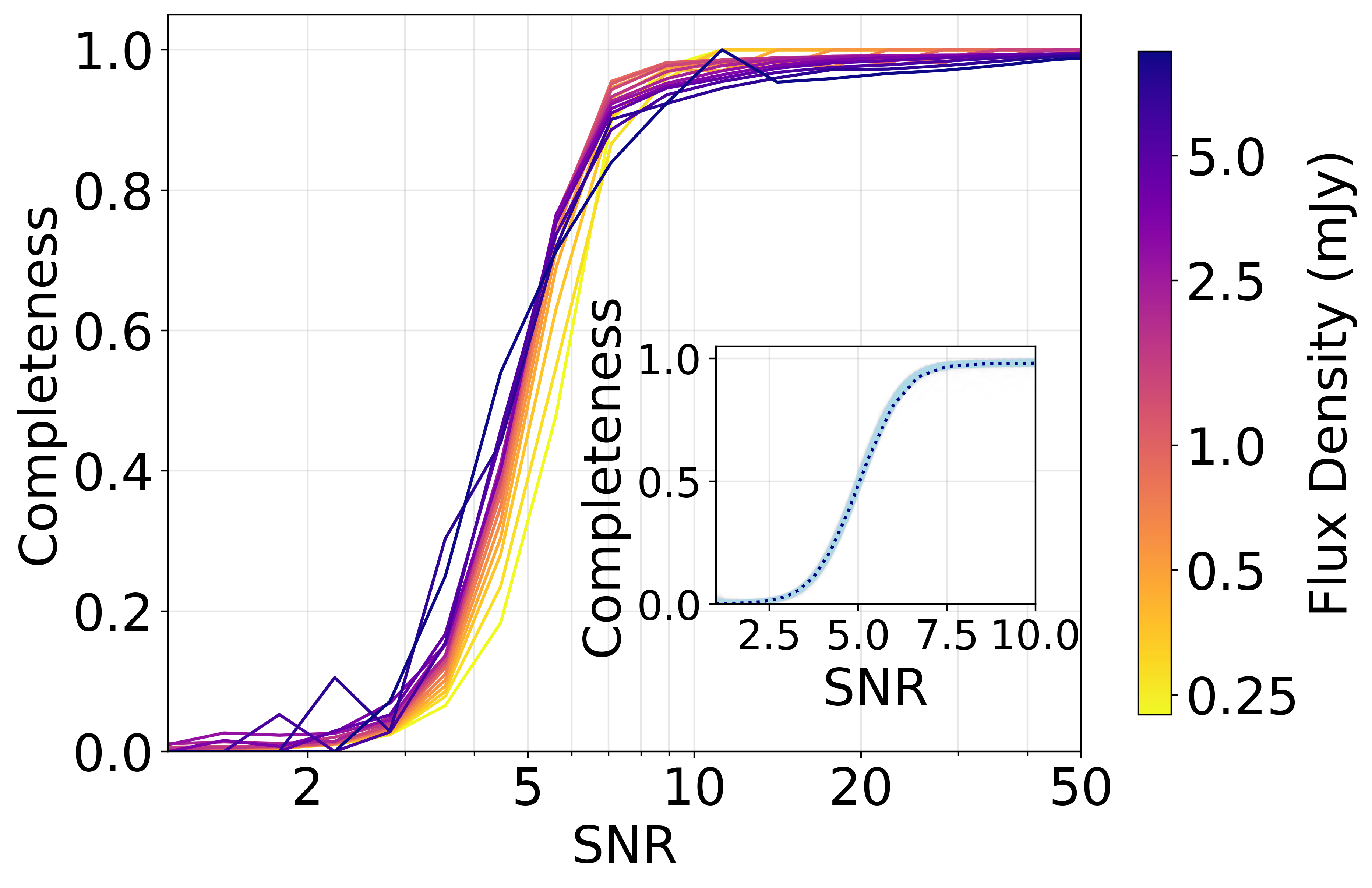}
    \caption{{Completeness as a function of {peak} SNR (x-axis) and as a function of flux density (see colourbar) for sources across the 841 pointings of {LoTSS-DR2}. {Inset: the completeness as a function of SNR only for each individual field (light blue) and the average across all fields (navy, dotted).} }}
    \label{fig:comp_flux_snr}
\end{figure}

{{Figure \ref{fig:smear_comparisons}, shows} {an increase} in smearing across the field of view as a function of distance from the pointing centre. {However,} there is also an apparent dependence on the declination and elevation of the field. The relationship with the right ascension of the observations is more complicated. If we first consider the effects of declination, the median flux density ratios {appear} to increase with declining declination, {whilst} for the two lowest declination bins considered there is similarity in {the trend of the} observed smearing as a function of separation. {If we consider the dependence on RA this does not appear to have a clear trend, but at the largest RA considered the smearing is minimized. However, we note that the comparison with FIRST does not have sufficient RA coverage to investigate {the full RA range observed} with LOFAR.} {Finally, if we investigated the elevation dependence of this smearing, {we see} increasing smearing with distance from the pointing centre, which also appears to {decrease} with elevation above an elevation of $\geq65^{\circ}$, and {to be} constant at elevations below this.} {As the elevation of an observation is related to the declination of the source combined with the time of observation, such smearing effects are likely correlated.} {For this work we {only} consider the elevation-dependent smearing to correct {the peak flux densities of the random sources, using} {for a model} of the form:}

\begin{equation}
    {\frac{S_I}{S_P} = C_1 + e^{-D_1 \times \theta}},
    \label{eq:smearcorr}
\end{equation}
{where $\theta$ is the angular separation (in degrees) from the pointing centre of the nearest pointing and $C_1$ and $D_1$ are values to be fit. We calculate the best fit values of $C_1$ and $D_1$  in bins of elevation and then model the average distribution of these parameters using a linear equation: }

\begin{equation}
    {C_1 = \alpha_C + \beta_C \times \epsilon,}
    \label{eq:fitlin}
\end{equation}
{and similarly for $D_1$}. Here $\alpha_C$ and $\beta_C$ are constants, and $\epsilon$ is the mid point of the elevation bin in degrees. {These are fit for elevation bins with {an} elevation $\geq$60\degree. For those elevations {$\leq62.5$\degree} we apply the same relation to that fit for the 60-65\degree elevation range.} {These models\footnote{{The model parameters that we find and use in this analysis are: $\alpha_C=0.506$, $\beta_C=-0.00428$, $\alpha_D=0.0557$ and $\beta_D=-0.000217$ (to 3 significant figures).}} are presented in Figure \ref{fig:smear_comparisons}.} {When applied to the random sources, angular separations are measured to the nearest pointing centre and the mean elevation is taken as that of the nearest pointing. As can be seen from Figure \ref{fig:smear_comparisons}, this functional form appears to be a good visual fit to the data. {This smearing shows that for those sources at the largest angular distances from the pointing centre have greater smearing and so would be less easy to detect than for a source with the same integrated flux density close to the pointing centre.}}

\subsubsection{Correcting the Simulations for Completeness and Source Measurement Effects}
\label{sec:comp_meas}
Once we have information for the flux density properties (both integrated and peak) for each simulated source, we consider the {likelihood a random source would be {detected, accounting} for completeness}. {Due to the variations in rms across the image and the source finder itself, the completeness will vary across the {sky} and not all sources with intrinsic peak flux densities above 5$\sigma$ will be detected by the source {finder, and some source with intrinsic SNR below the threshold will be pushed above the threshold. It} is then important to use this understanding of the completeness variation to determine which of our simulated randoms would be detected if they were observed through the LoTSS-DR2 survey.}

To {measure} this, we make use of the {image plane} completeness simulations which were presented and used in \cite{Shimwell2022} {and investigate the recovery of sources over a range of flux density and source shapes. We use the output from these simulations} in order to investigate completeness and the source counts for the survey. These simulations involved generating 10 simulated images for each field in which sources of varying flux densities and shapes\footnote{{We note these shapes are based on deconvolved source sizes, which may have smearing effects. We also note the SKADS models use elliptical based models, not Gaussians, and so this may lead to some residual differences when comparing the detection of extended sources.  We use these simulated sources from \cite{Shimwell2022}, though, as they are more appropriate than point sources, and allow some indication of the effect of non-point like objects.}} are injected within the residual images of the individual {pointings}. {This uses a source counts model from \cite{Mandal2021} to determine the number of sources to inject into a field}. \textsc{PyBDSF} is then used to re-extract the sources over the simulated images. This then allows the completeness to be {measured}, which is presented as a function of flux density in \cite{Shimwell2022} {for both point source completeness and using simulations which {include extended} sources, which} we use for this work. These simulations {can help quantify} which of our simulated sources are likely to be detected, but also to establish what the {``measured'' flux densities} of these sources may be, if they had theoretically been detected by the source finder. It is with a combination of accounting for these two effects that we generate our random catalogue of simulated sources. 

{Whilst the completeness is shown to have {a} large variation as a function of flux density for each LoTSS pointing \citep[see][]{Shimwell2022}}, the scatter is greatly reduced when its dependence on SNR is considered {(see Figure \ref{fig:comp_flux_snr})}. This smaller scatter is {due} to the fact that source finding with \textsc{PyBDSF} uses thresholding which is based on the {peak flux density} of pixels within a source, {compared to the local noise, i.e. SNR}. {Both the boundary of pixels which} contribute to a source {island} and the criteria which define which sources contribute to the catalogue both use a SNR threshold.} This is a 3$\sigma$ and 5$\sigma$ thresholding limit respectively for the two criteria defined. Therefore, while the rms values vary {between the different fields of} LoTSS-DR2{, so each field has a} different flux density dependence on completeness, the SNR dependence is more likely to be consistent across the fields. {This} can be seen in {the inset of} Figure \ref{fig:comp_flux_snr} {{which also} demonstrates that at} {a 5$\sigma$ limit}, which is used to generate the source catalogue, the completeness is in fact only $\sim$50\%, rising to $\sim$95\% at 7$\sigma$. Due to this consistency between fields, we therefore believe that using completeness as a function of SNR is a much more appropriate way to resample our simulated sources, instead of {using solely a flux density dependence}.

However, it is possible that while the average completeness as a function of SNR is consistent across the fields, it may be that {completeness has both a dependency on SNR and flux density}. This is {because} the intrinsic size distribution of sources {is likely to have a dependence on flux density}, such as AGN (which may have jets and be resolved) are likely to be brighter than star forming galaxies. For extended sources, these may be more likely to be detected at a given {peak} SNR as the larger sizes means that while the peak of the sources may be affected by a noise trough, pushing it below a detection limit, but the large {size} means that other neighbouring pixels could push the source above the detection limit, making it detectable. For smaller sources, they may be less likely to have a pixel above the detection threshold, given the smaller size. Therefore we also consider the flux density dependence of the completeness as a function of SNR (Figure \ref{fig:comp_flux_snr}). As can be seen in Figure \ref{fig:comp_flux_snr}, there does appear to be a {weak} flux density dependence of the completeness for the same SNR. For example at 5$\sigma$, there is a variation {in completeness from} $\sim$0.3 at $\sim$0.2 mJy to $\sim$0.65 at $\sim$5 mJy. This behaves in the way expected, as discussed above, {with larger sources better detected}. However, at $\sim$6-7$\sigma$ {for sources with the highest flux densities considered in Figure \ref{fig:comp_flux_snr}} there is the opposite behaviour, where the completeness appears to decrease with increasing flux density of the simulated sources.

\begin{figure}
    \centering
    \includegraphics[width=9cm]{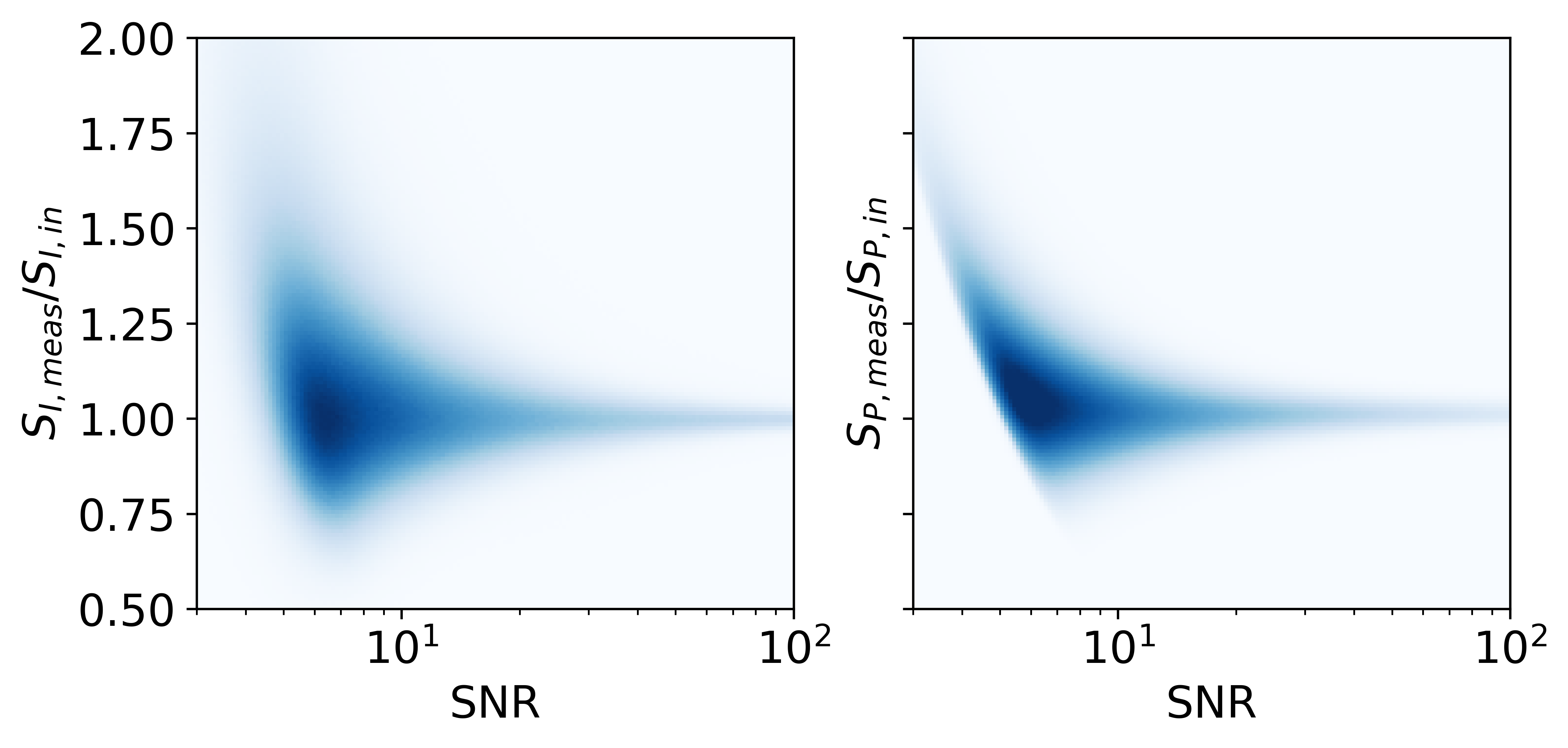}
    \caption{{Comparison of {the measured to input simulated} flux density as a function of {input} SNR for the simulated sources in \protect \cite{Shimwell2022} for {both} the integrated (left) and peak (right) flux densities.}}
    \label{fig:flrat_snr}
\end{figure}

Moreover, the simulations from \cite{Shimwell2022} allow us to also consider (i) the combined effects of Eddington bias \citep{Eddington1913}, where faint sources are preferentially boosted to higher flux densities, and (ii) {source} finder measurement errors. Combined, this allows sources which would {be inherently} fainter than 5$\sigma$ to be detected by \textsc{PyBDSF} but leads to sources at lower SNR to have {measured} integrated and peak flux densities at values different to their intrinsic values. Hence, we also consider the ratio of the measured to input flux density for each simulated source as a function of input SNR. This is shown for both the integrated and peak flux densities in Figure \ref{fig:flrat_snr}. As can be seen, at high SNR, the measured-to-input flux density ratio tends to a value of 1, indicating that these sources can be accurately characterised by the source finder. At lower SNR there is a {scatter for both} the integrated and peak flux density ratios which, at the lowest flux densities, {are} biased to measured flux densities that are larger than the intrinsic {flux densities.} 

\begin{figure}
    \centering
    \includegraphics[width=7.5cm]{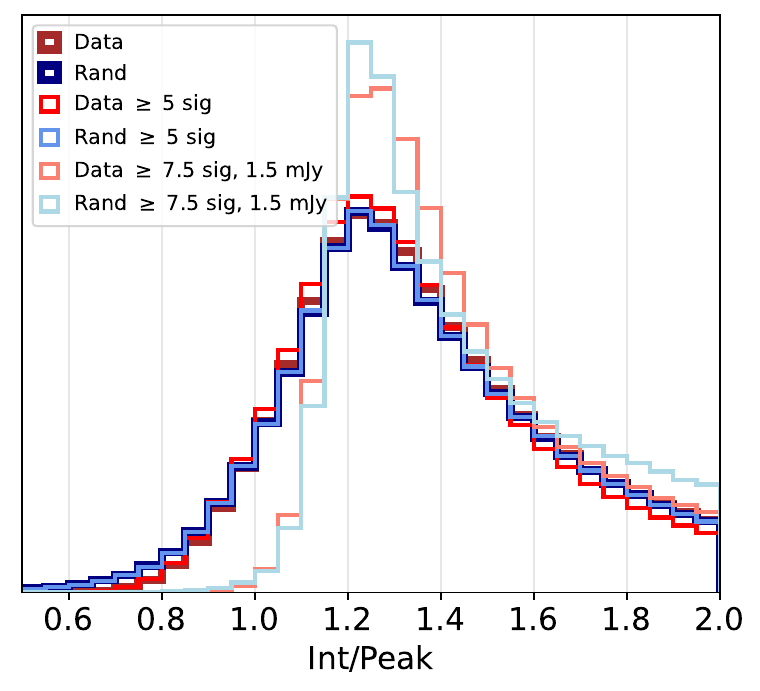}
    \caption{{Normalised distribution of the integrated (Int) to peak flux density ratio for the data (blue) compared to the random sources (red). This is shown for {all sources in LoTSS-DR2 and also those sources when a SNR cut of 5$\sigma$ is applied, and for the finally adopted cuts of 7.5$\sigma$, 1.5 mJy (see Sections \ref{sec:validate_rand} - \ref{sec:finaldata}). A lighter colour indicates a higher SNR cut.}}}
    \label{fig:int_peak_hist}
\end{figure}

{We therefore resample our randoms to correct for the effects of}: 
\begin{enumerate}
    \item The completeness as a function of both input SNR (peak flux density/rms) and integrated flux density;
    \item The ratio of the input simulated peak flux density ($S_{\textrm{P, in}}$) to the measured peak flux density ($S_{\textrm{P, meas}}$) as a function of input SNR {(to obtain a ``measured" peak flux density)}; 
    \item The ratio of the input {integrated-to-peak flux density ratio to the measured integrated-to-peak flux density ratio ($\frac{S_{\textrm{I, in}} S_{\textrm{P, meas}}}{S_{\textrm{I, meas}} S_{\textrm{P, in}}}$) as a function of input SNR} {(to obtain a ``measured" integrated flux density)}.
\end{enumerate}

\noindent {We use} {the simulations of \cite{Shimwell2022}} to take our input simulated catalogues and resample them to determine which sources are ``detected'' {based on} their expected completeness, {given their SNR and integrated flux density. For those sources which were considered to be detected, we} calculate a ``measured'' integrated and peak flux density for the simulated source.

{To} generate the final catalogue of randoms to be used to investigate the angular clustering we therefore take the input catalogue of random sources from SKADS discussed in Section \ref{sec:randoms_input} {and calculate} the peak flux densities that have been corrected for smearing ({see} Section \ref{sec:smearing}). We also apply a further constant smearing ratio by {dividing the peak flux densities by a ratio of 0.95}; this was found to be essential to {allow} the peak of the integrated-to-peak flux ratio {of the simulated sources to match that of} the data, see Figure \ref{fig:int_peak_hist}. {The value was chosen} to {align} the peak of these ratios and likely reflects a residual smearing issue from the data reduction processes such as from the effects of the ionosphere or residual calibration errors. Then, given the rms at the source location, it is possible to determine an input SNR. 

Using this input source SNR and integrated flux density {for an individual randoms source}, we then calculate {its completeness through interpolating from a 2D grid of completeness as a function of both SNR and flux density {which have been calculated from the simulations of \cite{Shimwell2022}}, across all fields\footnote{{Above 5 mJy there is more uncertainty due to the smaller number of simulated sources and so we assume the completeness variation with integrated flux density does not change above the maximum flux density shown.}}. For regions in SNR and flux density space {where} there is no or limited information from the simulations of \citet{Shimwell2022} to interpolate a completeness we extrapolate to reflect the detection.} For example, at high SNR ($\geq10$) and high flux densities where there is {limited} simulation information {(and so can be affected by smaller number statistics)}, we assume all sources will be detected, and at low SNR ($\leq1$), we assume the completeness is zero. From this 2D interpolation, we are able to calculate a {probability associated with the completeness} {which is compared to a randomly chosen probability and is considered to be ``detected'' if the completeness value is larger than the random probability.}

For these ``detected'' random sources, we then {determine the} ``measured'' peak and integrated flux densities for a source. This is important to consider {because} if we want to apply flux density or SNR cuts on the data (see Section \ref{sec:mask}) then such cuts would need to be applied to the random sources as {well. Therefore,} we again make use of the simulations of \cite{Shimwell2022} in order to generate a simulated ``measured'' peak and integrated flux density for each random source. To do this we again take the {simulations} from \cite{Shimwell2022} and construct a 2D histogram of the input SNR distribution vs. the ratio of the input to measured integrated flux density distribution (or similarly for peak flux density), {for each pointing observed in LoTSS-DR2}. {To generate the measured flux densities, we use the input SNR of each random source and use random sampling to obtain a measured peak flux-density input-to-output ratio and to obtain a ``measured'' peak flux density. For the integrated flux density we sample to find the ratio between the input-to-output peak flux density to integrated source flux density ratio, given the source SNR.} {Again, we make sensible extrapolations in those regimes where we have fewer sources, for example at high SNR.} Using this combined method means that we now have a distribution of random sources with not only positions, but also knowledge of the ``measured'' flux densities and SNR for the source. 

\subsubsection{Distribution of Randoms}
This methodology {{leads to a distribution of randoms that} can be seen in {the lower panel of} Figure \ref{fig:data_distribution}. This, in general, matches that of the data (Figure \ref{fig:data_distribution}) in that both under- and over-densities within the data are also apparent within the randoms in similar locations.} This highlights that the process we are using to generate the randoms appears to {broadly} represent the observational biases across the field of {view.} However, {as we believe {there is real} structure within the distribution of} galaxies, there will be differences between the distribution of data and randoms across the image. {There may, however, be additional SNR, flux density and positional cuts that need to be applied to the data to ensure the randoms reflect the data. We discuss such additional constraints in the next sub-section.}

\subsection{Additional Positional Constraints on the Data and Randoms}
\label{sec:mask}
While these randoms have been generated across the full field of view of the LoTSS-DR2 {survey, it is} important to apply additional { position-based} constraints in order to account for known observational systematics within the data. 

As discussed in Section {3.3.2 of \cite{Shimwell2022} and shown in their Figure 9}, {there} appears to be variations in the flux scale across an individual pointing within the LOFAR field. This appears to {be a} result of {differences in the model of the} primary beam across the field of view. Such flux scale variations {were seen to reduce by \cite{Shimwell2022} when pointings were} mosaiced together. {Therefore, we} only include regions where pointings have been mosaiced together {and by reducing} the area of observations for both the data and the randoms to remove the outer edges. Furthermore, and for a similar reason, we want to remove those areas where there are a large number of gaps within the {images} due to facets that failed the data reduction process. These often, though not exclusively, lie towards the outer edges of the {observations.}

\begin{figure}
    \centering
    \includegraphics[width=9cm]{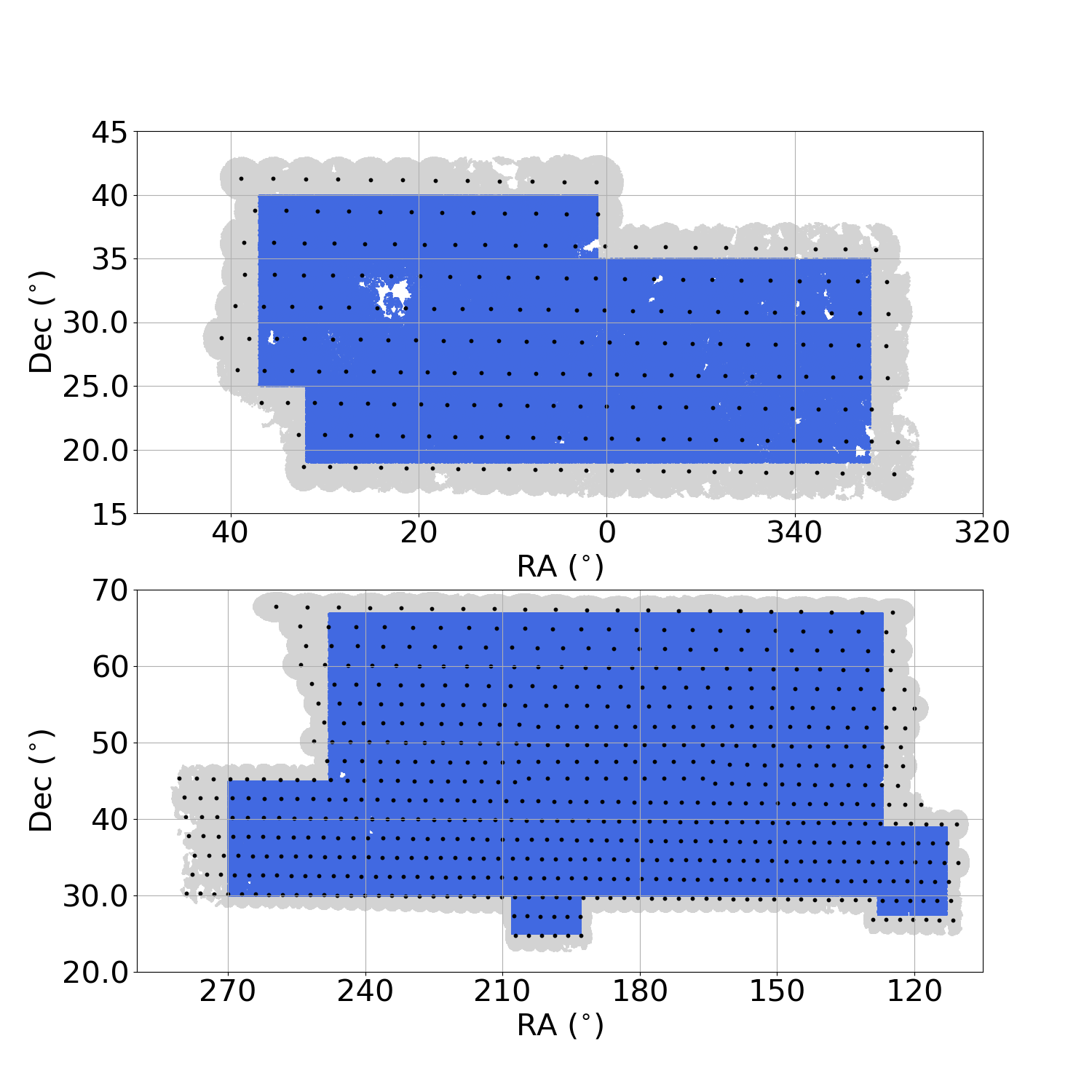}
    \caption{{Distribution of sources in the 1h ({top}) and 13h ({bottom}) fields of LoTSS-DR2 for the full area (grey) and inner masked region (blue) that is presented in Table \ref{tab:regions}. The black dots indicate the pointing centres for each of the 841 fields observed. {White regions indicate areas where the images are masked or outside the coverage of LoTSS-DR2.}}}
    \label{fig:maskedregions}
\end{figure}

{The reduced area is {defined in Table \ref{tab:regions}} and shown in Figure \ref{fig:maskedregions}, alongside the locations of the centres of the 841 {pointings which make up the DR2} region. The RA and Dec cuts are chosen to ensure that the {data is} at least a pointing radius from the outer edges of the observations. These cuts are employed to be conservative and remove regions where {uncertainty may be introduced in the flux scale across the image as the region is not mosaiced with neighbouring pointings. With these cuts applied, we have $\sim$80\% of the total area of LoTSS-DR2 remaining. This reduces the number of pointings which the data cover to 791.}}

\begin{table}
    \centering
    \begin{tabular}{c c c c c c c}
        Region & RA (\degree) & Dec (\degree) & & Region & RA (\degree) & Dec (\degree) \\ 
        \cline{1-3} \cline{5-7} 
        1 & [1, 37] & [25, 40] & & 5 & [127, 248] & [30, 67] \\
        2 & [1, 32] & [19, 25] & & 6 & [193, 208] & [25, 30]\\
        3 & [0, 1] & [19, 35] & & 7 & [248, 270] & [30, 45]\\
        4 & [113, 127] & [27.5, 39] & & 8 & [332, 360] & [19, 35]\\
    \end{tabular}
    \caption{Definition of inner regions used to mask both the data and random catalogues as described in {Section} \protect \ref{sec:mask}. }
    \label{tab:regions}
\end{table}

\subsubsection{Validation of Randoms}
\label{sec:validate_rand}
\begin{figure*}
    \centering
    \includegraphics[width=15cm]{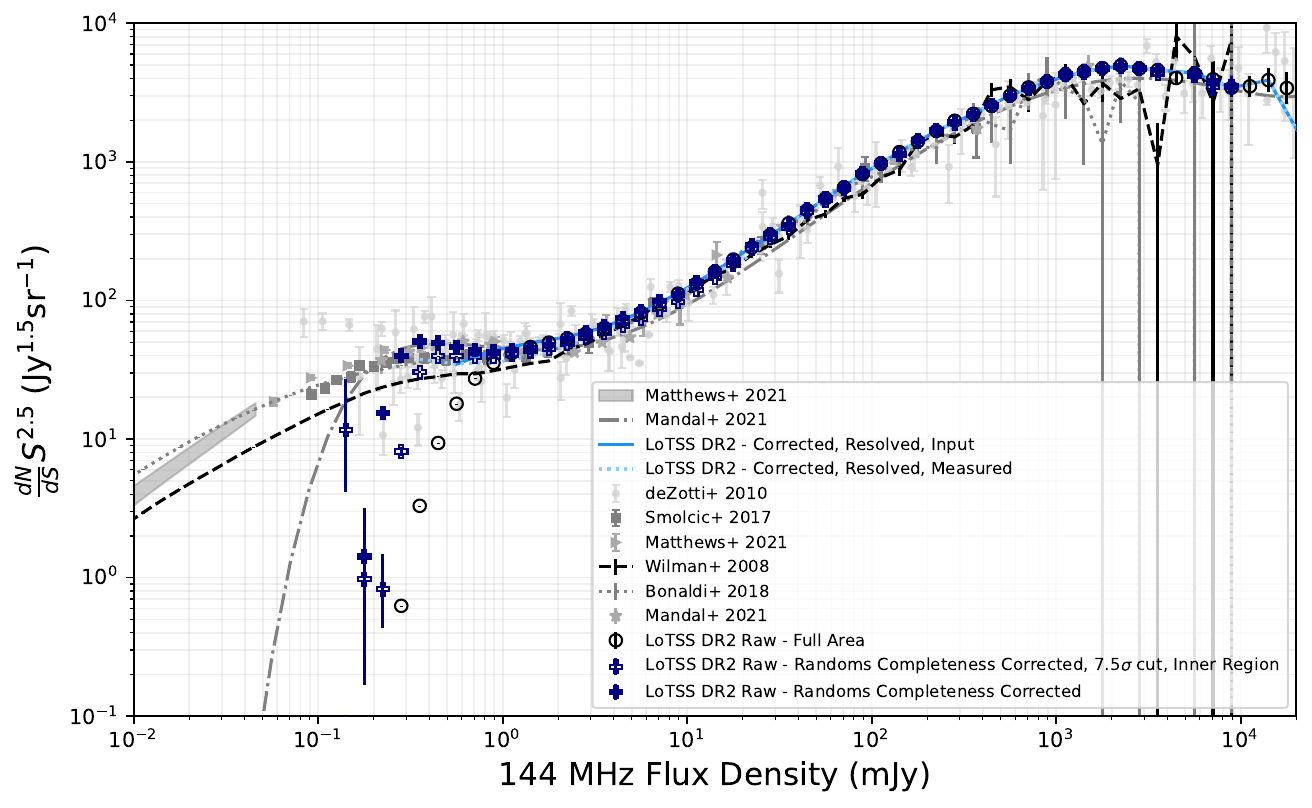}
    \caption{{Euclidean normalised source counts for the input and recovered randoms compared to that from previous data and simulated models. The randoms that are used as an input model (pink, right facing triangles) and recovered (red, left facing triangles) are shown, both scaled to reflect the larger ratio of randoms to data. The raw LoTSS-DR2 counts are also shown (black open circles) as well as {the} corrected source counts from the completeness derived from the recovered randoms (navy crosses) and the corrected source counts from the raw counts across DR2 using the completeness from the simulations of \protect\cite{Shimwell2022} both accounting for flux shifts between the simulated and detected flux density for a source (light blue dotted line) and not accounting for flux density shifts (blue solid line). {Also shown is previous data from the {LoTSS} Deep Fields \protect\citep[][{data - light grey stars and model - grey dot-dashed line}]{Mandal2021} and source counts converted to 144 MHz from \protect\citep[][{dark grey squares}]{Smolcic2017} and \protect\citep[][{grey triangles}]{Matthews2021}. Also compared is the source counts model from the {model of SKADS} \protect\citep[][{black dashed line}]{Wilman2008} and modified SKADS model used in this work (black dotted line). Errors associated with source counts not presented in previous papers are determined using the relations from \protect\cite{Gehrels1986}. When applying completeness corrections, we do not include uncertainty on the completeness as we only use a single randoms realisation.} We also include the LOFAR corrected source counts using the raw data and completeness corrections from randoms when a 7.5$\sigma$ cut is applied over the inner region described in Table \ref{tab:regions} (navy plus symbols, see Sections \ref{sec:validate_rand} - \ref{sec:finaldata}).}}
    \label{fig:scounts_all}
\end{figure*}

In order to validate that our randoms are accurate before using them and {to} determine any additional cuts to apply in order to study the angular clustering, we first make comparisons to check that the data and randoms have similar distributions, using those within the {region} defined {above (see Table \ref{tab:regions})}. First, we consider the apparent completeness produced by the random catalogues and what this implies for the ``intrinsic'' source counts that would be estimated based on this completeness. We present the Euclidean normalised source counts distribution in Figure \ref{fig:scounts_all}, where the {raw data are compared to the ``detected'' random sources}. As can be {seen,} there is good agreement between the raw source counts from the LoTSS-DR2 data and the ``detected" randoms to a {flux density of $\sim$0.3 mJy.}  {Below 0.3 mJy, deviations likely arise from the fact that the minimum flux density used for the random catalogues was 0.1 mJy. Therefore, below $\sim 0.3-0.4$mJy it is likely that the corrections are mis-estimated as the full effects of detection biases (e.g. measurement and Eddington biases) in the flux densities for low SNR sources will not be probed fully.  Further comparing the LoTSS random completeness corrected source counts to our input randoms sources, there are similar discrepancies below $\sim$0.3-0.4 mJy, which combines the resultant effects of not fully probing the correction for faint sources (as above) as well as the effect that the raw LOFAR data includes sources found from the wavelet fitting mode of \textsc{PyBDSF}, which is not modelled by the randoms. The effect of the wavelet fitting on the data can be better understood when we consider the SNR envelope of the data, which we discuss below.  }

\begin{figure*}
    \centering
    \includegraphics[width=16cm]{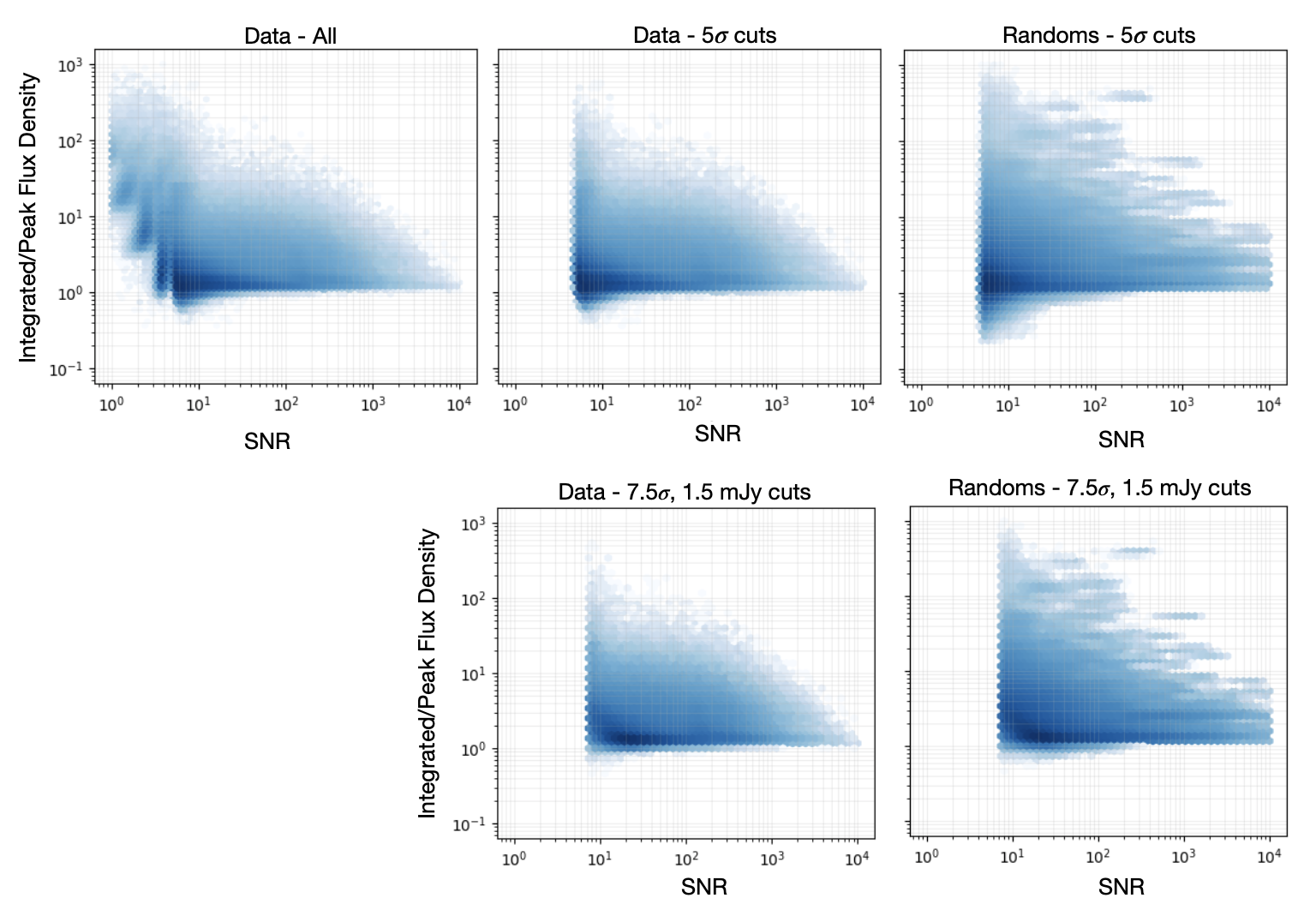}
    \caption{{Distribution {of integrated-to-peak flux density ratio (y-axis) as a function of measured SNR (x-axis)} for the full LoTSS-DR2 survey (upper left), for the data with a 5$\sigma$ cut applied (upper centre) and for the randoms with a 5$\sigma$ cut applied (upper right) and with the 1.5 mJy and 7.5$\sigma$ final cuts applied (lower panels, see Sections \ref{sec:validate_rand} - \ref{sec:finaldata}).}}
    \label{fig:snrenv_data}
\end{figure*}

{We compare the SNR envelope of our data to that of the randoms catalogue in Figure \ref{fig:snrenv_data}. This presents the integrated to peak flux ratio as a function of detected SNR (measured peak flux density/rms)}. In theory, this would consist of sources with an integrated to peak flux density ratio of 1 if they are unresolved or a ratio greater than 1 if they are resolved. In reality, an envelope distribution {is observed with increasing scatter in the ratio at low SNR.} {Figure \ref{fig:snrenv_data} also shows there are a wealth of {LoTSS-DR2} sources with SNR$<$5. These originate from \textsc{PyBDSF}'s wavelet fitting mode which was used during the source detection process}. This is {due to the fact that} a new rms map is recalculated for each wavelet fitting scale. This mode is {used} for finding larger extended sources. However, the simulations from \cite{Shimwell2022} use smooth models {for their simulated sources,} so do not employ the wavelet fitting mode when source finding with \textsc{PyBDSF}. Therefore, a SNR cut {of at least $5\sigma$} should be employed to ensure we use sources not detected through the wavelet fitting mode which have a different associated rms map that is not used here for the randoms. We present the comparison of the SNR envelope at $\geq 5\sigma$ for both the randoms and the data in Figure \ref{fig:snrenv_data}, which are in  better agreement {and for the final cuts to the data which are discussed in Section \ref{sec:validate_rand}-\ref{sec:finaldata}}. 

{Both of the comparisons presented in Figures \ref{fig:scounts_all} and \ref{fig:snrenv_data} examine the random} populations as a whole, not as a distribution across the field of view {and so} we also consider the distribution of randoms and data across the field of view, within the inner regions bounded by the {ranges listed} in Table \ref{tab:regions}. {In Figure \ref{fig:data_rand_dist} we} present the distribution of the ratio of {normalised number of data sources (normalising the number of sources in a bin to total number of sources)} to the {normalised number of randoms} as a function of declination with various SNR and integrated flux density cuts {applied. As} can be {seen,} the comparison of data to randoms is shown both when the randoms are uniformly distributed across the {sky} as well as the randoms generated from the resampling process discussed {in Section \ref{sec:randoms} above}. An accurate distribution of randoms which reflect the underlying observational systematics should show a ratio which is close to, or {scatters around,} a value of 1. 

{{Figure \ref{fig:data_rand_dist}, demonstrates} that up to a $5\,\mJy$ flux density limit, there is a clear difference between the uniform randoms and those which have the systematics of the data taken in to account. With just uniform randoms there is a clear declination dependence compared to the data, which likely reflects sensitivity variations across the sky. For example, the sensitivity becomes poorer at the lowest declination, therefore the uniform randoms will appear to be much more numerous than the sources observed in the data. However, the randoms generated for this work which account for sensitivity variations and observational systematics across the field of view show a more similar distribution to the data, oscillating around a value of 1. {For higher flux density cuts, the comparison between the data and randoms becomes more similar to a ratio of 1, staying within $\sim$5\% of a ratio of 1 above a flux density cut of 1 mJy}. }

{Given the comparisons presented, it is clear that a 5$\sigma$ SNR (at least) is needed {to avoid using those sources fit within the wavelet fitting mode of \textsc{PyBDSF}, whose rms maps will not reflect those used in this work}. Furthermore, from the source counts distribution it {has been discussed that at least a $0.3\,\mJy$ {integrated flux density} cut needs to be applied.}}

\subsubsection{Additional SNR and Flux Density Constraints}
\label{sec:additional_constraints}
\begin{figure*}
    \centering
    \includegraphics[width=19cm]{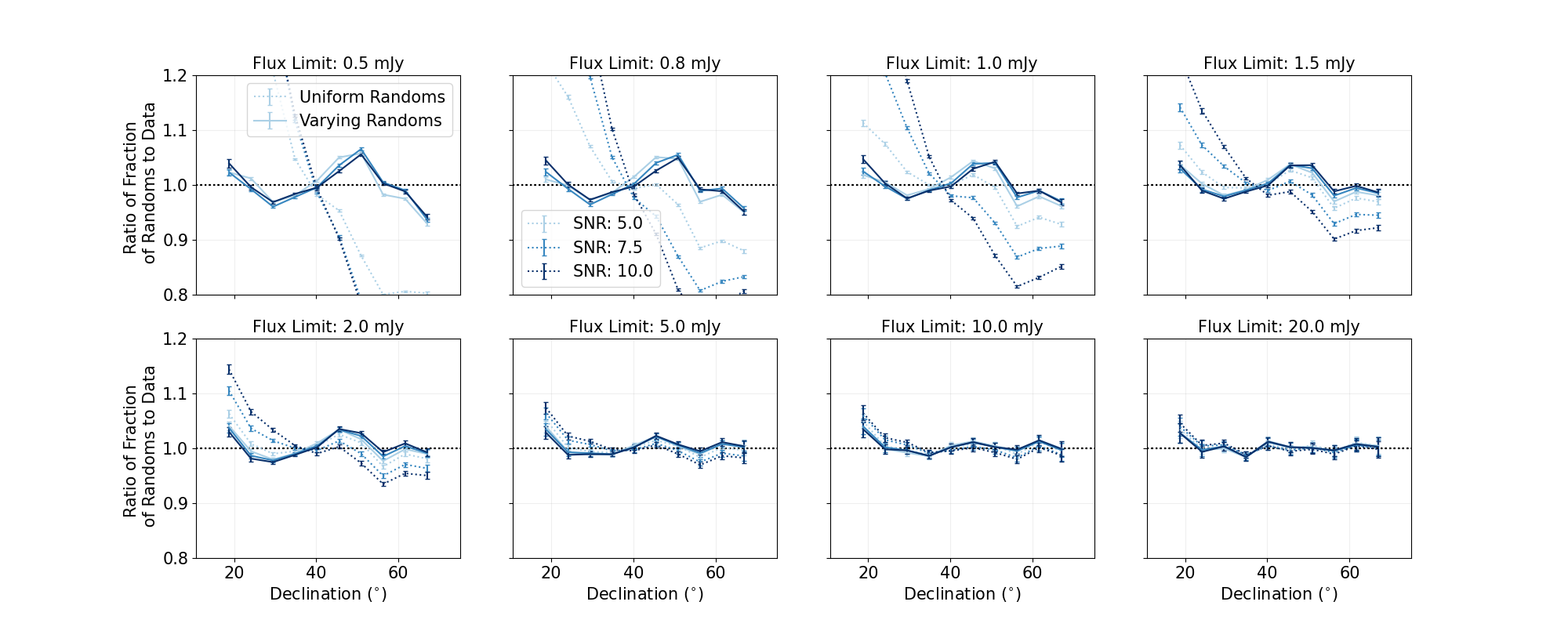}
    \caption{{Comparisons of the ratio of {the fraction of the total} random sources to {the fraction of the total} data as a function of declination (accounting for differences in sample sizes) for the randoms generated using the methods in Section \protect\ref{sec:randoms} (solid lines) {and for} {randoms generated uniformly across the sky area} (dotted {lines}) for sources {$\geq$}5$\sigma$ {(light blue)}, 7.5$\sigma$ {(blue)} and 10$\sigma$ {(dark blue)} {respectively in the regions defined by Table\protect \ref{tab:regions}.} This is shown with increasing flux density cut applied when moving from top left to bottom right.}}
    \label{fig:data_rand_dist}
\end{figure*}

Despite the more advanced random catalogues presented in this work compared to \cite{Siewert2020} for the clustering of sources in LoTSS-DR1, {we still may be limited by systematics in the data and may need to include additional cuts on the data and randoms.} While Figure \ref{fig:data_rand_dist} has demonstrated that {our}  randoms are smooth across the field of view as a function of declination, it cannot categorically show what flux density and SNR cuts to apply to the data and randoms in order to calculate the TPCF. {We therefore consider} the ratio across each pointing of the numbers of {real sources} to randoms {(both normalised by the total numbers of {real sources} and randoms respectively)} across the observations as a function of SNR and flux density cuts{, specifically how {the standard deviation in this ratio changes across each pointings}. We use standard deviation, as opposed to the mean values as the mean values will fluctuate around a constant {value,} but it is the deviations in these which illustrate the variation of fields which appear to have an over- or under-density of randoms compared to data around a mean value. If there are observational effects which are unaccounted for in the generation of our randoms, these would cause larger standard deviations in the normalised ratios of data to randoms across the sky coverage.} 

\begin{figure*}
    \centering
    \includegraphics[width=14.5cm]{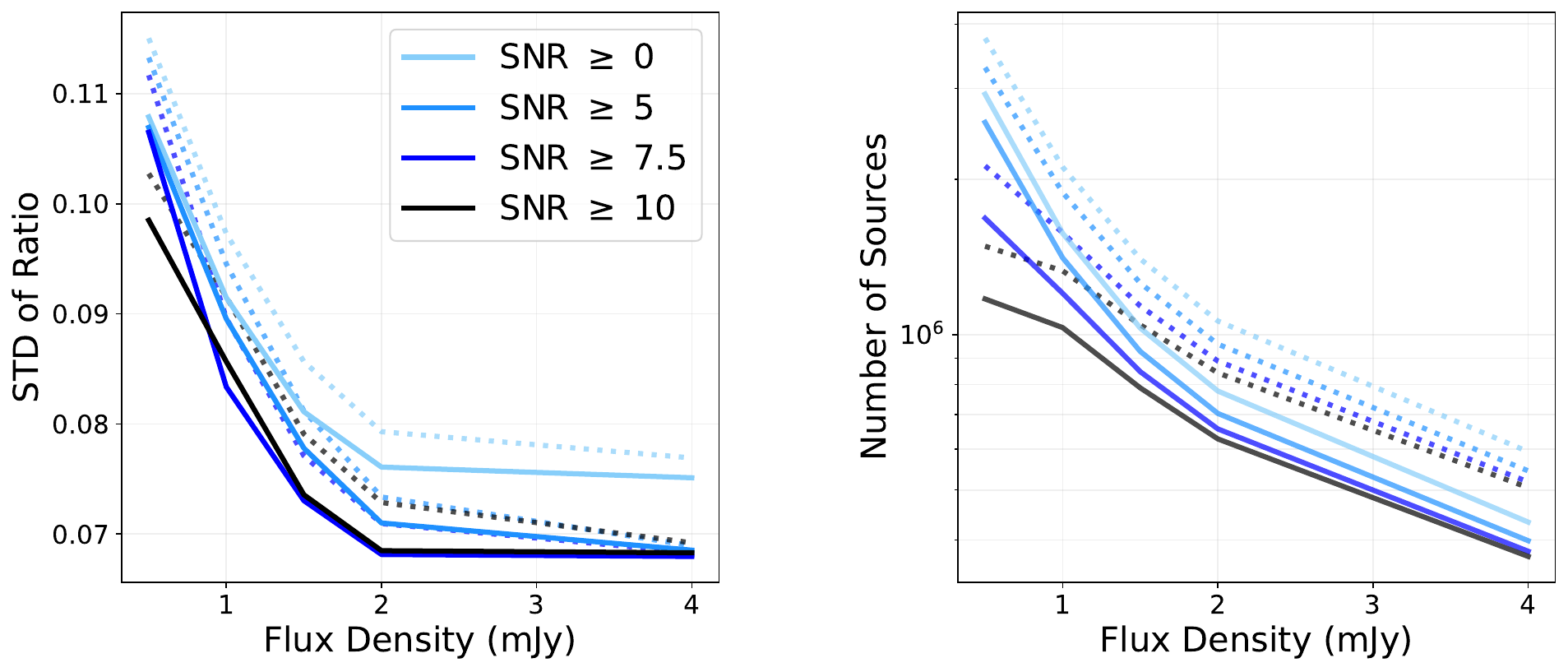}
    \caption{{Standard deviations {in the {field-to-field} scatter} of the ratio of the {LoTSS-DR2 sources} to randoms {across each individual pointings for different flux density and signal-to-noise cuts. Shown are the results for using the full field (dotted lines) and {for those pointings which are within the inner region of Table \ref{tab:regions} and contains at least 50\% of the data sources in that pointing contained within the inner region}} (solid lines). The right hand figure uses the same colour scheme, but instead indicates the number of {LoTSS-DR2} sources available for analysis.}}
    \label{fig:std_dat_to_rand}
\end{figure*}

In Figure \ref{fig:std_dat_to_rand} {we present the variation of this ratio {both across the full field of view (all 841 fields) and within the subset of pointings for which at least half of their sources lie within the inner region defined in Table 1 (where this limit is applied to avoid the effects of small number statistics).}} As can be seen, at a given SNR cut, the standard deviation declines with {increasing} flux density to $\sim2\,\mJy$, where it begins to flatten. {The right hand side of Figure \ref{fig:std_dat_to_rand} {shows} how the number of such sources in the data changes, given the cuts applied}. As a compromise {to balance both the number of sources we have as well as the variation in data compared to randoms, we apply} a flux density limit of $1.5\,\mJy$ and SNR cut of 7.5$\sigma$ for this work\footnote{{Given this higher flux density cut, we adopt a 0.2 mJy lower limit for our randoms as opposed to the 0.1 mJy described earlier.}}. Referring back to Figure \ref{fig:data_rand_dist}, it is clear that the distribution as a function of declination for such a SNR and flux density cut varies around a ratio of 1 within $\pm$5\%. Hence we believe this will be sufficient and {have a good reliability} for our clustering measurements. 

{Therefore, we are still limited in this work to a similar high flux density cut (1.5 mJy) which is $\sim 15-20\times$ {the typical point source sensitivity limit within the survey (70-100 $\muup$Jy)}, despite our additional investigations into generating accurate random sources. We believe that {contributing to this may relate to residual field-to-field systematics across the field of view. Whether this relates to flux scale differences between pointings, as presented in Figure 9 of \cite{Shimwell2022}, {imperfect primary beam models} or another residual observational systematic, remains unclear. Accounting for such residual systematics is something which is challenging to do within the simulations due to a lack of knowledge about, for example, these flux scale variations as a function of pointing.} {In order to assess any flux variations across the field of view, {the LoTSS-DR2} sources would need to be compared with similar large area, deep radio {surveys} across the field of view, using a catalogue with known high flux density accuracy. However, such a similar large area, high-resolution and moderately deep survey which allows a relatively large number of sources at a similar frequency for flux density comparison across the full field of view is not available at present. For those large area surveys that are currently available, applying SNR cuts, isolation criteria and other cuts to ensure accurate comparisons of source flux densities between the two catalogues would lead to too few sources to accurately study the flux variations across each pointing. We therefore are reliant on applying flux density and SNR cuts until we can fully understand and account for additional remaining observational systematics.}}

\subsubsection{Final data set}
\label{sec:finaldata}
\begin{figure*}
    \centering
    \includegraphics[width=12cm]{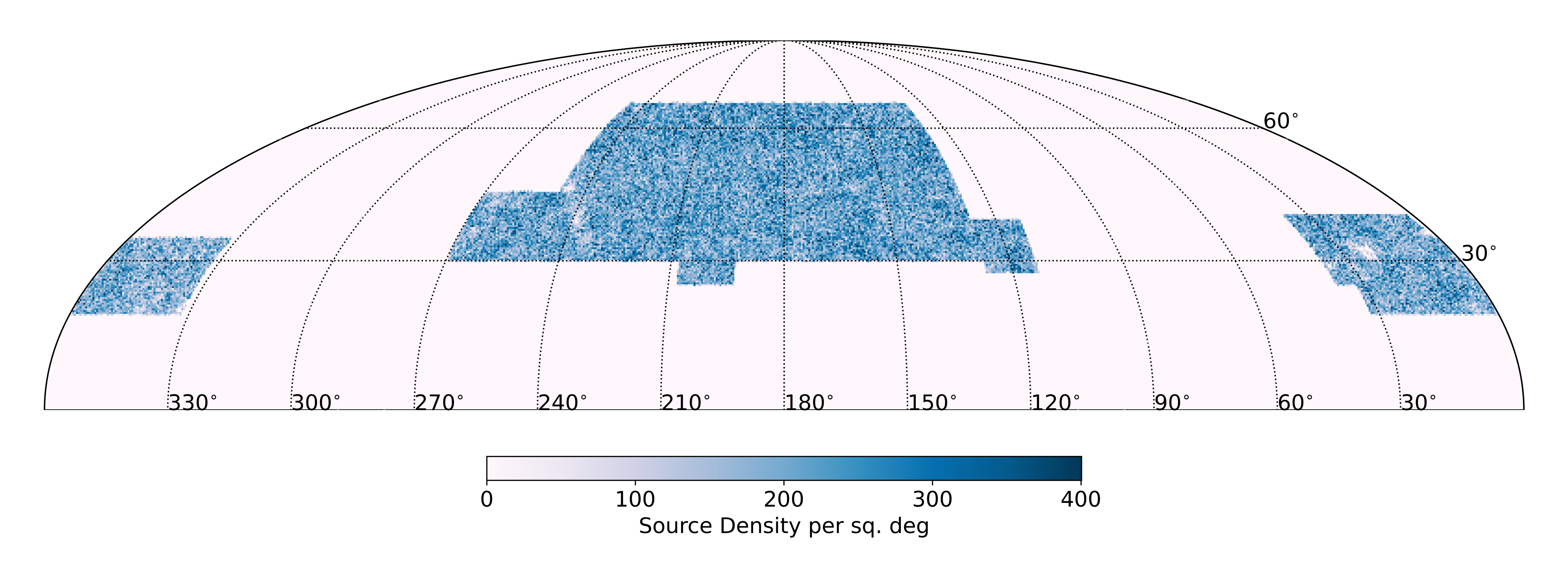}
    \includegraphics[width=12cm]{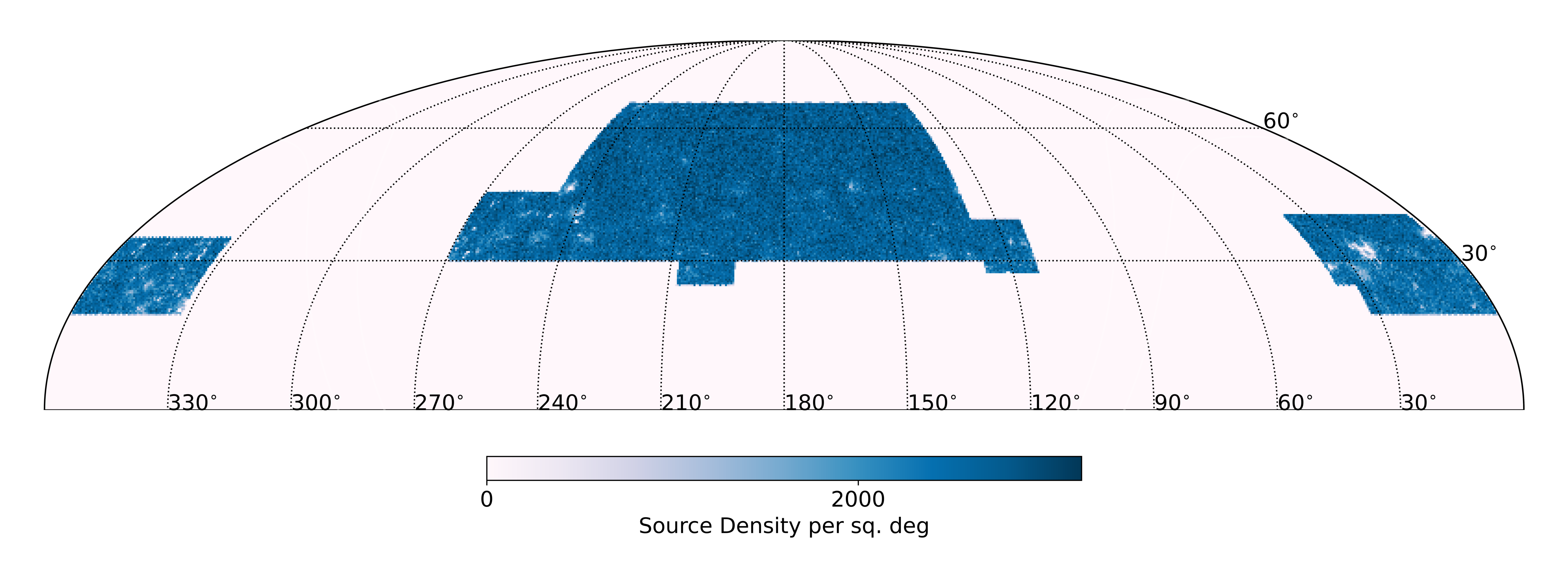}
    \caption{{Sky distribution of data (upper) and randoms (lower) used in this work after cuts are applied to the data. These are plotted using Healpy in the mollweide projection. {Note that the random sample is larger than the data sample, to minimise any Poisson errors associated with the randoms.}} }
    \label{fig:data_distribution_cut}
\end{figure*}

After applying the above SNR and flux density cuts as well as restricting to an inner region {and also flagging three HealPix pixels (using $N_{\textrm{side}}$=256) which were contaminated by a nearby spiral galaxy (see Pashapour-Ahmadabadi et al. in prep), {the number of sources which are used for these clustering studies is reduced}}. We present the number of data and random sources that are available after applying such cuts in Table \ref{tab:dataN}. {Such cuts help produce} a random catalogue which we believe is accurate to measure the intrinsic large scale structure. {The distribution of the final data and randoms used in this analysis can be seen in Figure \ref{fig:data_distribution_cut}.}

\begin{table*}
    \centering
    \begin{tabularx}{\textwidth}{X c c c c c }
       Cut Applied & N$_{\textrm{Data}}$ & \% of Initial Data Catalogue  & N$_{\textrm{Random}}$ & \% of Initial Random Catalogue  & N$_{\textrm{Randoms}}$/N$_{\textrm{Data}}$\\ \hline \hline
       No Cuts & 4,396,228 & 100 & {50,336,145}  & 100 & {11.4} \\
       Inner Region & 3,696,448 & 84 & {42,655,772} & 85 & {11.5} \\
       7.5$\sigma$ SNR cut & 2,160,232 & 49 & {27,364,838} & 54 & {12.7} \\
       $1.5\,\mJy$ Flux Density cut & 1,401,782 & 32 & {16,206,613} & 32 & {11.6} \\ \hline
       All cuts applied & 903,442 & 21  & {11,378,354} & 23 & 12.6 \\
       \hline 
    \end{tabularx}
    \caption{Number of data and random sources used when different cuts to the data are applied: using the inner region, a SNR cut and a flux density cut. {The effects of these cuts on the data are presented individually as well as their combined effect on the catalogues (alongside the masking of 3 Healpix pixels, see text), in the bottom row.} Presented are the number of data sources; the percentage of sources in the total catalogue that this consists of; the number of random sources; percentage of random sources compared to the initial (i.e.\ no cuts applied) random catalogue and the ratio of random sources to data sources with the same cuts applied.}
    \label{tab:dataN}
\end{table*}

\subsubsection{Changes in the process to create Randoms compared to LoTSS-DR1 and Remaining Limitations}
\label{sec:lims}
As this paper follows {on from} the clustering studies within the first data release of the LoTSS survey {(DR1)} \citep[{see cosmology analysis presented in}][]{Siewert2020}, we {briefly summarise} the {developments} in {random catalogues generated in this work compared to in \cite{Siewert2020} as well as the additional cuts applied to the data}. Firstly, in \cite{Siewert2020} the assumption was made that any sources {above 5$\sigma$ are detected}. However, as shown in {the inset of} Figure \ref{fig:comp_flux_snr}, at 5$\sigma$ the completeness is {$\sim$50\% on average}. This work, instead, uses the completeness curves as a function of SNR from \cite{Shimwell2022} which take into account the varying completeness with SNR and, therefore, do not use a hard cut off. This will result in fewer sources in the 5-10$\sigma$ range (based on input signal-to-noise) being included within the random sample, though with a 7.5$\sigma$ cut (on measured signal-to-noise), this will reduce the impact of such effects. Secondly, we also take into account the source sizes and do not assume all sources are point sources. {This} aims to take into account the effects of resolution bias, which will affect completeness within our {catalogue}{, though it does rely on a source shape model {which has uncertainties in the true distribution. Observations at higher angular resolution, such as sub-arcsecond LOFAR surveys \citep[see e.g.][]{Sweijen2022}, may aid with such knowledge but will be affected by resolution bias}}. Finally, we also calculate {more accurately, for each {random} source,} its ``measured'' peak and integrated flux densities. In \cite{Siewert2020} a flux {density} cut could be applied to the sources by ensuring the flux density added to the sampled noise associated with each source {(which provides an estimate for a measured flux density)} was greater than a given flux density limit. {However, this used the same noise term which would be applied to the peak flux density}. With this work, we are able to calculate the simulated to detected flux ratio as a function of SNR separately for the peak and integrated flux densities. This allows both SNR and flux density cuts to be applied on the appropriate ``measured'' flux density value. 

{While we have endeavoured to improve the generation of such random catalogues, residual caveats within the data still remain, which {we discuss} here for full clarity.} {Firstly,} as discussed above, {residual uncertainties in the beam model,  flux density scale across the field of view and other un-accounted for observational biases may impact the accuracy of the random catalogues}. We believe that these are a significant contribution to the {inability} {to use fainter flux density/SNR cuts.} While such flux offsets will average out when measuring e.g.\ source counts and declination dependencies over a full population, these will still exist on a field-to-field level. Furthermore, as we are not passing our randoms through a full end-to-end pipeline, there may be issues from the full LOFAR data reduction process, which we may not be fully able to {account for} the effects of. These include the effect of the ionosphere across each individual pointing, {astrometric errors, the direction dependent calibration introduced by} DDFacet or how {individual fields} are mosaiced together. The latter, especially, can lead to smearing of sources due to positional offsets within {overlapping areas, which cover a large fraction of the observations}. This smearing of sources may lead to a reduced sensitivity to detecting sources in the overlap regions {and may affect the smearing model used at the largest distances from the pointing centre}. {These effects are challenging to model{, as are the uncertainties in the intrinsic size distribution of radio sources}.} {Whilst full end-to-end simulations (starting from simulating sources in the \textit{uv}-data) could help such understanding, they are computationally expensive, especially for changes in the input source models considered.} \\

\noindent {With the methods discussed} we have aimed to characterize as {many} of the systematics {as possible} in order to generate accurate random catalogues. While the effectiveness of the detailed analysis when creating random catalogues through mimicking observational biases is reduced by the effect of the larger flux density and SNR cuts adopted in this work, our presentation of {a} {detailed discussion of the methods employed to generate the randoms as an example {of methods} which will be important for future {analyses} with deep radio surveys}.

\subsection{Errors on the TPCF}
\label{sec:jack}
{Once the randoms catalogues have been generated, it is possible to calculate $\omega(\theta)$ through Equation \ref{eq:LS} and attribute uncertainties to our measurements.} {We} consider several methods for quantifying the errors on the angular correlation function measurements. {Possible errors include those} from Poissonian {statistics} (i.e.\ just based on the number of sources observed within the data), bootstrap errors (where a random number of sources are replaced across the field of view) and jackknife errors (where {regions} are removed {one area at a time} and the scatter on the measured TPCFs assessed). Poissonian errors are known to underestimate the {true} errors \citep[see e.g.][]{Cress1996} and {do not take in to account} systematic variations in the data. {For the naive estimate of $\omega(\theta)$ given in Equation \ref{eq:tpcf1}, these Poissonian errors are given by:}

\begin{equation}
\delta\omega_\mathrm{Poisson} (\theta) = \frac{1 + \omega(\theta)}{\sqrt{DD(\theta)}}
\label{eq:poiserror}
\end{equation}

{However, when including the cross-terms ($DR$) in with the Landy-Szalay model, small changes to this are expected \citep[see e.g. the equations presented in][]{Landy1993, ChenSchwarz}. Either way, such estimates of the errors do not account for potential systematics in the errors across the field. Therefore, we consider several methods which resample the data to assess the errors more accurately across the field of view.} For bootstrap resampling, {$\sim$1/3} of sources are randomly removed from the data and randomly replaced with the same number of {randomly selected data sources}. This means that a source from the original catalogue {may} not be in the bootstrap sample, be in {it} a single time, or {multiple times}. This {process} is then repeated in order to make $N_B$ resamples. {For each resample, $\omega(\theta)$ is then calculated using \texttt{TreeCorr} as used for the original sample.} The errors are then calculated from these as in \cite{Barrow1984, Ling1986}:

\begin{equation}
\delta\omega_{B} (\theta) = \sqrt{\frac{1}{{N_{B} - 1}}\sum_{i=1}^{N_{B}} [\omega_i(\theta)-\omega_{{B}}(\theta)]^2},
\end{equation}

\noindent where $\omega_{B}$ is the mean value across the bootstrap samples. However, {bootstrap resampling randomly removes sources and is not able to trace systematic trends across the data. If such systematics exist {or {if there is significant variation in source density} across the field}, it is therefore possible that bootstrap resampling underestimates the errors on $\omega(\theta)$.}

We therefore, {also consider using jackknife errors \citep[see e.g.][]{Norberg2009} which {are} calculated by splitting} the field into a number of sub regions ($N_{J}$). {One sub-region is then removed in turn and we} measure the $\omega(\theta)$ from the remaining areas. The error is then calculated as:
\begin{equation}
    \delta\omega_{J}(\theta) = \sqrt{\frac{N_J-1}{N_{J}} \sum_{i=1}^{N_J} [\omega_i(\theta)-\omega_{J}(\theta)]^2},
    \label{eq:jackknife}
\end{equation}
where $\omega_{{J}}$ is the mean value of the angular two-point correlation function across the samples {where} a sub-region has been removed. 

\begin{figure}
    \centering
    \includegraphics[width=9cm]{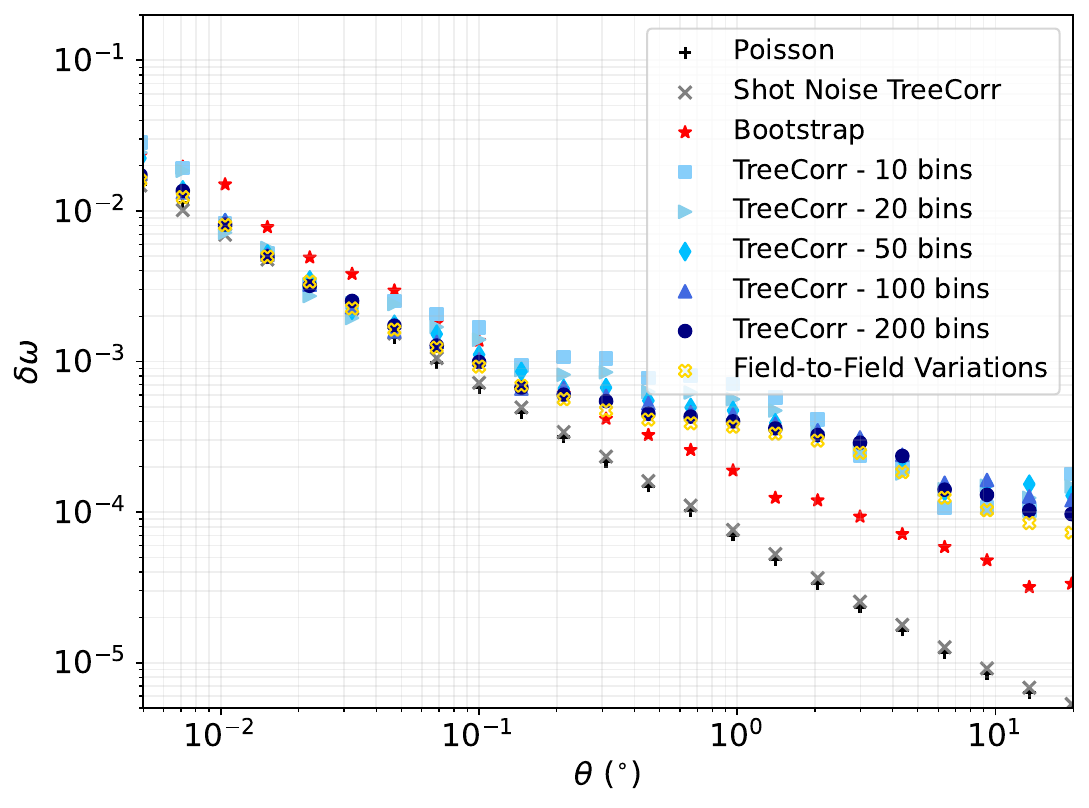}
    \caption{{Comparison of the ratio of errors from different resampling methods. Shown are the naive Poissonian errors (black crosses, {Equation \ref{eq:poiserror}}), the shot noise errors measured for the sample using the Landy-Szalay estimator in \texttt{TreeCorr} (grey crosses), bootstrap errors (red stars) and Jackknife errors for 10 {\texttt{TreeCorr} jackknife samples} (light blue squares), 20 {\texttt{TreeCorr}} {jackknife samples} (light blue right triangles), 50 {\texttt{TreeCorr}} {jackknife samples} (blue diamonds), 100 {\texttt{TreeCorr}} {jackknife samples} (blue triangles) and 200 {\texttt{TreeCorr}} {jackknife samples} (navy circles) {and field-to-field variation (yellow open crosses).}}}
    \label{fig:jackerrs}
\end{figure}

For completeness, we present the errors measured for the TPCF for {jackknife resampled errors}, {using \texttt{TreeCorr} to calculate the effect of changing the number of jackknife bins from 10 to 200}. Finally, we consider the effect of {field-to-field} variations between the individual pointings of LoTSS-DR2. {This method will directly probe the variations introduced from uncertainties between the different individual pointings of LoTSS-DR2.} We calculate the errors from this {using each pointing as} a jackknife sample. {We note that jackknife errors typically use regions of similar areas when calculating such errors, this will not be the case when calculating for the individual LoTSS-DR2 pointings being removed in turn. The internal pointings should be of roughly similar areas, but those towards the outside of the regions defined in Table \ref{tab:regions} could be significantly smaller. However, such jackknife scales are more relevant to understand the variation across the field of view.} {A comparison of these resampling errors {is} presented in Figure \ref{fig:jackerrs}, relative to the Poissonian errors. The {relative sizes of the} bootstrap and jackknife errors {varies at different angular scales}. At the smallest angles, $\theta \lesssim0.1-0.2^{\circ}$, bootstrap errors appear larger. At larger angular scales the jackknife errors are, as expected, significantly larger than found from bootstrap errors. This likely reflects variations in the data across the field of view either due to {real variation} across the field of view or systematics within the survey across the field of view. The bootstrap errors are a factor of $\sim$2 larger than the Poissonian errors at angles $\lesssim$1\degree, increasing to a factor of $\sim$5 at 10\degree. In contrast, the jackknife errors are similar to within a factor of 2 to the Poissonian errors for $\theta \lesssim$0.2\degree, rapidly increasing to a factor of $\sim$10 {larger at angles of $\sim$2\degree. }} {In general, since our fitting of $\omega(\theta)$ will focus on the largest angular scales, our comparison suggests we should use jackknife errors, compared to bootstrap errors, in order to not underestimate uncertainties at large angular scales $\gtrsim0.2$\degree. {These larger angular scales are important for fitting linear bias, see Section \ref{sec:b_pyccl}}.}

\begin{figure}
    \centering
    \includegraphics[width=9cm]{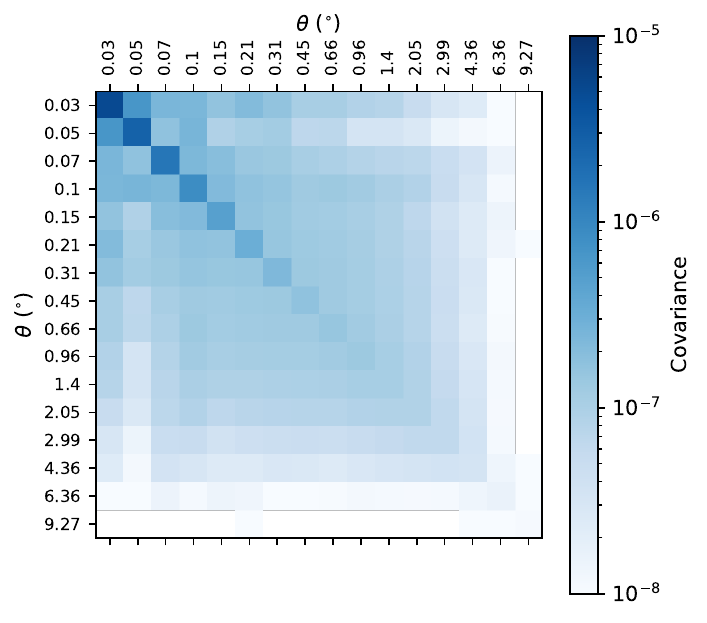}
    \caption{{Covariance matrix from resampling the errors using a Jackknife approach where each individual observed LOFAR pointing (791 within the inner region) is removed in turn.}}
    \label{fig:covariance}
\end{figure}

{{The errors from jackknife resampling {appear to be dependent} on the number of jackknife samples considered, with larger errors for smaller samples and more comparable errors for $\gtrsim$50 resamples}. {The errors generated using the individual field-to-field variations are comparable to those calculated using \texttt{Treecorr} when 100-200 resampling bins are used, which is expected as $\sim$800 pointings are used for the field-to-field variations.} {As the field-to-field sizes are the most physically motivated binning as they are based off scales of the pointings within the LoTSS-DR2 samples, we present result using such errors. The covariance matrix for such errors is presented in Figure \ref{fig:covariance}. We note that whilst the errors from \texttt{TreeCorr} compared to the field-to-field variation presented in Figure \ref{fig:jackerrs} appear similar for $N_{\textrm{Jack}}\geq100$, the covariance matrix using TreeCorr has a larger contribution of off-diagonal covariance values, especially for small $N_{\textrm{Jack}}$. As such off diagonal covariance values can affect the fitting of the source, we therefore will also briefly discuss the effect on the measured bias values of instead assuming 100 jackknife bins as well, in Section \ref{sec:discussion}.}}

\section{Angular Two-Point Correlation Function, $\omega(\theta)$}
\label{sec:tpcf_results}

\begin{figure}
    \centering
    \includegraphics[width=8cm]{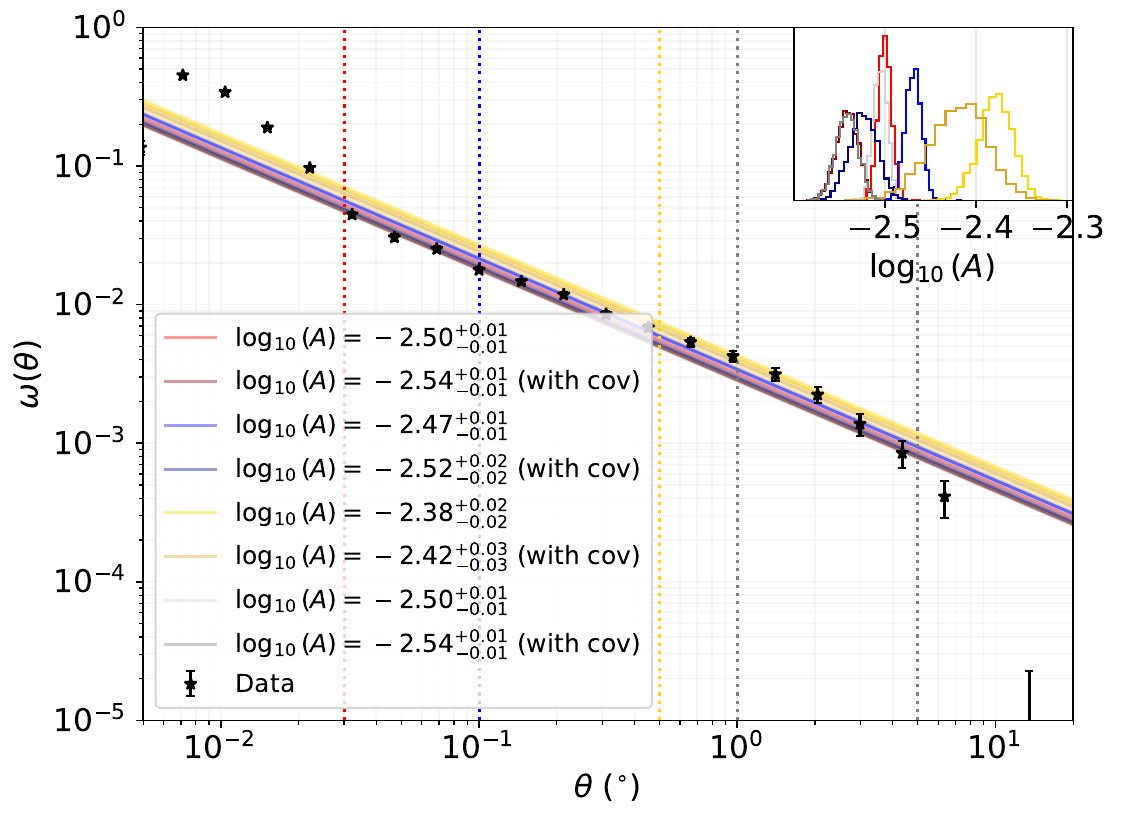}
    \caption{{Angular TPCF, $\omega(\theta)$ for {the final LoTSS-DR2 sample used in this work (black, see Section \protect \ref{sec:mask})} from the range of $\theta: 5\times 10^{-3} - 10^{2}$\degree. Also shown if the fit to $\omega(\theta)$ of the form $A\theta^{-0.8}$ and the probability distribution in the value of $A$ is shown in the figure inset (top right). These are shown for fitting over the angular ranges: 0.03-5$^{\circ}$ (red), 0.1-5$^{\circ}$ (blue), 0.5-5$^{\circ}$ (gold) as well as for the range where we reduce the largest fitting angle 0.03-1$^{\circ}$ (grey) both without (light colours) and with (dark colours) the full covariance matrix, {see Sections \ref{sec:tpcf_results} and \ref{sec:b_pyccl}.}}}
    \label{fig:tpcf_all}
\end{figure}

We present the angular two-point correlation function for LoTSS-DR2 sources with $S\geq1.5\,\mJy$ and SNR$\geq$7.5 in Figure \ref{fig:tpcf_all}. This is shown {above a minimum angular scale of $\sim$3$\times$ the PSF of the data {($\sim 3\times$6\arcsec \ $\sim 18$\arcsec)}}. As discussed in many previous studies \citep[e.g.][]{Peebles1975, Roche1999, Blake2002, Brodwin2008, Lindsay2014, Hale2018}, we can often describe the {angular clustering at small angular scales ($\theta \ll \pi$) as a power law distribution, given by:}
\begin{equation}
    \omega(\theta) = A \theta^{1-\gamma},
    \label{eq:omega}
\end{equation}
where $A$ is the amplitude, $\theta$ is measured in degrees and the power law slope is given by $1-\gamma$. Observations {suggest} $\gamma$ has a typical value of $\sim$1.8 \citep[see e.g.][]{Peebles1975, Peebles1980, Blake2002, Wilman2003}, meaning that $\omega(\theta)$ follows a power law of slope -0.8. 

\begin{figure}
    \centering
    \includegraphics[width=7cm]{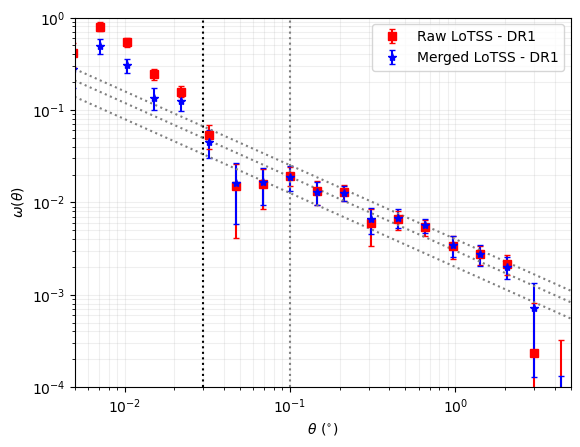}
    \caption{{Comparison of $\omega(\theta)$ for LoTSS-DR1 Data \protect \citep{Shimwell2019, Siewert2020} for the raw \textsc{PyBDSF} catalogue compared to the {source associated and cross-matched catalogue} described in \protect \cite{Williams2019} using a $1.5\,\mJy$ flux density cut and a $7.5\sigma$ SNR cut {and presented with bootstrapped {uncertainties}.}}}
    \label{fig:tpcf_dr1}
\end{figure}

\begin{figure}
    \centering
    \includegraphics[width=8cm]{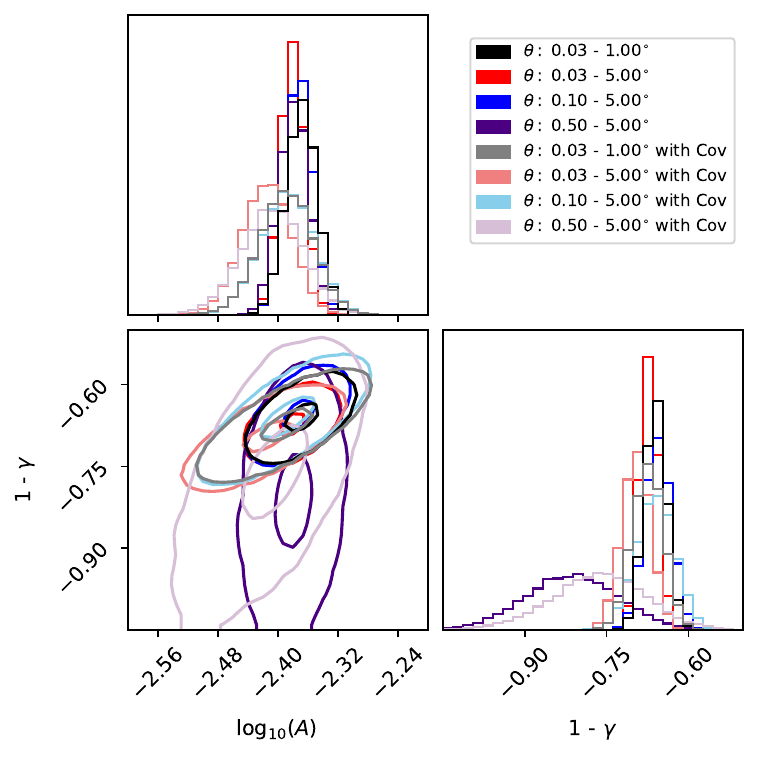}
    \caption{{Angular TPCF fitting parameter constraints for both $A$ and $\gamma$ {(with contours at 1 and 3$\sigma$)} for fitting over the angular ranges: 0.03-5$^{\circ}$ (red), 0.1-5$^{\circ}$ (blue), 0.5-5$^{\circ}$ (purple) as well as for the range where we reduce the largest fitting angle 0.03-1$^{\circ}$ (black) both without (dark colours) and with (light colours) the full covariance matrix, {see Sections \ref{sec:tpcf_results} and \ref{sec:b_pyccl}.}}}
    \label{fig:tpcf_all_2p}
\end{figure}

As can be seen in Figure \ref{fig:tpcf_all}, our results for $\omega(\theta)$ appear to follow a power law with $\gamma=1.8$ over a large range of angular scales {($0.03 \leq \theta < 1^{\circ}$)}, at larger angles ($\theta\gtrsim {10}^{\circ}$) there is more uncertainty {on} the value of $\omega(\theta)$ and so we do not present such scales in this work. At small angles ({$\theta \lesssim 0.03^{\circ}$}), there is a deviation from this power law distribution. This could arise from a combination of factors: (a) clustering of galaxies within the same dark matter halo and (b) the effect of multi-component sources.

{The first of these contributions {to the excess clustering at small angular scales} is related to whether the clustering of galaxies we are observing is from sources that are residing within the same dark matter halo \citep[{this is observed at small angular scales and is} known as the `1-halo' clustering, see e.g.][]{Zehavi2004}. {Measurements} of the `1-halo' clustering require observations which are both sensitive enough to observe multiple galaxies within the same dark matter halo and also have the resolution to ensure any galaxies within the same dark matter halo are not confused into a single source. In the radio, this `1-halo' clustering {has been challenging to observe due to the depths and resolutions of surveys previously observed}, however {it} will become increasingly possible with future deep, high-resolution radio surveys. When discussing clustering previously, we have instead focused on the clustering from galaxies in different dark matter haloes (known as the `2-halo' clustering) which presents as the power law behaviour given in Equation \ref{eq:omega} {on large angular scales}).}

The second contribution {to the excess clustering at small angular scales}, on the other hand, relates to the source detection within radio catalogues. For example, a {jetted} radio galaxy could be observed to have a core and two lobes separated from it. Depending on the separation of these lobes, conventional source finders \citep[e.g.][]{Selavy, Mohan2015, Aegean} may not be able to accurately characterise the components of the radio galaxy into a single source. As such, accurate cross matching of radio components relies on techniques such as visual identification \citep[see e.g.][]{Banfield2015, Williams2019}, or {machine learning/algorithm based techniques \citep[see e.g.][]{Galvin2020, Barkus2022, Alegre2022}}. If, in this example, the three components of the single radio source are {catalogued} to be different objects, then this will result in seeing an apparent excess angular clustering at small angular scales \citep[see e.g.][]{Blake2002, Overzier2003}, which can be described as a power law {with} a steeper slope. {To determine the angular scales below which such multi-component sources may become important in our work we consider the clustering in LoTSS-DR1 with both the raw \textsc{PyBDSF} catalogue and the value added catalogue of \cite{Williams2019}, where \textsc{PyBDSF} source components were combined into physical sources. We use the randoms generated for \cite{Siewert2020} and apply a 1.5 mJy and 7.5$\sigma$ cut, as used in this work, and present the clustering with and without source associations in Figure \ref{fig:tpcf_dr1}. This demonstrates a deviation between the raw and merged (source associated) catalogues, for which a deviation is seen at angles below 0.03\degree. This therefore suggests that the impact of multi-component sources is {likely} important below such an angular threshold and so we should not fit our $\omega(\theta)$ for LoTSS-DR2 below this scale.} 

We fit $\omega(\theta)$ {using} Equation \ref{eq:omega}, {with a maximum} angular separation of 5\degree \ and a minimum angular separation of either (i) 0.03\degree, below which multi-component source clustering becomes important; (ii) 0.5\degree \ {below which models that include both {1- and 2-halo clustering can diverge} \citep[see Section \ref{sec:b_pyccl} {for fitting} with the cosmology code \texttt{CCL},][]{Chisari2019}\footnote{which makes use of \texttt{CAMB} \cite{CAMB} and CLASS \cite{CLASS}} and (iii) 0.1\degree \  as a compromise between the two angular fitting ranges. Finally, we also include an angular fitting range of $0.03 \leq \theta <1$\degree to reflect the fact that the approximation of a power law model for $\omega(\theta)$ breaks down at large angles. } In our model we also include an extra term known as the integral {constraint} which accounts for finite field sizes \citep[see e.g.][]{Roche1999}. We therefore calculate the {$\chi^2$} through the difference between the observed data and the model {(with the integral constraint subtracted\footnote{{We note that the integral constraint will be very small due to the large field of observation in LoTSS-DR2, {on the scales considered}.}})}, {using two methods}. The first method, {that we adopt,} solely accounts for the diagonal elements of the errors ($\delta\omega$, as compared in Figure \ref{fig:jackerrs}), defining {$\chi^2$} as:
\begin{equation}
    \chi^2 =\sum_{i=1}^{N_{\theta}} \left( \frac{\omega(\theta_i)-\omega_M(\theta_i)}{\delta\omega_i} \right)^2, 
    \label{eq:chi1}
\end{equation}

\noindent where {$\omega_M(\theta_i)$} is the model for {the angular clustering, as in Equation \ref{eq:omega}, {for a given angular bin ($\theta_i$) and is fit across the $N_{\theta}$ bin in the angular range considered}. This {does not encapsulate the full systematic correlations between $\theta$ bins, but allows for a comparison to previous works who use such methods for fitting $\omega(\theta)$.} The second method uses the full covariance matrix, which allows correlations between $\theta$ bins to be accounted for. For this method, we calculate {$\chi^2$} as:}
\begin{equation}
    \chi^2=(\vec{\omega}-\vec{\omega}_M)^T{\rm Cov}^{-1}(\vec{\omega}-\vec{\omega}_M)
    \label{eq:chi2}
\end{equation}

\noindent where {$\textrm{{Cov}}$} is the associated covariance matrix for our measurements of $\omega(\theta)$, as calculated by \texttt{TreeCorr}. {The $T$ indicates that the transpose is being used.} {We fit a model for $\omega(\theta)$ using both Equations \ref{eq:chi1} and \ref{eq:chi2} to highlight the differences of accounting for the full covariance.}

{When fitting solely for $A$ (and fixing $\gamma$ to 1.8), we measure the variation in {$\chi^2$} when fitting the data using values of $\log_{10}(A)$ which are uniformly sampled from $-4$ to $-2$. From {the} {$\chi^2$} distribution we calculate a probability distribution {($P\propto {\rm e}^{-{\chi^2/2}}$)} and use a resampling method with {5000} samples to calculate a median value and associated error bars from this sample. {{The results are presented in Table \ref{tab:fit1} and Figure \ref{fig:tpcf_all}.} {As can be seen in Figure \ref{fig:tpcf_all}, the chosen angular scale below which we do not fit the data, $\theta <0.03^{\circ}$, appears to be an appropriate scale to restrict the fitting over.} Below these angular scales we observe a significant increase in $\omega(\theta)$, which we attribute {to the} contribution {of the combination of multi-component sources and 1-halo clustering.}}} {Figure \ref{fig:tpcf_all} shows the best fit models to the clustering amplitude, $\textrm{log}_{10}A$, of} {-2.50$\pm$0.01} (using {$\chi^2$} as in Equation \ref{eq:chi1}) and {-2.54$\pm$0.01} (using the full covariance) when fit over the largest angular range (0.03-5$^{\circ}$). {When fitting to the lower maximum angular scale ($0.03\leq \theta < 1$\degree) we find little difference to that when fitting in the range $0.03 \leq \theta <5$\degree. } {Whilst fitting $\omega(\theta)$ using Equation \ref{eq:chi1} shows a good fit to the data on a large range of angular scales, there is a deviation from such a power law around 1$^{\circ}$. This results in an increased clustering amplitude when fitting across the largest angular scales only 0.5-5$^{\circ}$, which then over-estimates clustering on smaller scales. This may suggest some excess residual systematics in the data, on the scale of $\sim1^{\circ}$. The fits using Equation \ref{eq:chi2} also appear to underestimate the values for $\omega(\theta)$ to more of an extent than with Equation \ref{eq:chi1}. }

{To test whether the assumed slope of -0.8 is suitable for this work, we also fit $\omega(\theta)$ for both $A$ and $1-\gamma$, using a fitting range of $-4$ to $-1$, for $\log_{10}(A)$ and $-2$ to $0$ for $1-\gamma$. We fit this using the Markov Chain Monte Carlo (MCMC) code, \texttt{emcee} \citep{emcee}. We fit using 100 walkers, each with 5000 chain steps and remove the first 90\% of chains as burn in. From this, we fit for $A$ and $\gamma$ using likelihoods based on the {$\chi^2$} described in Equations \ref{eq:chi1} and \ref{eq:chi2}. The results for such fitting across the angular ranges described above are presented in Figure \ref{fig:tpcf_all_2p} which, for the majority of angular scales, find a value of $1-\gamma \sim -0.6$ to $-0.75$, shallower than the $-0.8$ slope assumed when fixing $1-\gamma$. However, previous measurements of $1-\gamma$ using radio surveys \citep[see e.g.][]{Magliocchetti2017, Lindsay2014, Lindsay2014b} have found that such slopes ($1-\gamma$) observed for radio surveys are typically closer to $-1.2$ to $-0.8$. The differences observed here may therefore relate to a combination of factors, such as residual systematics in the data (as discussed above and in Section \ref{sec:lims}) as well as effects of combining multiple source populations in our measurement of $\omega(\theta)$. As such, we will predominately use our measurements where we fix the slope of $\gamma$ in order to measure bias, though in Section \ref{sec:bias} and \ref{sec:discussion} we will discuss the effect on the bias of assuming a variable slope.  }

\subsection{Variation with Location and Flux Density}
\label{sec:tpcf_variations}

\begin{figure}
    \centering
    \includegraphics[width=8cm]{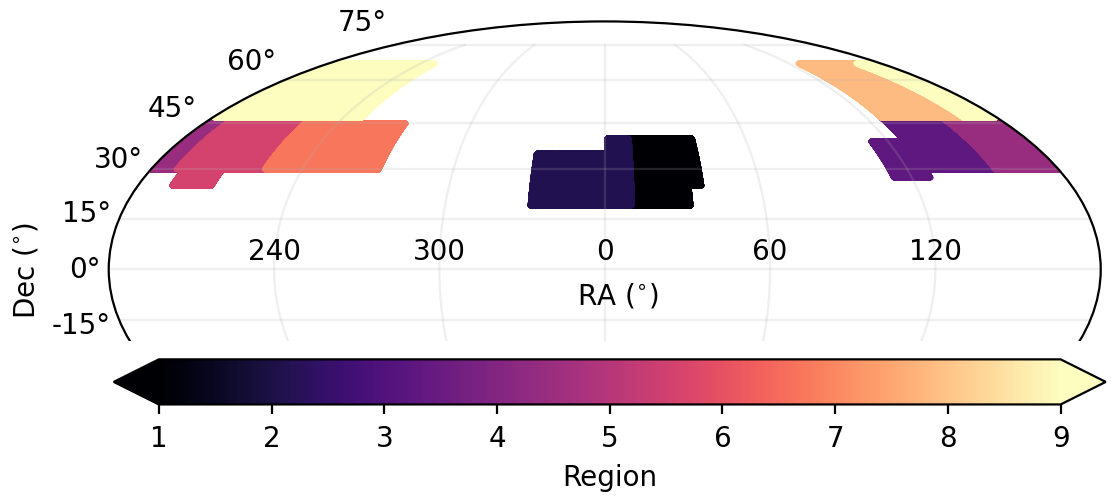}
    \caption{Regions used to investigate the TPCF variation as presented in Figure \protect \ref{fig:tpcf_comp}. Each colour indicates a different region used to quantify the TPCF. }
    \label{fig:region}
\end{figure}

In order to investigate the uniformity of $\omega(\theta)$ given the {possibility of systematics we are unable to correct for,} we also present comparisons of the angular clustering of the LoTSS-DR2 data as a function of Right Ascension, Declination and position within the full field of view. To do this, we consider the TPCF in RA angular ranges {spanning} 40\degree \ and declination in angular ranges {spanning} 10\degree \  and finally within nine different regions spread across the field of view in RA and Dec {bins} as presented in Figure \ref{fig:region}. {Uniform RA and Dec ranges are used to generate the RA and Dec bins, this will lead to significant differences in the number of sources in each of the bins which will have a more substantial impact on the measured $\omega(\theta)$ in regions where there are fewer sources.} This analysis, follows on from the comparisons of \cite{Siewert2020}, in which three regions were used to consider the variation in the angular clustering of {LoTSS-DR1}.

\begin{figure*}
    \centering
    \includegraphics[width=18cm]{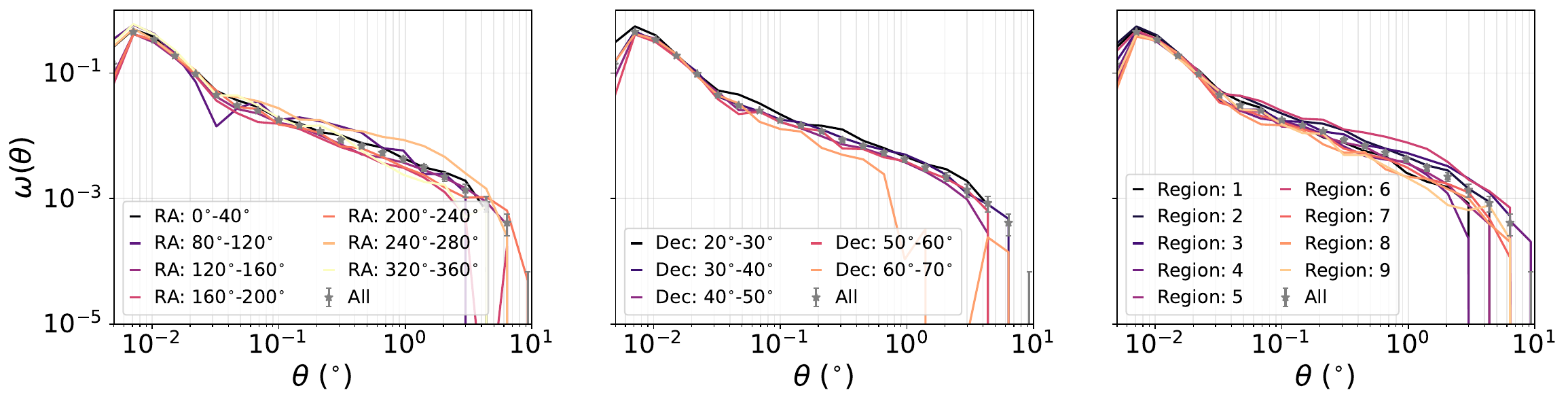}
    \caption{{The clustering variation between regions which are split based on their right ascension (left), declination (centre) and their location within the DR2 region as presented in Figure \protect \ref{fig:region} (right). The colour of the lines present the RA range, Dec range or region being considered and the results of the full area of the survey are shown in grey (stars). {Only the $\omega(\theta)$ value is presented for each subset, not the associated errors.}}}
    \label{fig:tpcf_comp}
\end{figure*}

The resulting variations in $\omega(\theta)$ are presented in Figure \ref{fig:tpcf_comp}. As can be seen, the variation of the angular clustering is typically restricted to larger angles $\theta \gtrsim 0.5^{\circ}$, whilst smaller angles are typically in much better agreement with one another. {Whilst there {are no apparent trends} with RA, there may be a suggestion of} a systematic trend in the angular clustering observed with declination, {with higher observed angular clustering at typically lower declinations}. {However,} this is not seen at all angular scales. {We also see there is more variation in the measured $\omega(\theta)$ when split into RA ranges and the regions presented in Figure \ref{fig:region}. As discussed in Sections \ref{sec:additional_constraints} and \ref{sec:lims}, we believe there are still limitations in the data which the randoms do not account for, such as individual flux shifts between pointings, uncertainty in the beam models and remaining systematics not modelled as full end-to-end simulations were not used to generate the random sources. It is {possible} that the effect of these can {be a cause} of the variation of $\omega(\theta)$ when split by these {sky regions however, true underlying large scale structure may also play a role}. The spread with declination is much smaller, with  $\omega(\theta)$ in the Dec: 60-70\degree \ bin showing the most variation, likely due to the smaller area and number of sources in this region. This smaller variation is likely due to the corrections implemented for elevation dependent smearing, which is related to the declination for fields observed with a good hour angle coverage. {If there are residual systematics relating to flux shifts between pointings \citep[as described in][]{Shimwell2022}, these are challenging to identify and model using available radio surveys. These effects and a combination of other residual systematics may relate to why there can be variations between $\omega(\theta)$ in different regions of the data. Identifying the cause of these and making further corrections may be possible in the future, with further understanding of the systematics. } }

\section{{Galaxy Bias}}
\label{sec:bias}
Whilst fitting a clustering amplitude, $A$, allows {for a} comparison with previous work, it is also challenging to compare with previous studies due to its dependence on flux density, {luminosity} and source type within the same sample \citep[see e.g.][]{Wilman2003, Overzier2003, Magliocchetti2017, Hale2018, Chakraborty2020}. {We} calculate the more physical parameter of bias, $b(z)$. {As discussed in Section \ref{sec:intro},} bias traces the clustering compared to matter and can be used to estimate the typical dark matter halo mass hosting a population of sources {\citep[see e.g.][]{Berlind2002, Zehavi2004}}. By calculating {the} bias, we not only calculate a more physical parameter, but also account for the redshift distribution of the sources being investigated. {However, this will also have a dependence on flux density, as the relative contribution of different source types to the overall population (e.g. AGN, SFGs) varies with flux density \citep[see][for a comparison of this in the LoTSS Deep Fields]{Best2023}. These populations can have different bias values and so will affect the bias measured for a full population \citep[see e.g.][]{Magliocchetti2017, Hale2018, Chakraborty2020}.} 

{In order to obtain measurements of the bias for the LoTSS-DR2 sources, knowledge of the redshift distribution, $p(z)$, for the data is required.} This is {because $\omega(\theta)$ is} a projected measurement of the clustering of galaxies over the sky, {and} to understand the bias, we need to understand the true spatial {distribution. Using a given $p(z)$ we then take two approaches to modelling the clustering}: (1) fitting using the cosmology code, {\texttt{CCL}} \citep{Chisari2019} and (2) using the {power law model fit for the amplitude, described in Section \ref{sec:tpcf_results},} and using Limber's inversion {\citep[see][assuming a power law model for $\omega(\theta)$ to calculate a clustering length, $r_0$, and subsequently a measurement of the bias]{Limber1953, Limber1954, Peebles1980}}, {as has been commonly employed in clustering studies for radio surveys} \citep[see e.g.][]{Magliocchetti2004, Lindsay2014, Magliocchetti2017, Hale2018, Chakraborty2020, Mazumder2022}. We will describe both approaches, below, however we first describe how the redshift distribution, $p(z)$, for the data is obtained, as this is critical for both approaches.

\subsection{Redshift distribution}
\label{sec:nz}
{In order to calculate the bias, we must assume a redshift distribution {for the sources in our sample, which is not possible from radio continuum measurements alone}}. {Instead,} a catalogue where radio data and multi-wavelength data have been cross-matched together \citep[{as with LoTSS-DR1, see}][]{Williams2019, Duncan2019}, may provide redshifts for some sources {however, redshifts are not currently available for a relatively complete population of LoTSS-DR2 sources}. Therefore, in order to {estimate} the expected redshift distribution of the sources observed in LoTSS-DR2, we make use of the {LoTSS} Deep Fields observations \citep{Tasse2021, Sabater2021, Kondapally2021, Duncan2021}. {The LoTSS Deep Fields data {are} more sensitive than in LoTSS-DR2 ({reaching an} rms $\sim20-40\,\uJy\,\mathrm{beam}^{-1}$) {over} three extragalactic fields {(see Section \ref{sec:data_deepfields} for details).} For the Deep Fields sources, 97\% have been cross-matched to a multi-wavelength host galaxy \citep{Kondapally2021} and have an associated redshift \citep{Duncan2021}. {A full probability distribution for the {photometric} redshift, $p_i(z)$, {of} those sources with an associated host galaxy is {presented} in \cite{Duncan2021}, which we use in this work.}}

\begin{figure}
    \centering
    \includegraphics[width=8cm]{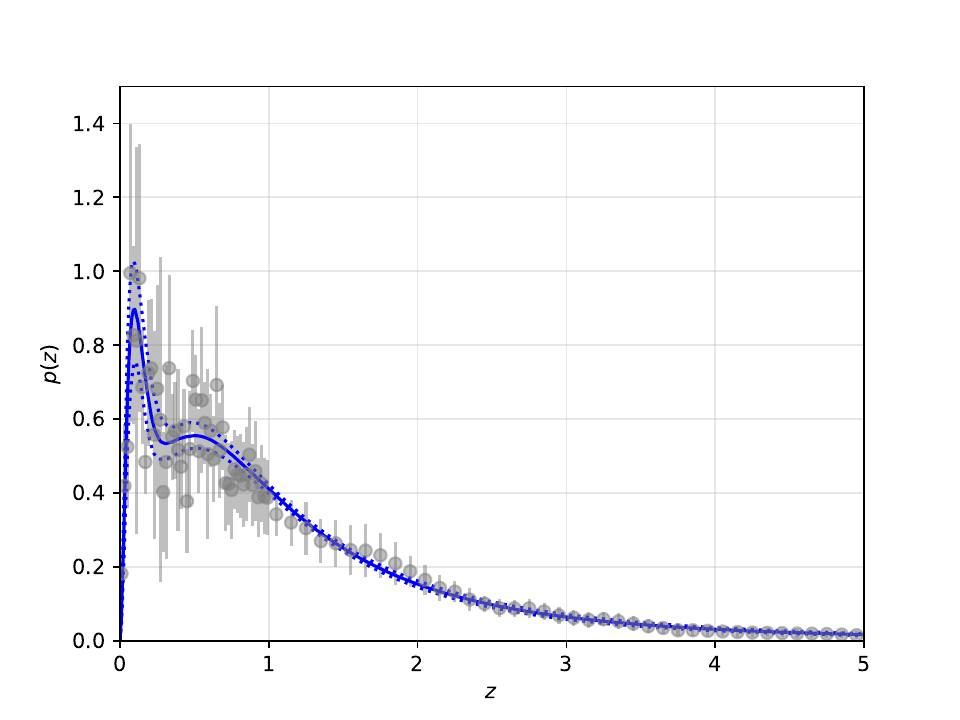}
    \caption{{Weighted redshift distribution generated from combining redshift distributions in the {LoTSS} deep fields (grey) as described in Section \ref{sec:nz}. The distribution of models fit to the resampled $p(z)$ as described in Section \ref{sec:bias} are presented as the median in blue (solid line) alongside the 16$^{\textrm{th}}$ and 84$^{\textrm{th}}$ percentiles respectively (as dashed lines).}}
    \label{fig:nz_fields}
\end{figure}

To determine the redshift distribution for the sources observed here, we {first apply a $1.5\,\mJy$ flux density cut to the cross-matched radio deep-field catalogues, matching that used here for LoTSS-DR2. {Specifically, we take an individual field and generate $N_f$ estimates for the redshift distribution, where $N_f$ across the three fields totals 1000 samples. The $N_f$ values are weighted for each field to gives more samples where there are larger number of $S\geq1.5$ mJy sources in the field.} To make a single resample within a field, we use those sources which have $S\geq1.5$ mJy and generate a resampled redshift for those sources {through the following process}. For those sources with a photometric redshift, {we sample from the full $p_i(z)$} distribution for the individual source. For those sources where a spectroscopic redshift exists, we instead consistently use the spectroscopic value. From the resampled redshifts for the $S\geq1.5$ mJy sources, we create a $p(z)$ by binning the redshifts and normalising the resultant distribution. When binning the redshift distribution, we use bins which have more frequent binning at low redshifts ($z\leq1$, using $\delta z = 0.02$, where we have more accurate spectroscopic information) and coarser binning at higher redshifts ($z>1$ , using $\delta z =0.1$)}\footnote{{We note that low redshifts also have an important contribution to $\omega(\theta)$ on larger angular scales ($\sim$O(1\degree)), and we found that averaging in larger redshifts bins affected the fitting of $\omega(\theta)$ on such scales.}}. {To generate the redshift distribution {across the fields}, we combine the samples from each field to produce 1000 resampled $p(z)$ distributions. From this we are able to determine a mean $p(z)$ distribution and associated errors from the standard deviations of the sample. The final $p(z)$ and errors is presented in Figure \ref{fig:nz_fields}.} 

{To use this $p(z)$ in our fitting and modelling of $b(z)$, we generate 1000 resampled $p(z)$ distributions using the mean and standard deviation across each redshift bin. We do this, as opposed to using the 1000 samples combined from the three fields, to avoid extreme models in each field that are driven by cosmic variance affecting such measurements, {as well as the effects of multi-wavelength data availability}. In order to ensure that such randomly sampled values does not lead to a highly varying $p(z)$ and satisfies $P(0)=0$. We model the resampled redshift distribution using a functional form given by:}

\begin{equation}
    p(z) \propto \frac{z^2}{1+z} \left[\exp\left(-\frac{z}{z_0}\right) + \frac{r^2}{(1+z)^a} \right],
    \label{eq:model_nz}
\end{equation}
{which we normalise such that it becomes a probability density function.} 

{Such a functional form is found to appropriately represent the redshift distribution, and was chosen to allow {contributions from AGN and SFGs} to the full redshift distribution. The form reflects the probed volume of a {$\Lambda$CDM} model at small redshifts with the exponential and power law terms representing the high luminosity cut-offs at large redshifts of SFGs and AGNs respectively (for more description see Nakoneczny et al. in prep). The model parameters ($z_0$, $r$ and $a$) are fit for each resample using \texttt{scipy}'s \texttt{curve\_fit} function.} {The range of the modelled redshift distribution from these resamples are presented in Figure \ref{fig:nz_fields}.}

{We note that with this method, the $\sim$5\% of LoTSS Deep Fields sources above} $1.5\,\mJy$ {which have} no associated redshift distribution cannot be included in $p(z)$. This may bias the results slightly, likely by missing some very high redshift AGN or SFGs {and those} which are {dust obscured}. {Furthermore, there are potential biases in the $p(z)$ due to the band selection and magnitude limits of the multi-wavelength data. For example, sources may not be detectable in all bands and there is differing availability of multi-wavelength data in the three deep fields, both of which will affect constraints which can be placed on their redshift distributions. Moreover, the deep fields are much smaller areas than the full LoTSS-DR2 survey, and so are more likely to be affected by variances in large-scale structures, however we mitigate this by averaging across the three fields. Finally, it is challenging to apply similar SNR cuts to the deeper LoTSS Deep Fields data, which may lead to residual systematics in the $p(z)$ models.} {However, this combined $p(z)$ is the best model available for a representative radio population and those {sources} without {any redshift information only {represent} a very} small fraction of sources {in the} data.}

\subsection{Measuring $b(z)$ using {\texttt{CCL}}}
\label{sec:b_pyccl}

In the first method to determine $b(z)$, we use {\texttt{CCL}} {to fit $\omega(\theta)$, assuming a bias model}. For this work we {follow the work of \cite{Alonso2021} and assume} two possible bias models either (i) a constant bias i.e.\ $b(z)=b_0$ or (ii) an evolving bias of the form $b_0$/$D(z)$, where $D(z)$ here is the normalised (to $z=0$) growth factor as described in e.g.\ \cite{Hamilton2001}. We also consider two matter power spectrum models (i) a `linear' model where only linear perturbation theory was assumed and (ii) a `HaloFit' {\citep{Smith2003, Takahashi2012}} model where non-linear effects within a dark matter halo are also accounted for. {Both models are considered {as we {may not expect to observe a strong contribution from `1-halo' clustering at the depth of this survey}, or that if such 1-halo contribution does exist that this may dominate predominately in the angular region where effects of multi-component sources is also important (see Figure \ref{fig:tpcf_dr1}).}} We use the {$\ell$} range {$1 \leq \ell \leq 10,000$} in 256 {logarithmically} spaced bins to generate the {$C_{\ell}$} power spectrum with {\texttt{CCL}} and then use this  to determine $\omega(\theta)$ over the $\theta$ range used in this work using a Legendre polynomial transform given by:
\begin{equation}
    \omega(\theta) = \frac{1}{4\pi} \sum_{{\ell}} (2\ell+1) C_{{\ell}} P_{{\ell}}(\cos\theta).
\end{equation}
{Such a conversion from $C_{{\ell}}$ to $\omega(\theta)$ was also used in}\ \cite{Siewert2020}. {To obtain $C_{{\ell}}$, we use the conversions in  {\texttt{CCL}} which convert the 3D power spectrum to $C_{{\ell}}$ using the equations in Section 2.4.1 of \cite{Chisari2019}, but assuming the redshift space distortion and magnification bias terms can be neglected:}
\begin{equation}
 {C_\ell = \int\frac{d\chi}{\chi^2}\,q^2(\chi)\,P\left(k=\frac{\ell+1/2}{\chi},z(\chi)\right),}
\end{equation}
{where {$\chi$} is the comoving radial distance, $P(k,z)$ is the matter power spectrum, and the radial kernel {$q(\chi)$} is:}
\begin{equation}
 {q(\chi)=\frac{H(z)}{c}\,b(z)\,p(z),}
\end{equation}
{with $H(z)$ the Hubble parameter. This relation relies on Limber's approximation {\citep{Limber1953, Limber1954}}, which is valid for the broad redshift distribution explored here.}

We fit for $b_0$ through calculating $\omega(\theta)$ with {\texttt{CCL}} {and fitting to the data using Equations \ref{eq:chi1} and \ref{eq:chi2}. Again, when fitting the data we consider three angular ranges: $0.03^{\circ}-5^{\circ}$, $0.1^{\circ}-5^{\circ}$ and $0.5^{\circ}-5^{\circ}$. We also consider {all possible combinations of linear and HaloFit models with the two bias evolutionary models.}} {To determine $b_0$ we use the 1000 redshift resamples described in Section \ref{sec:nz}. {Firstly, we calculate $\omega(\theta)$ for each resampled redshift distribution, assuming $b_0=1$ (denoted here as $\omega_{b_0=1}(\theta)$). Using this, we select random bias values within the range 0.5-3.5 and generate a model $\omega(\theta)$ through multiplying $\omega_{b_0=1}(\theta)$ by $b^2$. Using such a predicted model and comparing to the data we then calculate the associated {$\chi^2$} across the angular fitting ranges described above and calculate this both assuming only diagonal elements as well as using the full covariance matrix. The full covariance will highlight {if there are correlations in the $\omega(\theta)$ values at different $\theta$ which can affect the fitting of $b$.} In both cases we take the ``model'' to be the model produced from {\texttt{CCL}} with the integral constraint as modelled in \cite{Roche1999}, {though the contribution of an integral constraint will be negligible}.} Using such a {$\chi^2$} value we then calculate an associated probability for $b_0$ assuming {$P(b_0) \propto e^{-\chi^2/2}$} (which makes the assumption that errors on the data can be approximated as Gaussian). } 

{To determine final values of $b_0$ found from fitting our observations we then resample from $P(b)$. To do this, we consider two possibilities of how to include the redshift distribution to determine $b_0$. The first case assumes that the individual redshift resamples described in Section \ref{sec:nz} are all equally probable. In this case, any differences which may remain between the model and observations will reflect residual systematics in the data which are unaccounted for in the random catalogues or that a different bias evolution model is appropriate. For this method, we renormalise the $P(b)$ model from each redshift sample to 1. The second case assumes that there are no remaining systematics and so redshift resamples which better fit the data reflect the intrinsic $p(z)$ of our sample can be determined. In this case we do not normalise $P(b)$ for each sample to 1 before resampling and instead retain the difference in probabilities based on the magnitude of the {$\chi^2$}.} 

{Through resampling the data we determine $b_0$ accounting for the uncertainty in $p(z)$ models. In the first method, this means that the contribution of $p(z)$ samples from those models which satisfy the resampling criteria are approximately evenly distributed across the 1000 redshift resamples and, as such, some $p(z)$ samples may lead to large {$\chi^2$} values where the magnitude of the {$\chi^2$} for such a $p(z)$ was large. In the second method, there will instead be preferred $p(z)$ samples and others may not have any (or very little) contribution to the bias values which satisfy the resampling criteria, whilst other $p(z)$ models may substantially dominate the sample. This can lead to only a small fraction of $p(z)$ samples actually contributing to the fit, especially when the fit is poor. Due to this method, the associated {$\chi^2$} values of the fit will be lower to that of the previous method.} The $b_0$ values these are quoted as the median value with errors measured from the 16th and 84th percentiles and are presented in Table \ref{tab:fit2} and Figure \ref{fig:fit_models_evolve2}.  {To present associated models of $\omega(\theta)$ we use {10000 realisations} of the final $b_0$ sample to determine $\omega(\theta)$ models, this is shown in Figure \ref{fig:fit_models_evolve1} for the evolving and constant bias {models.} }

\subsection{Fitting $b(z)$ using Limber's equation {for a power law model of $\omega(\theta)$}}
\label{sec:limber}
The second commonly used method to infer the spatial clustering of galaxies from the angular clustering is by using Limber's equations {after assuming a power law model for $\omega(\theta)$} {\citep[see e.g.][]{Limber1953, Limber1954, Peebles1980}}. This method has been frequently employed in studies of the clustering of galaxies both at radio frequencies \citep[see e.g.][]{Lindsay2014, Hale2018, Chakraborty2020, Mazumder2022} and other frequencies {\citep[see e.g.][]{Puccetti2006, Starikova2012, Cochrane2017}}. To quantify $b(z)$, we use the fitting of $\omega(\theta)$ as described in Equation \ref{eq:omega}, discussed in Section \ref{sec:tpcf_results}, with the parameterisation of the spatial clustering:
\begin{equation}
    \xi_g(r) = \left( \frac{r}{r_0(z)} \right)^{-\gamma} = \left( \frac{r}{r_0} \right)^{-\gamma} (1+z)^{\gamma - (3+\epsilon)},
\end{equation}
where {$r_0$ is a spatial clustering length which parameterises the clustering of galaxies and} $\epsilon$ describes the evolving clustering model. {$\xi_g(r)$ is the spatial clustering of galaxies, as introduced in Section \ref{sec:intro}.} {We present $r_0$ and $b$ measurements using two assumptions for $\epsilon$: (i) assuming `comoving' clustering, where $\epsilon = \gamma-3$, {to make comparisons with previous studies \citep[e.g][]{Lindsay2014, Lindsay2014b, Hale2018, Mazumder2022} and (ii) assuming `linear' clustering\footnote{{We note that `linear' here does not refer to the mode used in \texttt{CCL} described earlier, but refers to an assumption of growth under linear perturbation theory, as discussed in \protect\cite{Lindsay2014}}.}, where $\epsilon = \gamma-1$, which probes a different range of bias evolution, see \cite{Lindsay2014}}. In order to determine the spatial clustering, we need both knowledge of $\gamma$ and $A$ from Equation \ref{eq:omega} as well as $p(z)$ to determine the spatial clustering length, $r_0$. {As discussed, {in the majority of cases we fix $\gamma$ to a value of 1.8, though we also consider the case for a variable $\gamma$ for comparison}}. The value of $r_0$, can then be calculated using Limber's equation {\citep[see e.g.][]{Limber1953, Limber1954, Peebles1980}}:}
\begin{equation}
    r_0 = \left(\frac{A_{\textrm{r}} \ c \ \left(\int_{0}^{\infty} p(z) dz\right)^2 }{H_{\gamma} H_0 \int_{0}^{\infty} E(z)^{\frac{1}{2}} p(z)^2 \chi(z)^{1-\gamma} (1+z)^{\gamma - (3+\epsilon)} dz}\right)^{\frac{1}{\gamma}},
    \label{eq:r0}
\end{equation}
\noindent {where $c$ is {the speed of light} in km s$^{-1}$}, {{$E(z) = \Omega_{m} (1+z)^3 +(1 - \Omega_m)$}} and {$\chi(z)$} is the comoving distance at redshift, $z$. $A_r$ is related to the amplitude ($A$) in Equation \ref{eq:omega} when $\theta$ is in the unit of radians. Finally, $H_{\gamma}$ is given by:
\begin{equation}
    H_{\gamma} = \frac{\Gamma(\frac 12) \Gamma(\frac{\gamma-1}{2})}{\Gamma(\frac{\gamma}{2})},
\end{equation}
\noindent where $\Gamma$ represents the gamma function. As described in {Section \ref{sec:intro} and Equation \ref{eq:bias}, the spatial clustering of galaxies can be related to that of matter to parameterise galaxy bias. Following analysis from \cite{Peebles1980} and discussed and used in works such as \cite{Koutoulidis2013, Lindsay2014, Hale2018, Mazumder2022}}, the bias can then be inferred from $r_0$ using:
\begin{equation}
    b(z) = \left(\frac{r_0(z)}{8 \textrm{Mpc} h^{-1} }\right)^{\gamma/2} \frac{J_2^{1/2}}{\sigma_8 D(z)/D(0)}, 
    \label{eq:bias_limber}
\end{equation}
where $D(z)$ is the growth factor, {and $J_2$ is given by $\frac{72}{2^{\gamma}(3-\gamma)(4-\gamma)(6-\gamma)}$} and $z$ is evaluated at the {median redshift of the redshift distribution (which is found here to be $z_{m}\approx 0.9$ for the full redshift distribution)}. 

In order to perform this fitting, we use the fit for {$\omega(\theta)$} described in Section \ref{sec:tpcf_results} {and the modelled resampled redshift distributions (using Equation \ref{eq:model_nz}) described in Section \ref{sec:b_pyccl}. We calculate $r_0$ and $b$ and their associated uncertainties} by using {5000} random values of {$\log_{10}(A)$ (and $\gamma$ for a two parameter model)} {from our sample which were generated to fit $A$ in Section \ref{sec:tpcf_results} and evaluate these using the random samples for the $p(z)$ distribution to then quantify $b(z)$}. {Using this method, we have no reason a priori to assume a certain redshift distribution and so use the 1000 modelled $p(z)$ resamples equally to calculate $b$. This is therefore most comparable to the first resampling method described in Section \ref{sec:b_pyccl}.} {From the $r_0$ and $b$ samples we then quantify the median value as well as the errors from the 16$^{\textrm{th}}$ and 84$^{\textrm{th}}$ percentiles. }

{We note though, that using Limber inversion used in this method does make assumptions, which could affect the results presented. These assumptions include that the angles considered are small. At larger angles, approximations in Limber's equation break down and $\omega(\theta)$ deviates from a power law. {{For the majority of angular fitting ranges considered (up to 5\degree), these use large scales where deviations from a power law are expected. Therefore, we also considered the fitting range for the power law fitting of $A$, $0.03 \leq \theta < 1^{\circ}$, as discussed in Section \ref{sec:tpcf_results} where such a power law distribution appears appropriate}.} Moreover, {assumptions are used to obtain Limber's equation, which can include} that $r_0$ is independent of luminsosity; this is likely not be the case \citep[see e.g.][]{Zehavi2011, Cochrane2017}, however without an ability to split by luminosity for our sources, {our analysis} will give an average value across the population. {We continue to present the bias measurements from this method} as a number of previous radio clustering papers \citep[as well as at other wavelengths, see e.g.][]{Lindsay2014, Magliocchetti2017, Hale2018, Chakraborty2020, Mazumder2022} all determine $r_0$ and bias through this {method and so allows for comparison with previous works}. }

{We note that CCL also uses Limber's inversion in order to obtain a measurement of the bias, but does not rely on assumptions about a power law functional form for $\omega(\theta)$ and $\xi_g(r, z)$ and accounts for the deviation from a power law at the largest angular scale. Therefore, different results for the bias may be obtained through these different models and we present results for measurements of $b$ from both methods to make direct comparison of the results obtained. } 

\subsection{$\omega(\theta)$ and $b(z)$ models}

\begin{figure*}
    \centering
         \begin{subfigure}[b]{\textwidth}
         \centering
         \includegraphics[width=0.95\textwidth]{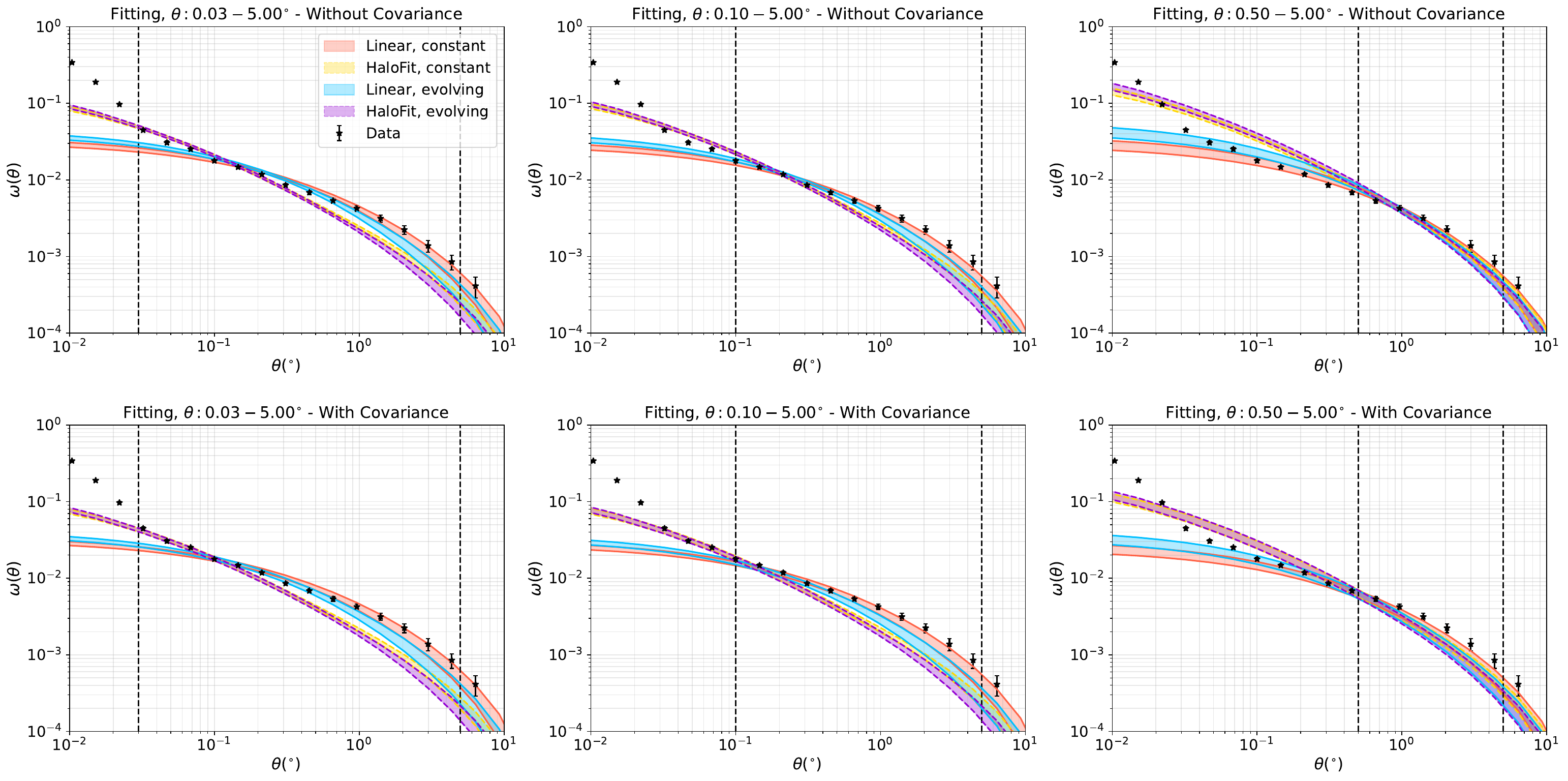}
         \caption{{Using method 1 - when each $p(z)$ sample is equally weighted.}}
         \label{fig:fit_models_evolve1a}
     \end{subfigure}
    \begin{subfigure}[b]{\textwidth}
         \centering
         \includegraphics[width=0.95\textwidth]{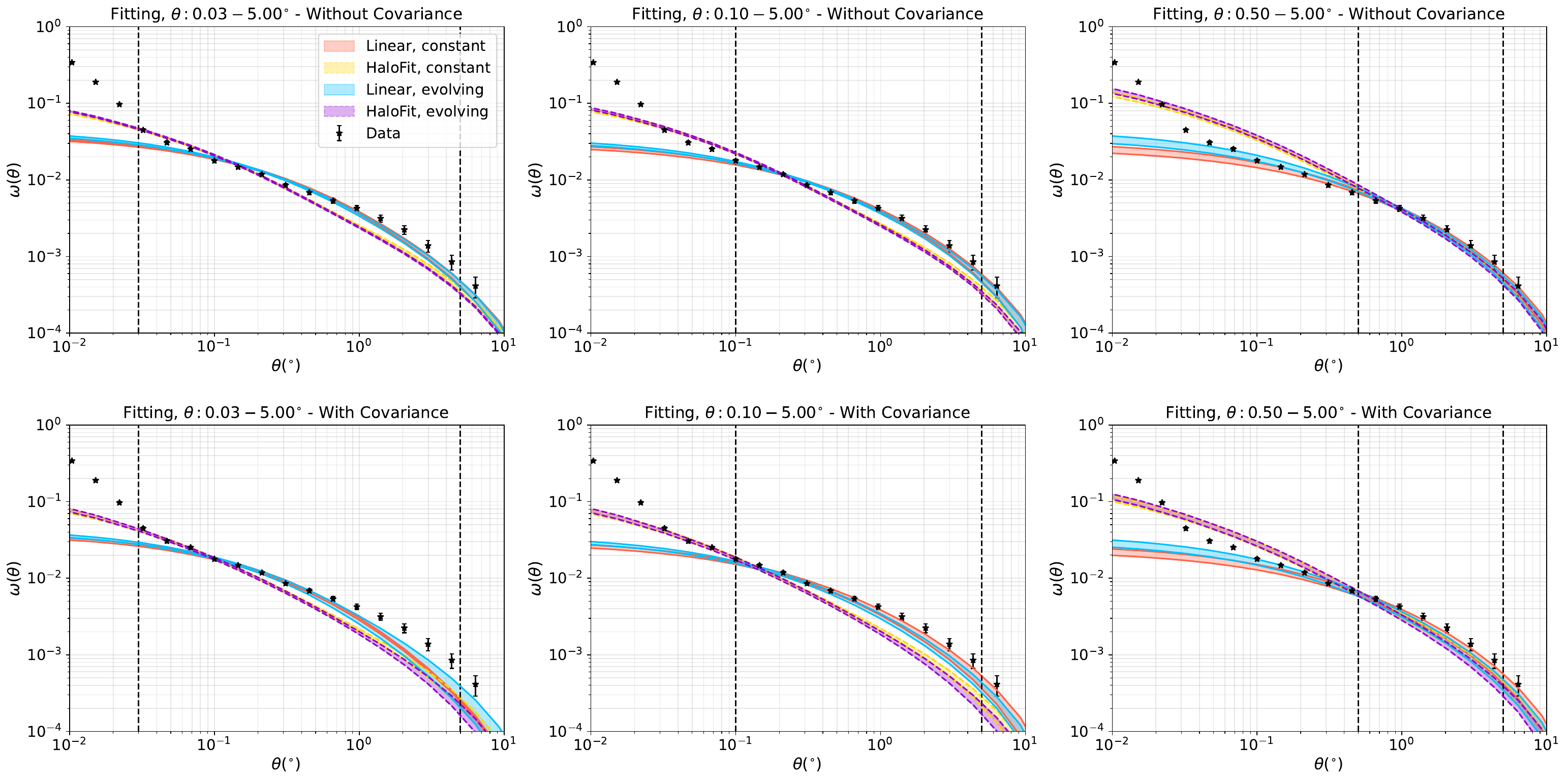}
        \caption{{Using method 2 - when each $p(z)$ sample is not equally weighted.}}
        \label{fig:fit_models_evolve1b}
    \end{subfigure}         
     \caption{{Comparisons of {$\omega(\theta)$ for LoTSS-DR2 and their modelled fits (subtracting the integral constraint) assuming errors without accounting for covariance between $\theta$ bins (upper row of each sub figure) and using the full covariance matrix is shown (lower row of each sub figure). These models are shown for the angular fitting ranges $0.03-5\,\deg$ (left), $0.1-5\,\deg$ (centre) and $0.5-5\,\deg$ (right), with the dashed vertical lines indicating the angular scales used for fitting. Black stars correspond to the measurements from LoTSS-DR2, and the shaded regions correspond to (i) the linear constant bias model (red), (ii) the HaloFit constant bias model (yellow), (iii) the linear evolving bias model (blue) and (iv) the HaloFit evolving bias model (purple). The upper panel presents the results when all redshift resamples are weighted equally, whilst the lower panel allows preferential $p(z)$ resamples to be weighted preferentially. }}}
    \label{fig:fit_models_evolve1}
\end{figure*}

\begin{figure*}
    \centering
         \includegraphics[width=0.95\textwidth]{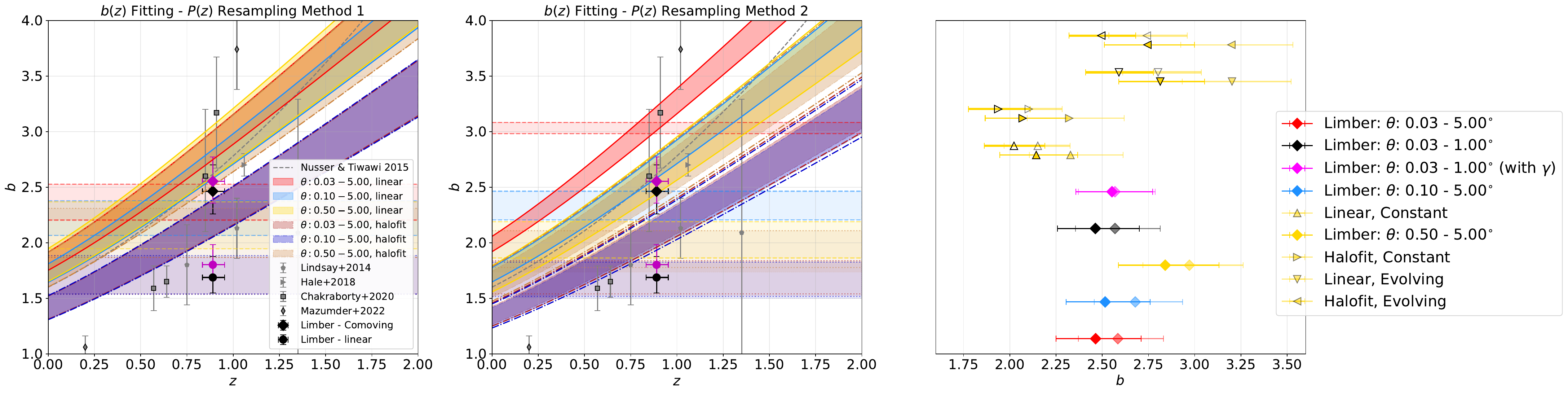}
    \caption{{Comparisons of the bias models fit {(using the full covariance)} for the data for a constant bias model and evolving bias model for the three angular fitting ranges: $0.03-5\,\deg$ (red), $0.1-5\,\deg$ (blue) and $0.5-5\,\deg$ (yellow) for linear (lighter colours) and HaloFit (darker colours) models. The left panel shows {the results when each $p(z)$ resamples is weighted as equally probable (method 1, Section \protect\ref{sec:b_pyccl}) and the centre panel shows the results when preferential $p(z)$ models are upweighted (method 2, Section \protect\ref{sec:b_pyccl}).} This is presented alongside previous measurements from \protect\citet{Nusser2015} (grey dashed line), \protect\citet{Lindsay2014} (grey pentagons), \protect\citet{Hale2018} (grey triangles), \protect\citet{Chakraborty2020} (grey squares) and \protect\citet{Mazumder2022} (grey diamonds). Also shown are the fitting of $b(z)$ from Equation \ref{eq:bias_limber} using the angular fitting range 0.03-1\degree \ {(evaluated at the median redshift of the sample)} for the fixed slope ($\gamma$) model (black) and 2 parameter model (magenta) for both the comoving (diamond) and linear (circle) Limber models. The right hand panel shows a comparison of the bias values (evaluated at $z_{med} \approx 0.89$) from \texttt{CCL} (in the 0.5-5\degree \ fitting range) using the linear constant (up facing triangle), HaloFit constant (right facing triangle), linear evolving (down facing triangle) and HaloFit evolving (left facing triangle) with and without covariance (indicated by a fainter symbol). {The filled markers for the CCL fitting represent those models where the $p(z)$ samples are uniformly weighted and open markers indicate where a preferential $p(z)$ model was preferentially selected. These are presented alongside} the Limber comoving linear models across the three angular fitting ranges. {Values on the right hand panel} are shown with an arbitrary offset on the y-axis to highlight the differences in the values.}}
    \label{fig:fit_models_evolve2}
\end{figure*}

We present the results from fitting $\omega(\theta)$ assuming the evolving bias and constant bias model in Figure \ref{fig:fit_models_evolve1}. For each model we present the fits using the {three} different angular ranges described above, for both the diagonal only errors and also the full covariance array. The associated bias models are {then presented in Figure \ref{fig:fit_models_evolve2} along with the values from the Limber method assuming a power law distribution of $\omega(\theta)$, with additional comparisons} to previous results from analysis of the large area NVSS survey \citep{Nusser2015} as well as other individual measurements of bias evaluated at specific redshifts from \cite{Lindsay2014, Hale2018, Chakraborty2020, Mazumder2022}. The results of such fitting for both the power law amplitude, spatial clustering length ($r_0$) and bias for both the Limber and {\texttt{CCL}} derived bias models are also provided in Tables \ref{tab:fit1} and \ref{tab:fit2}. A comparison of the amplitude fit assuming a power law distribution as in Equation \ref{eq:omega} is also presented in Figure \ref{fig:amp_comp} compared to the work of \cite{Lindsay2014, Hale2019, Siewert2020, Bonato2021, Mazumder2022}. As these surveys are at different frequencies and flux density limits {(shown in the inset)}, {this may affect the populations observed and hence the estimated biases for such sources, and so {an equivalent survey limit scaled to 144 MHz is used}. We note that Figure \ref{fig:fit_models_evolve2} includes the bias values from the 2 parameter fitting model compared to the fixed slope model, which appear in good agreement} 

\begin{table*}
    \centering
    \begin{tabular}{ccccccccccc}
$\theta$ Range  & Fitting Type & $\textrm{log}_{10}(A)$ & $r_{0,c}$ (Mpc) & $b_c(z_m)$ & $r_{0,l}$ (Mpc)  & $b_l(z_m)$ & \\ 
  (\degree) &  &  &  &  &  & & \\  \hline \hline
0.03 - 5.00 & Without Cov &$-2.50^{+0.01}_{-0.01}$  &$11.55^{+0.92}_{-0.77}$ &$2.58^{+0.25}_{-0.21}$ &  $15.41^{+1.99}_{-1.44}$ & $1.77^{+0.20}_{-0.14}$  \\ 
0.10 - 5.00 & Without Cov &$-2.47^{+0.01}_{-0.01}$  &$12.02^{+0.96}_{-0.81}$ &$2.68^{+0.26}_{-0.22}$ &  $16.04^{+2.06}_{-1.50}$ & $1.83^{+0.20}_{-0.15}$  \\ 
0.50 - 5.00 & Without Cov &$-2.38^{+0.02}_{-0.02}$  &$13.51^{+1.11}_{-0.96}$ &$2.97^{+0.29}_{-0.25}$ &  $18.03^{+2.33}_{-1.74}$ & $2.04^{+0.23}_{-0.17}$  \\ 
0.03 - 1.00 & Without Cov &$-2.50^{+0.01}_{-0.01}$  &$11.48^{+0.92}_{-0.77}$ &$2.57^{+0.25}_{-0.21}$ &  $15.32^{+1.97}_{-1.43}$ & $1.76^{+0.20}_{-0.14}$  \\ 
0.03 - 5.00 & With Cov &$-2.54^{+0.01}_{-0.01}$  &$10.96^{+0.88}_{-0.75}$ &$2.46^{+0.24}_{-0.21}$ &  $14.63^{+1.88}_{-1.38}$ & $1.69^{+0.19}_{-0.14}$  \\ 
0.10 - 5.00 & With Cov &$-2.52^{+0.02}_{-0.02}$  &$11.22^{+0.91}_{-0.78}$ &$2.51^{+0.24}_{-0.21}$ &  $14.97^{+1.93}_{-1.43}$ & $1.72^{+0.19}_{-0.14}$  \\ 
0.50 - 5.00 & With Cov &$-2.42^{+0.03}_{-0.03}$  &$12.83^{+1.13}_{-1.01}$ &$2.84^{+0.29}_{-0.25}$ &  $17.14^{+2.26}_{-1.76}$ & $1.95^{+0.22}_{-0.18}$  \\ 
0.03 - 1.00 & With Cov &$-2.54^{+0.01}_{-0.01}$  &$10.96^{+0.88}_{-0.75}$ &$2.46^{+0.24}_{-0.21}$ &  $14.62^{+1.88}_{-1.38}$ & $1.69^{+0.19}_{-0.14}$  \\ 
\end{tabular}
    \caption{{Results from fitting $\omega(\theta)$ for models across a range of angular fitting ranges. Presented is the fitting range, Fitting type, amplitude of power law ($A$) as in Equation \ref{eq:omega}, clustering length, $r_0$, and bias, $b_L$, from Limber inversion using both a comoving (c) and linear (l) assumption. Bias values are evaluated at the median value  of the median redshifts ($z_m$) from the $p(z)$ resamples, as in Figure \ref{fig:nz_fields}, {$z_{m}\approx 0.89$}. This is for both the case where the full covariance matrix is (With Cov) and is not (Without Cov) used.}}
    \label{tab:fit1}
\end{table*}

\begin{table*}
    \centering
    \begin{tabular}{ccccccccccccccc}
$\theta$ Range  & Fit  & $b_{0,L}$ & $b_{L}(z_{m})$ & {$\chi_{L}^2$/} & $b_{0, H}$ & $b_{H}(z_{m})$ & {$\chi_{H}^2$/}  & $b_{0,L}$ & $b_{L}(z_{m})$ & {$\chi_{L}^2$/} & $b_{0,H}$ & $b_{H}(z_{m})$& {$\chi_{H}^2$/} \\ 
  (\degree)  & Type &  &  & DOF &  &  & DOF & Cov& Cov& DOF & Cov & Cov & DOF\\ 
  &  &  &  &  &  &  &  & & & Cov &  &  & Cov\\  \hline \hline
0.03 - 5.00 & E/U & $1.90^{+0.10}_{-0.09}$  &$2.97^{+0.15}_{-0.15}$ & 9.34 &$1.51^{+0.12}_{-0.10}$ &  $2.37^{+0.19}_{-0.16}$ & 13.69 & $1.83^{+0.08}_{-0.08}$  &$2.87^{+0.13}_{-0.13}$ & 10.50 &$1.41^{+0.11}_{-0.10}$ &  $2.21^{+0.18}_{-0.15}$ & 4.43\\ 
0.10 - 5.00 & E/U & $1.83^{+0.10}_{-0.10}$  &$2.87^{+0.16}_{-0.15}$ & 4.12 &$1.57^{+0.13}_{-0.11}$ &  $2.46^{+0.21}_{-0.17}$ & 14.58 & $1.73^{+0.08}_{-0.08}$  &$2.71^{+0.13}_{-0.13}$ & 2.73 &$1.41^{+0.12}_{-0.10}$ &  $2.21^{+0.18}_{-0.16}$ & 5.52\\ 
0.50 - 5.00 & E/U & $2.04^{+0.20}_{-0.17}$  &$3.20^{+0.32}_{-0.27}$ & 3.53 &$2.04^{+0.21}_{-0.17}$ &  $3.20^{+0.33}_{-0.27}$ & 4.49 & $1.79^{+0.15}_{-0.14}$  &$2.81^{+0.24}_{-0.22}$ & 3.18 &$1.75^{+0.16}_{-0.15}$ &  $2.74^{+0.25}_{-0.23}$ & 4.05\\ 
0.03 - 5.00 & C/U & $2.37^{+0.19}_{-0.17}$  &-& 12.74 &$1.79^{+0.20}_{-0.15}$ & -& 11.40 & $2.36^{+0.17}_{-0.15}$  &- & 14.01 &$1.68^{+0.19}_{-0.14}$ & - & 3.95\\ 
0.10 - 5.00 & C/U & $2.27^{+0.19}_{-0.16}$  &- & 2.24 &$1.87^{+0.22}_{-0.16}$ &  -& 11.62 & $2.21^{+0.16}_{-0.15}$  &- & 3.05 &$1.69^{+0.20}_{-0.15}$ &  - & 4.81\\ 
0.50 - 5.00 & C/U & $2.33^{+0.28}_{-0.22}$  &- & 1.76 &$2.32^{+0.30}_{-0.23}$ &  - & 2.63 & $2.14^{+0.22}_{-0.20}$  &- & 1.81 &$2.07^{+0.24}_{-0.20}$ & - & 2.79\\  \\  \hline \\
0.03 - 5.00 & E/W & $1.98^{+0.05}_{-0.06}$  &$3.11^{+0.07}_{-0.10}$ & 7.81 &$1.18^{+0.01}_{-0.01}$ &  $1.84^{+0.02}_{-0.02}$ & 10.27 & $1.97^{+0.09}_{-0.05}$  &$3.09^{+0.14}_{-0.08}$ & 9.07 &$1.35^{+0.08}_{-0.08}$ &  $2.11^{+0.13}_{-0.13}$ & 4.19\\ 
0.10 - 5.00 & E/W & $1.69^{+0.04}_{-0.07}$  &$2.66^{+0.06}_{-0.11}$ & 1.46 &$1.21^{+0.06}_{-0.02}$ &  $1.90^{+0.09}_{-0.03}$ & 11.49 & $1.71^{+0.07}_{-0.06}$  &$2.68^{+0.11}_{-0.10}$ & 2.15 &$1.33^{+0.08}_{-0.09}$ &  $2.09^{+0.13}_{-0.13}$ & 5.18\\ 
0.50 - 5.00 & E/W & $1.81^{+0.15}_{-0.12}$  &$2.84^{+0.24}_{-0.19}$ & 1.46 &$1.78^{+0.14}_{-0.13}$ &  $2.79^{+0.22}_{-0.21}$ & 2.59 & $1.67^{+0.12}_{-0.12}$  &$2.62^{+0.19}_{-0.18}$ & 1.86 &$1.62^{+0.12}_{-0.11}$ &  $2.54^{+0.19}_{-0.18}$ & 3.11\\ 
0.03 - 5.00 & C/W & $2.77^{+0.17}_{-0.15}$  &- & 9.54 &$1.49^{+0.22}_{-0.18}$ &  - & 9.20 & $3.04^{+0.05}_{-0.06}$  &- & 10.63 &$1.67^{+0.17}_{-0.13}$ &   & 3.83\\ 
0.10 - 5.00 & C/W & $2.28^{+0.13}_{-0.11}$  & & 1.68 &$1.57^{+0.15}_{-0.20}$ &  - & 10.56 & $2.33^{+0.13}_{-0.13}$  &- & 2.59 &$1.65^{+0.18}_{-0.13}$ &  - & 4.65\\ 
0.50 - 5.00 & C/W & $2.15^{+0.17}_{-0.18}$  & - & 0.74 &$2.10^{+0.18}_{-0.18}$ &  - & 1.67 & $2.02^{+0.17}_{-0.16}$  &- & 1.03 &$1.94^{+0.17}_{-0.16}$ &  - & 2.24\\ 
\end{tabular}
    \caption{{Results from fitting bias with \texttt{CCL} across a range of angular fitting scales, with both the linear ($b_{L}$) and HaloFit ($b_{H}$) models of \texttt{CCL}. These are both given by their value at $z=0$ and, for the evolving bias model, are evaluated at the median value  of the median redshifts ($z_m$) from the $p(z)$ resamples, as in Figure \ref{fig:nz_fields}, {$z_{m}\approx 0.89$}. These are given for both the case where the full covariance matrix is not used and where it is included (denoted by Cov). For each model the median reduced 2{{$\chi^2$} ({$\chi^2$}/DOF)} from the resampled bias values is also given. This will be larger than the best fit model found across the samples, but is provided to show representative values for the fit. A fit type is given by the combination of the bias evolution type (E=evolving, C=constant) and redshift resampling method (U = unweighted i.e. all $p(z)$ samples weighted equally and W = weighted i.e. preferential $p(z)$ resamples are selected).}}
    \label{tab:fit2}
\end{table*}

\section{Discussion}
\label{sec:discussion}

In this section we shall discuss our results in context of the different models used to fit the data as well as comparing to previous studies of the angular clustering of radio sources. 

\subsection{Comparing {\texttt{CCL}} derived models for $\omega(\theta)$ and $b(z)$}
First we compare the fitting of $\omega(\theta)$ using the linear and HaloFit models of {\texttt{CCL}}. As can be seen in Figure \ref{fig:fit_models_evolve1}, the fit of $\omega(\theta)$ using the linear model appears to have {relatively} good agreement with {the data} across the angular range 0.06-1\degree \ using all three angular fitting ranges considered in this work {when using the more simplistic {$\chi^2$} for both the evolving and constant bias models.} {Above 1\degree, the evolving bias model appears to underestimate slightly $\omega(\theta)$, compared to the constant bias model, especially when using fitting ranges that cover the largest angular range and the full covariance is considered. As the full covariance accounts for correlations between different angular bins, this allows the model to under-predict $\omega(\theta)$ on these scales relative to what might be expected by simply looking at minimizing {$\chi^2$} using the diagonal errors on $\omega(\theta)$ only. However, such an effect is less notable in Figure \ref{fig:fit_models_evolve1b} where we allow the $p(z)$ resamples to be preferentially selected to best fit the model. Below 0.06\degree, the measured value for $\omega(\theta)$ appears to be larger than expected from the linear model for both the evolving and constant bias models, with an even larger discrepancy for $\theta<0.03$\degree, where we believe the effect of multi-component sources within the LoTSS-DR2 survey is important. On the contrary, {the HaloFit model, {shows} greater agreement with $\omega(\theta)$ for $\theta\leq0.06$\degree \ when fitting with minimum angular scales $\theta \leq 0.1^{\circ}$. However, in doing so these models greatly underestimate $\omega(\theta)$ on the majority of larger angular scales ($\theta \geq 0.1$\degree), which is where linear bias is dominating. This results in significantly larger reduced {$\chi^2$} values compared to the linear models. For the narrowest angular fitting range (fitting between 0.5-5\degree), instead, there is much better agreement with the measured $\omega(\theta)$ on the largest angular scales (comparable to that when using a linear model), but the model significantly over predicts the clustering at angles $\lesssim0.5^{\circ}$.}}

{This comparison suggests that neither the linear or HaloFit models {can completely} {reproduce} the measured $\omega(\theta)$ across the full range of angular scales presented in Figure \ref{fig:fit_models_evolve1}, though above the angular scale where we believe the effects of multi-component sources is negligible ($\theta\geq0.03$\degree), the linear models are able to much more accurately fit the data across a wider range of angular scales using both $p(z)$ resampling methods}. The linear and HaloFit models should agree on the largest angular scales and only deviate at small angular scales due to the `1-halo' clustering from sources within the same dark matter halo. When measuring the linear bias, where we measure the `2-halo' clustering relating to galaxies in different dark matter haloes, it is important that {the model $\omega(\theta)$ from the fitting} be an accurate representation on the largest angular scales. {Therefore, the bias measured by the HaloFit models using the angular ranges 0.03-5\degree \  and 0.1-5\degree \ {appears} to underestimate $\omega(\theta)$ on the largest angular scales compared to the linear models and so will underestimate the linear bias. These {should therefore not be used to draw conclusions of $b_0$}. {When fitting for angular scales of $\theta\geq 0.5$\degree \ there is better agreement between the linear and HaloFit models and so measurements of bias from such models are more likely to represent the true bias.}}

{Given that cross-matched data for the LoTSS-DR2 is not currently {available for the full LoTSS-DR2 sample, {and instead cross-matching} is only complete above 8 mJy {\citep{Hardcastle2023}}}}, it is not possible to conclusively determine whether we do have a {significant} contribution of {1-halo clustering to $\omega(\theta)$ in this work.} {However,} from the LoTSS-DR1 clustering measurements shown in Figure \ref{fig:tpcf_dr1}, the correction for multi-component sources is relatively small and would be insufficient to explain the excess clustering seen here at small angular scales {($\theta \lesssim$0.03\degree)}. {This therefore suggests that we are indeed observing some 1-halo clustering within LoTSS-DR2. Given the uncertainty in the effect of multi-component sources, however, we are also unable to do a full halo occupation distribution modelling \citep[HOD; see e.g.][]{Berlind2002, Zheng2005} in order to determine properties of the haloes which allow them to host multiple radio sources of the type observed in this data.}

{At the largest angular scales, we note that the linear and HaloFit models are slightly lower than the measured $\omega(\theta)$ from the data when the full covariance is used (especially when uniform weighting is used for each $p(z)$ resample). This may suggest that residual systematics remain within the data which are not fully captured by the randoms but are accounted for by the covariance. Alternatively, it could also represent a contribution of the radio dipole to the observed TPCF, which can cause an excess clustering at larger angular scales \citep[see][]{ChenSchwarz}, but is not included in our models. More likely, these differences could suggest the assumed bias models used in this analysis may be too simplistic for the sources observed in this work. Our sample is a combination of different sources types and luminosities which dominate at different redshift ranges and so contribute differently across the redshift distribution. Such sub-populations have different bias evolution models \citep[see e.g.][]{Magliocchetti2017, Hale2018, Chakraborty2020, Mazumder2022}, which are complex to combine when considering only a single population. As we are unable to separate the LoTSS-DR2 sources into different source classes we rely on more simplistic models to probe the population as an average population, until the time where such sources can be studied in greater detail, split by source type. {Such studies which account for differences in bias models are more beneficial for those data where sources have been associated with a galaxy host, assigned a redshift and source classification has been undertaken to identify the source type. This will be aided in future over such large sky areas with WEAVE-LOFAR \citep{Smith2016}, where spectra can be used to attribute redshifts to sources and to classify the source type. At present, though, such studies should focus on deep, multi-wavelength fields, as in the recent works of \cite{Hale2018, Chakraborty2020, Mazumder2022}.}}

\begin{figure}
    \centering
    \includegraphics[width=8.cm]{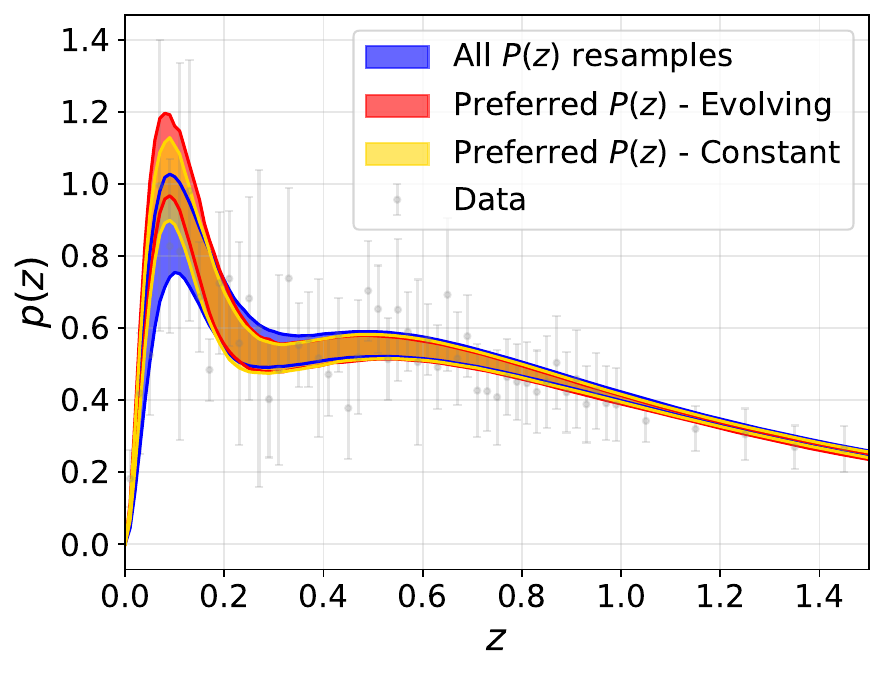}
    \caption{{$p(z)$ for the data (grey) compared to the range of $p(z)$ models when uniformly sampling the data (blue) compared to allowing the $p(z)$ resamples to preferentially selected in the fitting process (see Section \protect\ref{sec:b_pyccl}) for a linear model across the angular range 0.5-5\degree \ using the full covariance array with an evolving (red) and constant (yellow) bias model. All models are shown in the range given by the 16th and 84th percentiles.}}
    \label{fig:nz_withresamp}
\end{figure}

{Alternatively, if the systematics within the data have been fully accounted for it could imply that the true $p(z)$ is different from that currently estimated from the LoTSS Deep Fields. Figure \ref{fig:nz_withresamp} shows the preferred $p(z)$ models (using a linear model, fit over the angular range 0.5-5\degree), which favour a model with a greater fraction of sources at these low redshifts. As discussed, this provides a much better fit to the data at the largest angular scales than using a uniform weighting of our resampled $p(z)$ models, reflected in the smaller average {$\chi^2/DOF$} values for our samples. For other angular fitting ranges which may give poorer fits to the data, the preferred $p(z)$ may shift to higher or lower redshifts, however we present the 0.5-5\degree \ range which we believe is the most trustworthy to measure linear bias. We note that over the 0.5-5\degree \ fitting range, the measured bias values presented in Table \ref{tab:fit2} are lower using the weighted $p(z)$ resampling, but are consistent with one another within $\sim$ 1$\sigma$. Discerning between whether we expect a $p(z)$ with a stronger preference to low redshift sources or that there are residual systematics in our data is challenging, but will be aided with future spectroscopic surveys such as WEAVE-LOFAR \citep{Smith2016}.  }

Next, we consider the comparison between the evolving, $b(z) = b_0/D(z)$, and constant, $b(z)=b_0$, bias models for our data, as presented in Figure \ref{fig:fit_models_evolve2}. {The model {used in} analysis of NVSS in \cite{Nusser2015} {was an} evolving bias model and we also note that for previous measurements using Limber inversion, the choice of comoving clustering assumes a non-evolving $r_0$ and so an evolving bias model inversely proportional to the growth factor, as can be seen in Equation \ref{eq:bias_limber}}. As can be seen in Figure \ref{fig:fit_models_evolve1}, both the evolving and constant bias model appear to accurately recreate the observed angular TPCF across a diverse range of angular scales ($\sim 0.07-1^{\circ}$). However, whilst at $\theta \sim 1-5^{\circ}$ the model for $\omega(\theta)$ using the constant bias model (and assuming equal weighting for our $p(z)$ resamples, see Figure \ref{fig:fit_models_evolve1a}) can be seen to {better model $\omega(\theta)$ at the largest angular scales,} the evolving bias model under predicts the observed angular-two-point correlation function. This would therefore imply that a constant bias model appears to more accurately represent the measurements made in this work. {However, in the literature, bias models which evolve and increase with redshift have typically been assumed due to expectations that at higher redshifts a halo of the same mass represents a more extreme fluctuation from the average, and so is more biased. In SKADS \citep{Wilman2008}, the authors used an assumption of a constant mass haloes for each different source population, these result in an evolving bias model for such an assumption. These models have been used in numerous cosmology forecasts {\citep{Raccanelli2012, Ferramacho2014, Bacon2020}}. The model used in the analysis presented in this work, however, includes a more simplistic evolving bias model, inversely proportional to the growth factor, and more complicated evolutionary models taking into account the contributions of different source populations are likely more appropriate.} {If the $p(z)$ resamples are allowed to be preferentially chosen to best fit the data (see Figure \ref{fig:fit_models_evolve1b}), the constant and evolving bias models both appear to become more similar compared to the measurements of $\omega(\theta)$. }

Finally, comparisons can be made for the results when using the full covariance matrix, compared to errors based on the diagonals of the covariance matrix. Work such as \cite{Lindsay2014} and \cite{Hale2018} have followed methods where only the uncertainties on a $\theta$ bin and not the full covariance matrix was assumed, which could affect the measurements of bias. As can be seen in Figures \ref{fig:fit_models_evolve1}-\ref{fig:fit_models_evolve2} and in Tables \ref{tab:fit1}-\ref{tab:fit2}, there do exist differences in the measured bias and $\omega(\theta)$ models depending on whether or not the full covariance matrix is provided. These often find a lower bias value when the full covariance matrix is {used, although the values are typically consistent} within 1-2$\sigma$. Differences between the results with and without the full covariance imply a correlation between angular scales which needs to be accounted for in the fitting of $\omega(\theta)$. We therefore use the models in which the full covariance is incorporated for drawing conclusions. {We also note that when weighting all $p(z)$ resamples equally (and modelling these as in Equation \ref{eq:model_nz}), the results when using the {covariance} matrix from \texttt{TreeCorr} {(with $N_{\textrm{Jack}}=100$)} were consistent within $\sim1\sigma$ and using a $\delta z=0.1$ binning for the $p(z)$ from the LoTSS Deep fields also resulted in $b_0$ values consistent within $\sim1-1.5\sigma$ to those presented in this work. }

\subsection{{Comparison of $b(z)$ to other surveys}}

We next present comparisons to the results made from previous measurements with similar large area surveys. As this work follows from the previous work of LoTSS-DR1 presented in \cite{Siewert2020}, we first make comparisons to the results found in {that} work. In \cite{Siewert2020}, redshifts were not available for the full population of LoTSS-DR1 sources and no redshift data for LOFAR sources in the Deep Fields were available at that time. Therefore for bias measurements this relied on those sources which had cross-matched hosts \citep[from][]{Williams2019} and redshifts \citep[from][]{Duncan2019}. This meant that approximately 50\% of sources had redshifts available, but that measurements of bias in redshift bins were skewed to those sources. Therefore it is challenging to make direct comparisons to that work. However, it is possible to make comparisons to the fitting parameters for $\omega(\theta)$ provided in \cite{Siewert2020}.

\begin{figure}
    \centering
    \includegraphics[width=8cm]{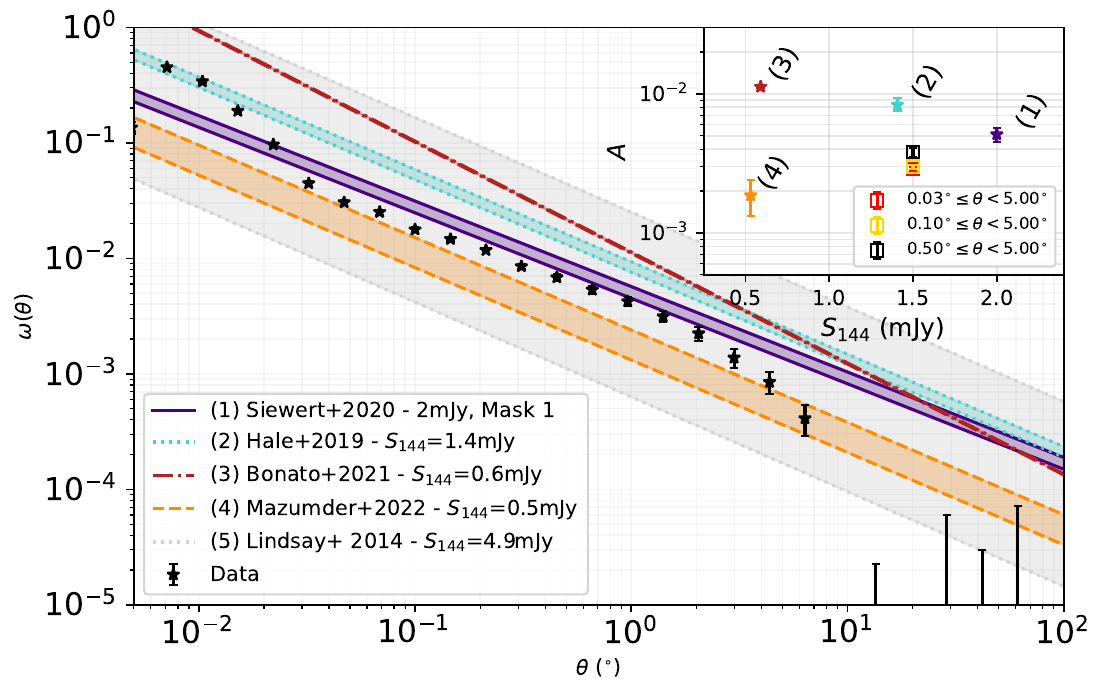}
    \caption{{Comparison of the $\omega(\theta)$ from this work (black stars) compared to previous power law fitting from the studies of \protect \cite{Siewert2020} using a $2\,\mJy$ (solid purple) cut as well as the works of \protect \cite{Hale2019} (turquoise dotted), \protect \cite{Bonato2021} (dark red dot-dashed), \protect \cite{Mazumder2022} (orange dashed) and \protect \cite{Lindsay2014} (light grey dotted). {Inset: Amplitude variation as a function of flux density compared to the fitting here using the simple {$\chi^2$} method across the three fitting ranges: 0.03-5.00$^{\circ}$ (red),  0.10-5.00$^{\circ}$ (yellow) and  0.50-5.00$^{\circ}$ (black). {The quoted flux limits are scaled to 144 MHz to allow more equivalent comparisons.}} }}
    \label{fig:amp_comp}
\end{figure}

In Figure \ref{fig:amp_comp} we present comparisons of the best fit models {to \cite{Siewert2020}} as well as a number of other previous works from \cite{Lindsay2014, Hale2019, Bonato2021} and \cite{Mazumder2022}. {For these} works we include an indication of the equivalent flux limit used, scaled to 144 MHz. For those with fainter populations we note that differences in the populations being observed, which will be increasingly dominated by SFGs below $1\,\mJy$, will affect the comparison of such measurements. As can be seen from Figure \ref{fig:amp_comp}, our work {finds a smaller clustering} amplitude to that found in Mask 1 used in \cite{Siewert2020} at $2\,\mJy$ (their best model from their paper). {We do note that our result} is in excellent agreement to that of \cite{Siewert2020} using their $2\,\mJy$ cut in Mask d {(not shown in Figure \ref{fig:amp_comp})}, which used a less conservative {masking of} regions they considered to have `good' sensitivity. {As discussed though in Section \ref{sec:lims}, there are differences introduced in this work for the method of generating random sources compared to that in \cite{Siewert2020}, which may also affect comparisons of the measurements, as systematics in the data were accounted for using some different methods.} 

{At both similar flux densities and a similar frequency to this work} is the clustering presented in \cite{Hale2019}. In their work, the clustering of sources within the XMM-LSS field as observed with LOFAR was presented, and \cite{Hale2019} found a clustering amplitude approximately three times larger to the work presented here. These difference could arise from cosmic variance {as the XMM-LSS field} covers a much smaller area ($\sim$25 sq. deg) compared to the $\sim$5000 sq. deg used in this work. However, we also note that \cite{Hale2019} discuss the fact that the corrected source counts appear to suggest that the completeness corrections applied are an underestimation. This could affect the measurement of $\omega(\theta)$ in their work. Our work is consistent with that of \cite{Lindsay2014} who study the clustering of sources in FIRST \citep{Becker1995,Helfand2015} with an equivalent limit at 144 MHz of $\sim5\,\mJy$, yet there are large uncertainties in their work. {We derive a larger amplitude} than that of \cite{Mazumder2022}, who use 325 MHz observations of the Lockman Hole field which are the equivalent of $\sim$3$\times$ more sensitive {than for LoTSS-DR2,} but restricted over smaller areas. Whilst {previous} work {has investigated} how the amplitude of clustering changes with flux density \citep[see e.g.][]{Wilman2003, Overzier2003}, who find a typical declining amplitude at smaller flux densities, the complication between the different populations introduced and changes in redshift distribution as flux limits decrease means that discussion of the power law amplitude is complicated to make direct comparisons. {We provide the inset in Figure \ref{fig:amp_comp} to show the flux density dependence in context with the other work presented}.

Next, comparing the bias evolution models implied from this work to those from other works, we note that again there exists challenges when making comparisons due to the variety of radio populations, and their variation with flux density. {Radio surveys} are dominated by AGN at the brightest flux {densities,} with SFG dominating at fainter flux densities \citep[see e.g.][]{Smolcic2017b, Algera2020, Hale2023} and \cite{Best2023}. For example, {\cite{Nusser2015} used a quadratic polynomial model to investigate an evolving bias model for NVSS sources with} $S_{1.4 \textrm{GHz}}\geq2.5\,\mJy$. This is an equivalent flux density limit of $\sim12.5\,\mJy$ at $144\,\MHz$, approximately 8$\times$ the flux density limit used in this work. These sources will be dominated by AGN and have very little contribution of SFG, whereas we expect a much larger contribution of SFGs within this work. As shown in radio clustering studies such as \cite{Magliocchetti2017, Hale2018} and \cite{Mazumder2022}, these two populations are believed to have different biases and so by investigating the bias for a source population as a whole, the bias measured will be {an average} between the bias of the two populations. Moreover, if such previous studies use comoving clustering, these should be compared to the evolving bias models instead of a constant bias model. Therefore the results shown for the Limber derived bias values for comoving clustering in this work are only comparable for the evolving bias model and not the constant model. Our measurements of bias with Limber's equation {(when assuming a power law spatial clustering model)} can {underestimate} the bias model (if comparing to those from {\texttt{CCL}}), though these are typically consistent within 1-2$\sigma$. {The remaining differences {highlight the challenges} when making comparisons of bias evolution models using these different approaches. }.

{Evolving} bias models (with the covariance) are consistent with some of the measured values from \cite{Chakraborty2020} and \cite{Hale2018} as well as the evolving bias model from NVSS \citep{Nusser2015}, especially when the linear model is assumed. We note that whilst for \cite{Hale2018} we present results for the full population in Figure \ref{fig:fit_models_evolve2}, the results for \cite{Chakraborty2020, Mazumder2022} are separated by source type, with those for SFGs found to have lower bias values. Therefore our agreement with \cite{Chakraborty2020} is to their AGN population measurements and similarly, as discussed, NVSS will also be dominated by AGN at the flux densities applied. {Recent work from {\cite{Best2023}} for the LoTSS Deep Fields, suggests $\sim$20\% of SFGs and $\sim$6\% of radio quiet quasars \citep[RQQs, which become more important at faint flux densities, see e.g.][]{Jarvis2004} at the limiting flux density used in this work. } 

{It is also important to compare to the results of \cite{Alonso2021} who used a combination of LoTSS-DR1 and CMB measurements to jointly constrain both $p(z)$ and $b(z)$ (for sources $\geq$2 mJy).} Their results suggested that for an evolving bias model the {value of $b_0$ is} expected to be $\sim$1.2-1.7, {assuming} a redshift distribution similar to that of \cite{Smolcic2017b} using an appropriate flux density cut. Our {measurements over the 0.5-5\degree \ angular fitting range using the full covariance matrix to determine $b_0$ are slightly larger than {the results of \cite{Alonso2021} (when the $p(z)$ samples are equally weighted), though our results are consistent with their upper limits within our 1$\sigma$ uncertainties. However, when we allow more preferential $p(z)$ models to be weighted, we find $b_0\sim1.6-1.7$, consistent with the work of \cite{Alonso2021}. In their work, \cite{Alonso2021} fit for both the $p(z)$ and $b(z)$ model, and so are more comparable to when we allow preferential selection of the $p(z)$ samples.} For the constant bias models, on the other hand, our $b_0$ values are typically lower than those found in {\citep[][who find $b_0\sim 2.3-4$]{Alonso2021}}.} However, their redshift distribution which {they find for such a constant bias} model is skewed to a much higher redshift than shown in Figure \ref{fig:nz_fields}. Our redshift distribution peaks significantly below $z\sim1$, similar to the evolving bias model of \cite{Alonso2021}, whereas their constant bias model predicts a redshift distribution peaking at $z\sim 1-2$. {From Figure \ref{fig:nz_withresamp}, we see that the LoTSS Deep Fields data do not indicate such a peak at higher redshifts. Therefore, to have agreement between this work and that of \cite{Alonso2021} this suggests a preference towards an evolving bias model for LoTSS sources {assuming a redshift distribution similar to that of the LoTSS Deep Fields}.}

\begin{table*}
    
\end{table*}

\section{Conclusions}
\label{sec:conclusions}
The LOFAR Two-metre Sky Survey Data Release 2 \citep[LoTSS-DR2;][]{Shimwell2022} provides a catalogue of $\sim$4.4 million low frequency radio sources over $\sim 5600\,\deg^2$, making it an ideal data set for radio cosmology studies of the large-scale structure of the Universe. In this work we provided analysis of the angular clustering of sources in the LoTSS-DR2 survey and comparison of the bias models implied for such sources. We provide a comprehensive description of the methods used to improve upon the accuracy of the random catalogues generated in this work compared to those used in the LoTSS{-}DR1 clustering analysis of \cite{Siewert2020}. Our random catalogues account for a variety of observational biases within the data including: rms sensitivity variations across the field of view; resolution bias; smearing variations across the observations; detection completeness of \textsc{PyBDSF}; and the effect of Eddington and measurement biases on the measured flux density properties of sources.

Using the random catalogues generated we measure the angular two-point correlation function, $\omega(\theta)$, for sources with SNR$\geq$7.5 and integrated flux density $\geq$$1.5\,\mJy$, which shows an approximate power law behaviour ($\omega(\theta) \propto \theta^{1-\gamma}$) over {the angular scales between 0.03 and 2\degree}. We model $\omega(\theta)$ using a variety of models which account for both an evolving and constant bias model as well as using matter power spectrum models which account for linear effects only (`linear') or with non-linear effects also included (`HaloFit'). Our results show that in order to best model the $\omega(\theta)$ measured from LoTSS-DR2 across a range of angular scales {($\sim0.1-1^{\circ}$), the linear model is {preferred}}, which suggests that at the sensitivities probed by this work, we are typically only observing a single radio source per dark matter halo, and do not have a strong contribution from `1-halo' clustering. However, we note that the {linear model} underestimates the clustering at smaller angular scales, where a combination of 1-halo clustering and multi-component source clustering may play a role.

Comparing bias evolutionary models {with the linear halo model}, assuming the models based on the redshift distributions from the LoTSS Deep Fields accurately represent that of our data, our work suggests that for an evolving bias model of the form $b(z) = b_0/D(z)$, the best fit value of $b_0 \sim 1.7-1.8$ over the angular scales which we believe are most accurate for measuring bias (0.5-5\degree). Instead for a constant bias model, of the form $b(z) = b_0$, we find $b_0 \sim 2.1$. At the largest angles ($\geq 1^{\circ}$), we see that the constant bias model provides a slightly better fit to the observed data {when we use equally weighted $p(z)$ models from the LoTSS Deep fields to measure bias}. {Such differences are reduced if we allow our models to have preferential $p(z)$ models, based on the fit to the data. Where we allow our $p(z)$ model to be preferentially selected, the bias values in both the constant and evolving bias models also reduced slightly, to $b_0 \sim 1.6-1.7$ in an evolving model, and $b_0\sim 2.0$ for a constant model.} Assuming an evolving bias model and taking into account the full covariance matrix, we find good agreement with the results from NVSS of \cite{Nusser2015} up to $z\sim 1$ and previous results from \cite{Hale2018, Chakraborty2020}, though we note that these probe different populations at both different frequencies and different equivalent sensitivities to {that used in} this work. 

{Moreover, in} comparison with work from LoTSS-DR1 of \cite{Alonso2021} who used both CMB and LOFAR measurements to jointly constrain the redshift distribution and bias evolution model of LoTSS-DR1 sources ($\geq$2 mJy), we find that given the greater knowledge of the redshift distributions contributed by the LoTSS Deep Fields \citep[see][]{Sabater2021, Tasse2021, Duncan2021}, an evolving model from \cite{Alonso2021} is necessary to reflect the redshift distribution found in their work. We find that the bias values presented from \cite{Alonso2021} for their evolving model is similar to that of the evolving bias models {presented in this work, especially when we allow $p(z)$ models to be preferentially determined during the fitting process. Using a linear model {for the matter power spectrum} to fit across the largest angular scales (0.5-5\degree) and equally weighting $p(z)$ models from the LoTSS Deep Fields, we find, for an evolving bias model, a value of {$b_{0}= 1.79^{+0.15}_{-0.14}$} which is equivalent to {$b_E= 2.81^{+0.24}_{-0.22}$} at the median redshift of our sample, {$z_\mathrm{m} \approx 0.9$} when we do not show a preference to the $p(z)$ models, reducing to {$b_{0, E}= 1.67^{+0.12}_{-0.12}$} which is equivalent to {$b_E= 2.62^{+0.19}_{-0.18}$} and {$b_{0, C}= 2.02^{+0.17}_{-0.16}$} when we allow our measurements to suggest preferential $p(z)$ models\footnote{{Note, whilst the different $p(z)$ models preferred may result in a different median redshift, we evaluate the bias values at the same redshift (the median suggested by the 1000 $p(z)$ resamples) to allow a consistent comparison between the values.}}, which are found to peak more strongly at lower redshifts.}

{Observations from future spectroscopic surveys such as WEAVE-LOFAR \citep{Smith2016} will allow us to more accurately determine the redshift distribution of LOFAR sources at low redshifts and allow more understanding of the $p(z)$ models we expect for the sources observed in this work. This will allow us to disentangle whether small systematics remain within our data or we have a population of radio sources which are more highly skewed to low redshifts (e.g. from SFGs). {As the low redshift $p(z)$ appears important for this work in modelling $\omega(\theta)$ at the larger angular scales, such accurate redshifts at $z<1$ are important for constraining the results of future studies.} This work has highlighted how a number of observational systematics can be corrected for future deep radio cosmology studies, whilst also demonstrating that the understanding of systematics in wide field mosaiced images is complex, and needs deep understanding for use in cosmological studies.  }

\section*{Acknowledgements}

{We thank the referee for their helpful comments to improve the clarity of this manuscript.} {CLH acknowledges support from the Leverhulme Trust through an Early Career Research Fellowship. PNB and RK are grateful for support from the UK STFC via grant ST/V000594/1. LB acknowledged support of Studienstiftung des Deutschen Volkes. DJS and NB acknowledge support of Deutsche Forschungsgemeinschaft (DFG) grant RTG 1620 'Models of Gravity'.  CSH's work is funded by the Volkswagen Foundation. CSH acknowledges additional support by the Deutsche Forschungsgemeinschaft (DFG, German Research Foundation) under Germany’s Excellence Strategy – EXC 2121 ``Quantum Universe'' – 390833306 and EXC 2181/1 - 390900948 (the Heidelberg STRUCTURES Excellence Cluster). {SJN is supported by the US National Science Foundation (NSF) through grant AST-2108402, and the Polish National Science Centre through grant UMO-2018/31/N/ST9/03975.} MB is supported by the Polish National Science Center through grants no. 2020/38/E/ST9/00395, 2018/30/E/ST9/00698, 2018/31/G/ST9/03388 and 2020/39/B/ST9/03494, and by the Polish Ministry of Science and Higher Education through grant DIR/WK/2018/12. DA acknowledges support from the Beecroft Trust, and from the Science and Technology Facilities Council through an Ernest Rutherford Fellowship, grant reference ST/P004474. MJJ acknowledges support of the STFC consolidated grant [ST/S000488/1] and [ST/W000903/1], from a UKRI Frontiers Research Grant [EP/X026639/1] and the Oxford Hintze Centre for Astrophysical Surveys which is funded through generous support from the Hintze Family Charitable Foundation. JZ acknowledges support by the project ``NRW-Cluster for data intensive radio astronomy: Big Bang to Big Data (B3D)'' funded through the programme ``Profilbildung 2020'', an initiative of the Ministry of Culture and Science of the State of North Rhine-Westphalia. KJD acknowledges funding from the European Union's Horizon 2020 research and innovation programme under the Marie Sk\l{}odowska-Curie grant agreement No. 892117 (HIZRAD) and support from the STFC through an Ernest Rutherford Fellowship (grant number ST/W003120/1).}

LOFAR is the Low Frequency Array designed and constructed by ASTRON. It has observing, data processing, and data storage facilities in several countries, which are owned by various parties (each with their own funding sources), and which are collectively operated by the ILT foundation under a joint scientific policy. The ILT resources have benefited from the following recent major funding sources: CNRS-INSU, Observatoire de Paris and Université d'Orléans, France; BMBF, MIWF-NRW, MPG, Germany; Science Foundation Ireland (SFI), Department of Business, Enterprise and Innovation (DBEI), Ireland; NWO, The Netherlands; The Science and Technology Facilities Council, UK; Ministry of Science and Higher Education, Poland; The Istituto Nazionale di Astrofisica (INAF), Italy. 

This research made use of the Dutch national e-infrastructure with support of the SURF Cooperative (e-infra 180169) and the LOFAR e-infra group. The Jülich LOFAR Long Term Archive and the German LOFAR network are both coordinated and operated by the Jülich Supercomputing Centre (JSC), and computing resources on the supercomputer JUWELS at JSC were provided by the Gauss Centre for Supercomputing e.V. (grant CHTB00) through the John von Neumann Institute for Computing (NIC). This research made use of the University of Hertfordshire high-performance computing facility and the LOFAR-UK computing facility located at the University of Hertfordshire and supported by STFC [ST/P000096/1], and of the Italian LOFAR IT computing infrastructure supported and operated by INAF, and by the Physics Department of Turin university (under an agreement with Consorzio Interuniversitario per la Fisica Spaziale) at the C3S Supercomputing Centre, Italy.

This research made use of a number of tools and python packages: \texttt{Astropy}, community developed core Python package for astronomy \citep{astropy1, astropy2} hosted at \url{http://www.astropy.org/}; \texttt{healpy} \citep{healpy} and \texttt{HEALPix} \citep{healpix} package; \texttt{TOPCAT} \citep{topcat1, topcat2}; \texttt{matplotlib} \citep{matplotlib}; \texttt{NumPy} \citep{numpy1,numpy2}; \texttt{SciPy} \citep{scipy}; \texttt{TreeCorr} \citep{TreeCorr}, {\texttt{tqdm} \citep{tqdm}, \texttt{emcee} \citep{emcee} and \texttt{corner} \citep{corner}.}

For the purpose of open access, the author has applied a Creative Commons Attribution (CC BY) licence to any Author Accepted Manuscript version of this paper.

\section*{Data Availability}

 LoTSS-DR2 catalogue data is available to the community; for more details on the data used in this work please see \cite{Shimwell2022}. LoTSS Deep Fields data is also available publicly, with further information available in \cite{Sabater2021, Tasse2021, Kondapally2021, Duncan2021}. Further information can also be found at \url{https://lofar-surveys.org/index.html}. Other data presented in this article can be made available upon reasonable request to the author.



\bibliographystyle{mnras}
\bibliography{lotss-dr2-clustering} 








\bsp	
\label{lastpage}
\end{document}